\pdfoutput=1

\documentclass[10pt,letterpaper]{book}

\usepackage[top=1in, bottom=1.25in, left=1.25in, right=1.25in]{geometry}

\usepackage[utf8]{inputenc}
\usepackage{graphicx}
\usepackage{rotating}
\usepackage{lscape}
\usepackage{booktabs}
\usepackage{adjustbox}

\usepackage{amsmath}
\numberwithin{equation}{section}
\usepackage{caption}

\usepackage{amsthm}

\usepackage{xcolor}
\usepackage{multirow}

\newtheoremstyle{examplestyle}
  {} 
  {} 
  {\addtolength{\leftskip}{1.5em}} 
  {} 
  {\bfseries} 
  { } 
  {\newline} 
  {} 

\newtheoremstyle{commentstyle}
  {} 
  {} 
  {\addtolength{\leftskip}{1.5em}} 
  { } 
  {\bfseries} 
  { } 
  {.5em} 
  {} 

\theoremstyle{examplestyle}
\newtheorem{myexample}{EXAMPLE}[section]

\theoremstyle{commentstyle}
\newtheorem{mycomment}{Comment}

\usepackage{hyperref}
\usepackage{natbib}

\author{W.J. Conover, Texas Tech University, Lubbock, Texas \and V. G. Tercero-G\'omez,Tecnol\'ogico de Monterrey M\'exico \and A. E. Cordero-Franco, Universidad Aut\'onoma de Nuevo Le\'on,M\'exico}
\title{An Approach to Statistical Process Control that is New, Nonparametric, Simple, and Powerful}
\begin{document}

\frontmatter

\maketitle

\chapter*{Abstract}
To maintain the desired quality of a product or service it is necessary to monitor the process that results in the product or service. This monitoring method is called Statistical Process Management, or Statistical Process Control. It is in widespread usage in industry. Extensive statistical methodology has been developed to make it possible to detect when a process goes out of control, while allowing for natural variability that occurs when the process is in control. This paper introduces nonparametric methods for monitoring data, whether it is univariate or multivariate, and whether the interest is in detecting a change of location or scale or both. These methods, based on sequential normal scores, are much simpler than the most popular nonparametric methods currently in use and have good power for detecting out of control observations. Sixteen new statistical tests are introduced for the first time in this paper, with 17 examples, 33 tables and 48 figures to complete the instructions for their application.
\mainmatter
\tableofcontents
\chapter{Introduction}\label{ch:c1}

The methods discussed in this paper are new nonparametric methods based on {\em sequential normal scores} (SNS), designed for sequences of observations, usually time series data, which may occur singly or in batches, and may be univariate or multivariate. These methods are designed to detect changes in the process, which may occur as changes in {\em location} (mean or median), changes in {\em scale} (standard deviation, or variance), or other changes of interest in the distribution of the observations, over the time observed. They usually apply to large data sets, so computations need to be simple enough to be done in a reasonable time on a computer, and easily updated as each new observation (or batch of observations) becomes available. Most methods are too complex to be computed by hand or even on a hand calculator, but can be easily computed using \textsc{Microsoft Excel}, \textsc{R}, or other popular software programs.

The methods are selected on the basis of ability to detect changes ({\em statistical power}) as well as relative simplicity to apply. In most cases they are compared with some of the best available existing nonparametric methods, as illustrated in the excellent reference book by \citet{qiu_2014}. Many examples are from the data sets available free of charge from Peihua Qiu’s home web page \url{http://users.phhp.ufl.edu/pqiu/research/book/spc/data/}.

Because sequential normal scores behave in many ways as standard normal random variables, many parametric statistical methods designed for standard normal random variables are easily adapted to become nonparametric methods, as shown in the following cases. Also the need for special tables, or simulation programs for the computer, to obtain limits is eliminated for many of these procedures based on SNS, because tables based on normal random variables may be used. Many of the best nonparametric methods involve ranks, which means that each time new observations are obtained all of the observations, new and old, need to be re-ranked, which is time-consuming. Sequential normal scores are based on sequential ranks \citep{parent_1965}, which involves ranking only the new observations as they are obtained, and all of the previous observations retain their original sequential ranks, thus saving much computing time.

Methods based on sequential ranks, such as by \citet{mcdonald_1990}, are awkward to use, which may be why they never became popular. However, a simple transformation to scores that behave like standard normal random variables was proposed by \citet{conover_etal_2017}, which enabled existing normal-theory methods to be used. Advantages of sequential rank methods over ordinary rank methods are detailed and illustrated by \citet{conover_etal_2018}. The purpose of this paper is to illustrate the wide range of opportunities for expanding the use of sequential normal scores in the field of {\em statistical process management} (SPM), also known as {\em statistical process control} (SPC). The encyclopedic book by \citet{qiu_2014} covers the wide range of statistical procedures developed for SPC, both parametric and nonparametric. This paper relies heavily on the book by \citet{qiu_2014} and on more recent papers by \citet{conover_etal_2017, conover_etal_2018} to present some of the many ways sequential normal scores can result in convenient, yet powerful, nonparametric methods for SPC.

\section{Sequential Normal Scores}\label{sec:c1s1}
The data being analyzed by SPC methods consist of batches of independent observations distributed according to an unknown distribution function $F(x)$, $n$ in each batch, denoted by $\{X_{i,1}, X_{i,2}, \ldots , X_{i,n}\}$, for $i = \{1, 2, \ldots \}$. The batch size $n$ can be as small as 1, representing individual observations in sequence, or larger than 1, representing groups of observations, called batches. The batches do not need to be all the same size, but for simplicity in presentation we will assume all the batches are the same size except possibly for the first batch, which may be regarded as a {\em reference sample} or sometimes called {\em in-control sample}, or {\em Phase 1}. If the Phase 1 sample consists of several batches, they may be treated as several batches in sequence, or as one large batch, whichever is more convenient. If there are no observations known to be {\em in control} then these methods are {\em self starting} with the first batch.

The observations in the first batch are considered the Phase 1 batch, the {\em in control} batch, or the reference batch. They can be ranked relative to the other observations in that batch, but it is not necessary for them to be ranked, for reasons explained later. They are the reference observations against which the later observations receive ranks. For individual observations (not in batches) the same instructions are followed, but the batch size is $n=1$, so the first observation simply gets $\text{rank} = 1$.

All observations in subsequent batches, after the first batch, are ranked relative to only the previous batches, not to other observations in the subsequent batch. The rank of the $j^\text{th}$ observation in the $i^\text{th}$ batch is denoted by $R_{i,j}$ and is called the {\em sequential rank} of $X_{i,j}$. The number of observations involved in the ranking of the $i^\text{th}$ batch is denoted by $N_i$, and equals the number of observations in the previous $(i-1)$ batches, plus one for the observation being ranked in the $i^\text{th}$ batch.

The {\em rankits} $P_{i,j}$ are obtained by dividing $(R_{i,j} – 0.5)$ by the appropriate number of observations involved in the ranking, $N_i$. These rankits are converted to pseudo normal random variables by taking the inverse of the standard normal distribution, $Z_{i,j} = \Phi^{-1}(P_{i,j})$. This $Z_{i,j}$ is the sequential normal score SNS of $X_{i,j}$. It can be used to graph the sequence, and behaves approximately like a standard normal random variable.

However with a batch of $n$ observations, it is more convenient to graph the statistic $Z_i = \sum_j Z_{i,j}/\sqrt{n}$ which is also approximately standard normal. It is always true that $Z_1 = 0$, and that is why it is not necessary to rank the observations in the first batch among themselves. In our examples we will usually display the ranks of the first batch for completeness, although it is not required for the statistical methods that are introduced in this work.

It is shown in  \citet{conover_etal_2017} that for $i>1$ the sequential ranks $R_{i,j}$ are mutually independent random variables, and therefore so are the sequential normal scores $Z_{i,j}$ and the summary statistics $Z_i$.

\section{Conditional Sequential Normal Scores}\label{sec:c1s2}
The data consist of batches of independent observations distributed according to an unknown distribution function $F(x)$, $n$ in each batch, denoted by $\{X_{i,1}, X_{i,2}, \ldots , X_{i,n}\}$, for $i=\{1, 2, \ldots \}$ just as in the previous section, except one piece of information is known, or assumed to be known, regarding the unknown distribution function $F(x)$. The target $p^\text{th}$ quantile $\theta$, is given, so $F(\theta) = p$ is all that is known about $F(x)$, such as when $\theta$ is the median (known) and thus $p = 0.5$.

The observations in the first batch are ranked relative to the other observations on the same side of $\theta$. That is, observations less than or equal to $\theta$ are compared only with the other observations less than or equal to $\theta$, and observations greater than $\theta$ are compared only with the other observations greater than $\theta$.

Each observation in subsequent batches, after the first batch, is ranked relative to only observations in the previous batches on the same side of $\theta$ as the new observation. That is, new observations less than or equal to $\theta$ are ranked relative to only observations less than or equal to  $\theta$ in all previous batches, and observations greater than $\theta$ are ranked relative to only observations greater than $\theta$ in all previous batches. The rank of the $j^\text{th}$ observation in the $i^\text{th}$ batch is denoted by R$_{i,j|\theta}$, and is called the {\em conditional sequential rank of $X_{i,j}$ given $\theta$}.

The number of observations involved in the ranking of the $i^\text{th}$ batch is denoted by $N_i^{-}$ for the observations less than or equal to $\theta$, and by $N_i^{+}$ for the observations greater than $\theta$. The {\em conditional rankits} $P_{i,j|\theta}$ are obtained by dividing $(R_{i,j|\theta} – 0.5)$ by the appropriate number of observations involved in the ranking, $N_i^{-}$ or $N_i^{+}$. These conditional rankits are converted to {\em unconditional rankits} $P_{i,j}$ by using
\begin{equation}\label{eq:p1}
    P_{i,j} = 
    \begin{cases}
      F(\theta) P_{i,j|\theta}, & \text{if the observation on } X_{i,j} \leq \theta, \\
      F(\theta) + (1-F(\theta))P_{i,j|\theta}, & \text{if the observation on } X_{i,j} > \theta,
    \end{cases}
\end{equation}

Then the unconditional rankits are converted to pseudo normal random variables by taking the inverse of the standard normal distribution, $Z_{i,j} = \Phi^{-1}(P_{i,j})$. This $Z_{i,j}$ is the conditional sequential normal score of $X_{i,j}$. It can be used to graph the sequence, and behaves approximately like a standard normal random variable. However with a batch of $n$ observations, it is more convenient to graph the statistic $Z_i = \sum_j Z_{i,j} / \sqrt{n}$ which is also approximately standard normal. It is shown in  \citet{conover_etal_2017} that for $i>1$ the conditional sequential ranks $R_{i,j|\theta}$ are mutually independent random variables, and therefore so are the conditional sequential normal scores $Z_{i,j}$ and the summary statistics $Z_i$.
\chapter{Methods for Improving Sensitivity}\label{ch:c2}

\section{The Advantage of Using Batches}\label{sec:c2s1}

Sequential normal scores may be used to monitor single observations in a time series. However they are basically rank-based statistics, so they depend on a reference sample to obtain their ranks. In a Shewhart chart \citep{shewhart_1931} the bounds are frequently set at plus or minus 3 sigma relative to the mean. Sequential normal scores have a mean of zero and a standard deviation of about 1, so the bounds in SNS charts are often plus 3 and minus 3. In order for a new observation to exceed those limits, it must be the largest or the smallest of 371 observations, so the reference sample must be at least 370 observations in length. Reference samples are not always this large, so alternatives are available for determining if an observation is {\em out of control}.

One alternative is to gather the data in batches, or to group the observations already obtained into batches. The sum of sequential normal scores of several new observations divided by the square root of the number of observations is also approximately standard normal, and it can exceed the Shewhart limits with a much smaller reference sample. For example, the sum of two SNS values, divided by the square root of 2, can exceed the $\pm 3.0$ limits with a reference sample as small as 29. A batch of three observations can exceed the $\pm 3.0$ limits with only 12 prior observations, or four previous batches of 3 each, and so forth. This is a distinct advantage when monitoring a short sequence of observations. Table \ref{tab:c2s1t1} gives the minimum number of observations required for various batch sizes $m$ and exceedance probabilities $p$. In SPC it is customary to use the average run length between exceedances $(ARL)$, which equals $1/p$ in a Shewhart type control chart, as a measure of the ability of the chart to detect out-of-control observations, so these also appear in Table \ref{tab:c2s1t1}.
\begin{table}[t!]
    \centering
     \caption{The minimum number of observations in the reference sample needed to have significance at the level $p$ and an {\em Average Run Length} (ARL), with a batch size $m$. The upper and lower bounds in a Shewhart Chart are also given for each $p$ for SNS.}
    \begin{tabular}{|r|r|r|r|r|r|r|r|}
    \hline
    ARL   & 20    & 50    & 100   & 200   & 370   & 500   & $1\,000$ \\ \hline
    p     & 0.0500 & 0.0200 & 0.0100 & 0.0050 & 0.0027 & 0.0020 & 0.0010 \\ \hline
    bounds & $\pm1.9600$ & $\pm2.3263$ & $\pm2.5758$ & $\pm2.8070$ & $\pm3.0000$ & $\pm3.0902$ & $\pm3.2905$ \\ \hline
    \multicolumn{1}{|c|}{$m$} &       &       &       &       &       &       &  \\ \hline
    1     & 20    & 50    & 99    & 200   & 370   & 500   & $1\,000$ \\
    2     & 6     & 10    & 14    & 21    & 29    & 34    & 50 \\
    3     & 3     & 5     & 7     & 9     & 12    & 13    & 17 \\
    4     & 3     & 4     & 5     & 6     & 7     & 8     & 10 \\
    5     & 2     & 3     & 4     & 4     & 5     & 5     & 7 \\
    6     & 2     & 2     & 3     & 3     & 4     & 4     & 5 \\
    7     & 2     & 2     & 3     & 3     & 3     & 4     & 4 \\
    8     & 2     & 2     & 2     & 3     & 3     & 3     & 4 \\
    9     & 1     & 2     & 2     & 2     & 3     & 3     & 3 \\
    10    & 1     & 2     & 2     & 2     & 2     & 3     & 3 \\
    11    & 1     & 2     & 2     & 2     & 2     & 2     & 3 \\
    12    & 1     & 1     & 2     & 2     & 2     & 2     & 2 \\
    13    & 1     & 1     & 2     & 2     & 2     & 2     & 2 \\
    14    & 1     & 1     & 2     & 2     & 2     & 2     & 2 \\
    15    & 1     & 1     & 1     & 2     & 2     & 2     & 2 \\
    16    & 1     & 1     & 1     & 2     & 2     & 2     & 2 \\
    17    & 1     & 1     & 1     & 2     & 2     & 2     & 2 \\
    18    & 1     & 1     & 1     & 1     & 2     & 2     & 2 \\
    19    & 1     & 1     & 1     & 1     & 2     & 2     & 2 \\
    20    & 1     & 1     & 1     & 1     & 1     & 2     & 2 \\
    21    & 1     & 1     & 1     & 1     & 1     & 1     & 2 \\
    22    & 1     & 1     & 1     & 1     & 1     & 1     & 2 \\
    23    & 1     & 1     & 1     & 1     & 1     & 1     & 2 \\
    24    & 1     & 1     & 1     & 1     & 1     & 1     & 1 \\
    $>$24    & 1     & 1     & 1     & 1     & 1     & 1     & 1 \\ \hline
    \end{tabular}%
    \label{tab:c2s1t1}
\end{table}

\section{Cumulative Sums}\label{sec:c2s2}

The previous method concentrates on identifying individual batches that indicate a change in the location. It may not be sensitive to a slight but persistent change in location, a change in location so slight that individual batches are not able to detect the change, but cumulative batches may be able to detect a more subtle change. It is a simple adjustment to adapt the {\em cumulative sum} (CUSUM) procedure of \citet{page_1961} to the SNS method of analysis.

Using a tabular CUSUM scheme (just CUSUM from here), the CUSUM procedure for sequential normal scores uses two statistics, one to detect upward trends, and one to detect downward trends. Because sequential normal scores have a mean of zero and a standard deviation of 1 the CUSUM statistics are given by $C^{+}$ and $C^{-}$, defined as
\begin{align}
    C_i^{+}  =  \max(0, C_{i-1}^+ + Z_i – k) \text{ for } i > 0,    & & C_0^+ = 0,
\end{align}
and
\begin{align}
C_i^{-}  =  \min(0, C_{i-1}^- + Z_i + k) \text{ for } i > 0,    & & C_0^- = 0,
\end{align}
for a suitably chosen {\em allowance} factor $k$. For example, if we choose $k=0.5$, which is a popular choice, this results in upper and lower limits of about 4.095 to achieve an $\text{ARL}=370$, the same value obtained with an exceedance probability of 0.0027 associated with 2-sided $3\sigma$-Shewhart charts, for independent standard normal random variables (see Table 4.1 on page 130 of \citet{qiu_2014}).  Other bounds are given in Table \ref{tab:c2s2t1}, copied from Table 4.1 on page 130 of \citet{qiu_2014}, for different popular values of $k$ and ARL. These are upper bounds $U$. Lower bounds $L$ are simply the negative of these numbers, that is, $L=-U$. All in-control ARL values used in CUSUM charts in this paper correspond to zero-state ARL calculations, i.e., the average number of observations until the first signal when the CUSUM is initialized at zero.

\begin{table}[t!]
\caption{Upper bounds for CUSUMs for a given allowance parameter $k$, to achieve a target zero-state ARL for standard normal random variables. This table is given on page 130 of \citet{qiu_2014} as his Table 4.1. Lower bounds are simply the negative of the upper bound. They can be used to obtain approximate bounds on a sequence of SNS, or $Z_i$.}
    \centering
    \begin{tabular}{|r|r|r|r|r|r|r|r|}
    \cline{2-8}
    \multicolumn{1}{c|}{} & \multicolumn{7}{|c|}{$k$} \\
\hline      ARL& \multicolumn{1}{|c|}{0.10}  & \multicolumn{1}{|c|}{0.25}  & \multicolumn{1}{|c|}{0.50}  & \multicolumn{1}{|c|}{0.75}  & \multicolumn{1}{|c|}{1.00}  & \multicolumn{1}{|c|}{1.25}  & \multicolumn{1}{|c|}{1.50} \\
    \hline
    50    & 4.567 & 3.340 & 2.225 & 1.601 & 1.181 & 0.854 & 0.570 \\
    100   & 6.361 & 4.418 & 2.849 & 2.037 & 1.532 & 1.164 & 0.860 \\
    200   & 8.520 & 5.597 & 3.501 & 2.481 & 1.874 & 1.458 & 1.131 \\
    300   & 9.943 & 6.324 & 3.892 & 2.745 & 2.073 & 1.624 & 1.282 \\
    370   & 10.722 & 6.708 & 4.095 & 2.882 & 2.175 & 1.709 & 1.359 \\
    400   & 11.019 & 6.852 & 4.171 & 2.933 & 2.214 & 1.741 & 1.387 \\
    500   & 11.890 & 7.267 & 4.389 & 3.080 & 2.323 & 1.830 & 1.466 \\
    $1\,000$  & 14.764 & 8.585 & 5.071 & 3.538 & 2.665 & 2.105 & 1.708 \\ \hline
    \end{tabular}%

    \label{tab:c2s2t1}
\end{table}

It is possible for a CUSUM to exceed its upper or lower bound for smaller reference sample sizes than in a sequence without using the CUSUM. For example, even though it takes a reference sample of at least 370 observations for a new observation to have a SNS greater than 3 in absolute value, a CUSUM using $k = 0.5$ can exceed its bound of 4.095 with a sequentially evaluated reference sample of only 7 observations. That is, if each observation is smaller than all previous observations in a sequence of 7 observations, the $8^\text{th}$ observation, if it is smaller than the previous 7 observations, has a CUSUM of $-4.956$, which is less than $-4.095$. Thus a tabular CUSUM chart can detect a trend in location changes long before a simple Shewhart chart is able to. Table 2.3 gives the minimum reference sample size necessary to have a signal in a CUSUM, for various values of the allowance parameter $k$ and the average run length ARL.

\begin{table}[t!]
\caption{The minimum reference sample size necessary to have a signal in a tabular CUSUM chart of SNS, for various values of the allowance parameter $k$ and the corresponding average run length ARL as given in Table 2.2. A batch size of 1 is assumed.}
    \centering
    \begin{tabular}{|r|r|r|r|r|r|r|r|}
    \hline
          & \multicolumn{7}{|c|}{$k$} \\
\cline{2-8}    \multicolumn{1}{|c|}{ARL} & 0.1   & 0.25  & 0.5   & 0.75  & 1     & 1.25  & 1.5 \\
    \hline
    50    & 5     & 5     & 5     & 5     & 6     & 8     & 11 \\
    100   & 7     & 6     & 5     & 6     & 7     & 9     & 12 \\
    200   & 8     & 7     & 6     & 6     & 8     & 10    & 13 \\
    300   & 9     & 7     & 6     & 7     & 8     & 10    & 14 \\
    370   & 9     & 8     & 7     & 7     & 8     & 10    & 14 \\
    400   & 10    & 8     & 7     & 7     & 8     & 10    & 14 \\
    500   & 10    & 8     & 7     & 7     & 8     & 10    & 14 \\
    1000  & 12    & 9     & 8     & 8     & 9     & 11    & 15 \\
    \hline
    \end{tabular}
    \label{tab:c2s2t2}
\end{table}

\section{Exponentially Weighted Moving Averages}\label{sec:c2s3}

Like the CUSUM method, the {\em exponentially weighted moving average} (EWMA) method was designed to detect small persistent changes in location, as opposed to large changes by individual batches. It is a simple adjustment to adapt the EWMA procedure to the SNS method of analysis.

Unlike the CUSUM procedure for sequential normal scores which uses two statistics, the EWMA method uses one moving average to detect location changes in both directions. Because the sequential normal scores charting statistic $Z_i$ has a mean of zero and a standard deviation of 1, the EWMA is given by
\begin{align}
E_i  =  \lambda Z_i + (1-\lambda)E_{i-1} \text{ for } i>0,
\end{align}
and
\begin{align}
E_0 = 0,
\end{align}
for a suitably chosen smoothing constant $\lambda$. This applies equally well to individual observations as to the graphing statistics in batches. Steady state upper and lower control limits $U$ and $L$ depend on $\lambda$ and another parameter $\rho$, which govern the value of the ARL, in the steady state equations $U = \rho \sqrt{(\lambda/(2-\lambda))}$, and $L = -U$. See Table 5.1 on page 188 of \citet{qiu_2014} for some choices of $\lambda$ and $\rho$ to obtain various ARL values for standard normal random variables. In Table \ref{tab:c2s3t1}, the values in Table 5.1 on page 188 of \citet{qiu_2014} have been converted to Shewhart chart bounds for sequential normal scores. Table \ref{tab:c2s3t1} gives only the upper bound $U$ for a desired choice of $\lambda$ and average run length ARL, but by symmetry the lower bound $L$ is simply the negative of the tabled value, that is, $L=-U$.

\begin{table}[t!]
\caption{The steady state upper bound $U$ of an exponentially weighted moving average EWMA for a desired choice of $\lambda$ and corresponding zero-state ARL in a sequence of standard normal random variables. These are approximate limits for SNS. By symmetry the lower bound $L$ is simply the negative of the tabled value, that is, $L=-U$.}
    \centering
    \begin{tabular}{rrrrrrrrr}
    \hline
          & \multicolumn{8}{c}{$\lambda$} \\
\cline{2-9}    \multicolumn{1}{c}{ARL} & 0.01  & 0.05  & 0.10  & 0.20  & 0.30  & 0.40  & 0.50  & 0.75 \\
    \hline
    50    & 0.060 & 0.243 & 0.415 & 0.685 & 0.910 & 1.115 & 1.309 & 1.793 \\
    100   & 0.082 & 0.301 & 0.493 & 0.787 & 1.030 & 1.252 & 1.463 & 1.989 \\
    200   & 0.106 & 0.355 & 0.563 & 0.878 & 1.140 & 1.377 & 1.603 & 2.170 \\
    300   & 0.121 & 0.384 & 0.601 & 0.928 & 1.199 & 1.445 & 1.681 & 2.270 \\
    370   & 0.129 & 0.399 & 0.620 & 0.953 & 1.229 & 1.480 & 1.719 & 2.321 \\
    400   & 0.132 & 0.404 & 0.627 & 0.962 & 1.239 & 1.492 & 1.733 & 2.339 \\
    500   & 0.140 & 0.419 & 0.646 & 0.987 & 1.270 & 1.527 & 1.773 & 2.391 \\
    1000  & 0.163 & 0.462 & 0.702 & 1.062 & 1.360 & 1.632 & 1.892 & 2.548 \\
    \hline
    \end{tabular}  
    \label{tab:c2s3t1}
\end{table}

The EWMA, like the CUSUM, is also capable of exceeding its bound with a much smaller reference sample than with individual observations in a Shewhart chart. For example, to exceed a bound of 3.0 for individual observations, which has an $\text{ARL} = 370$, a EWMA using $\lambda = 0.1$ can exceed its bound of 0.620 with a reference sample of only 7 sequentially evaluated observations. That is, if each observation is smaller than all previous observations, the $8^\text{th}$ observation has a $\text{EWMA} = -0.659$, which exceeds the lower bound given in Table 2.4 as $-0.620$. Thus the EWMA can declare significance with a reference sample as small as 7. Other minimum reference sample sizes are given in Table 2.5.

\begin{table}[t!]
    \caption{The minimum number of observations in a reference sample to achieve statistical significance for a EWMA with parameter $\lambda$ and target average run length ARL using the bounds given in Table 2.4. These apply to a EWMA on SNS as well as on $Z_i$.}
    \centering
    \begin{tabular}{rrrrrrrrr}
    \hline
          & \multicolumn{8}{c}{$\lambda$} \\
\cline{2-9}    \multicolumn{1}{c}{ARL} & 0.01  & 0.05  & 0.10  & 0.20  & 0.30  & 0.40  & 0.50  & 0.75 \\ \hline
    50    & 6     & 5     & 5     & 5     & 5     & 5     & 6     & 14 \\
    100   & 7     & 6     & 6     & 6     & 6     & 5     & 8     & 21 \\
    200   & 9     & 7     & 7     & 6     & 6     & 7     & 10    & 33 \\
    300   & 10    & 8     & 7     & 7     & 7     & 8     & 11    & 43 \\
    370   & 11    & 8     & 7     & 7     & 7     & 9     & 12    & 49 \\
    400   & 11    & 8     & 7     & 7     & 7     & 9     & 13    & 53 \\
    500   & 11    & 8     & 7     & 7     & 8     & 9     & 14    & 59 \\
    1000  & 13    & 9     & 8     & 8     & 9     & 11    & 18    & 92 \\
    \hline
    \end{tabular}
    \label{tab:c2s3t2}
\end{table}

In some cases the interest in a sequence of observations is to detect a change in spread, or variance, or standard deviation. In the parametric framework this usually involves the $(z_{\text{score}})^2 = (X-\mu)^2/\sigma^2$ which has a chi-squared distribution for $N(\mu, \sigma^2)$ random variables. Sequential normal scores have a mean of zero and a variance close to 1.0 so $\text{SNS}^2$ is very close to a chi-squared random variable with 1 degree of freedom. Therefore $\text{SNS}^2$ is a natural statistic to use for nonparametric tests of spread.

\citet{qiu_2014} developed a table for using the EWMA on random variables with a $\chi^2$ distribution with 1 degree of freedom, using computer simulation. It is Table 5.4 found on page 202 in his book, and is reproduced in Table 2.6. The center line in a Shewhart chart is the mean which is 1.0, and the steady state upper and lower bounds are given by $U = 1 + \rho_U\sqrt{(2\lambda/(2-\lambda))}$ for the upper bound $U$, and $L = 1 - \rho_L \sqrt{(2\lambda/(2-\lambda))}$ for the lower bound. The values of $\rho_U$ and $\rho_L$ are not equal to each other as before because of the skewed nature of a chi-squared random variable. The values of $\rho_U$ and $\rho_L$ are given in Table 2.6 for selected values of $\lambda$ and ARL. The last two entries were not given in the table, hence the \textsc{NA}.

\begin{table}[t!]
    \caption{Values of the parameters $\rho_U$ and $\rho_L$ needed to find the upper limit and lower limit of a EWMA with the selected values of $\lambda$ and average run length ARL for a chi-squared random variable with 1 degree of freedom. These can be used as approximate values for SNS$^2$ also. This table is taken from Table 5.4 in \citet{qiu_2014}.}
    \centering
    \begin{tabular}{rrrrrrrrr}
    \hline
          & \multicolumn{4}{c}{$\rho_U$ for values of lambda} & \multicolumn{4}{c}{$\rho_L$ for values of lambda} \\
\cline{2-9}    \multicolumn{1}{l}{ARL} & 0.050 & 0.100 & 0.200 & 0.300 & 0.050 & 0.100 & 0.200 & 0.300 \\
    \hline
    50    & 0.901 & 1.380 & 1.916 & 2.259 & 0.865 & 1.100 & 1.195 & 1.166 \\
    100   & 1.455 & 1.988 & 2.606 & 2.996 & 1.201 & 1.366 & 1.360 & 1.273 \\
    200   & 2.017 & 2.595 & 3.258 & 3.702 & 1.510 & 1.580 & 1.480 & 1.349 \\
    300   & 2.342 & 2.944 & 3.655 & 4.132 & 1.670 & 1.682 & 1.538 & 1.384 \\
    370   & 2.518 & 3.133 & 3.854 & 4.354 & 1.746 & 1.731 & 1.563 & 1.401 \\
    400   & 2.588 & 3.199 & 3.935 & 4.440 & 1.774 & 1.750 & 1.574 & 1.407 \\
    500   & 2.796 & 3.419 & 4.184 & 4.722 & 1.862 & 1.808 & 1.605 & 1.426 \\
    1000  & 5.122 & 5.822 & 7.788 & 8.467 & 2.569 & 2.452 & NA    & NA \\
   \hline
   \end{tabular}
    \label{tab:c2s3t3}
\end{table}

For the reader’s convenience the steady state upper and lower bounds are computed based on the numbers given in Table 2.6, and appear in Table 2.7. That is, for a EWMA using $\lambda = 0.1$ and for a desired $\text{ARL} = 370$ the upper bound for $\text{SNS}^2$ is 2.016 and the lower bound is 0.438.

\begin{table}[t!]
  \caption{Some steady state upper limits and lower limits for a EWMA computed on a chi-squared random variable with 1 degree of freedom, for selected values of the parameter $\lambda$ and target average run length ARL. These limits are approximate limits for a EWMA on SNS$^2$.}
    \centering
    \begin{tabular}{rrrrrrrrr}
    \hline
          & \multicolumn{4}{c}{$\lambda$ for upper EWMA limits} & \multicolumn{4}{c}{$\lambda$ for lower EWMA limits} \\
\cline{2-9}    \multicolumn{1}{c}{ARL} & \multicolumn{1}{c}{0.05} & \multicolumn{1}{c}{0.10} & \multicolumn{1}{c}{0.20} & \multicolumn{1}{c}{0.30} & \multicolumn{1}{c}{0.05} & \multicolumn{1}{c}{0.10} & \multicolumn{1}{c}{0.20} & \multicolumn{1}{c}{0.30} \\
    \hline
    50    & 1.204 & 1.448 & 1.903 & 2.342 & 0.804 & 0.643 & 0.437 & 0.307 \\
    100   & 1.329 & 1.645 & 2.228 & 2.780 & 0.728 & 0.557 & 0.359 & 0.244 \\
    200   & 1.457 & 1.842 & 2.536 & 3.199 & 0.658 & 0.487 & 0.302 & 0.199 \\
    300   & 1.530 & 1.955 & 2.723 & 3.455 & 0.622 & 0.454 & 0.275 & 0.178 \\
    370   & 1.570 & 2.016 & 2.817 & 3.587 & 0.605 & 0.438 & 0.263 & 0.168 \\
    400   & 1.586 & 2.038 & 2.855 & 3.638 & 0.598 & 0.432 & 0.258 & 0.164 \\
    500   & 1.633 & 2.109 & 2.972 & 3.805 & 0.578 & 0.413 & 0.243 & 0.153 \\
    1000  & 2.160 & 2.889 & 4.671 & 6.030 & 0.418 & 0.204 & NA    & NA \\
    \hline
    \end{tabular}
   
    \label{tab:c2s3t4}
\end{table}

As with batches and CUSUMs, the use of EWMAs makes it easier to detect changes than by using individual observations. For example, to cross a 3-sigma bound in either direction of a Shewhart chart using sequential normal scores requires a reference set of at least 370 prior observations. Similarly, to cross an upper bound for SNS$^2$, which is $9 = 3^2$, also requires at least 370 prior observations. However, for a EWMA to cross the bound for $\lambda=0.1$, for $\text{ARL}=370$, given in Table 2.7 as 2.016, only 15 prior observations are required, as seen in Table 2.8. Similarly, crossing the lower bound for SNS$^2$, which is given in Table 2.7 for $\lambda=0.1$ and $\text{ARL}=370$ as 0.438, requires only 9 previous observations. Thus the EWMA is able to detect changes in spread much sooner than by using a Shewhart chart on individual SNS$^2$ values.

\begin{table}[t!]
    \caption{The minimum number of observations in a reference sample to achieve statistical significance for a EWMA with parameter $\lambda$ and target average run length ARL. These apply to a EWMA on SNS$^2$.}
    \centering
    \begin{tabular}{rrrrrrrrr}
    \hline
          & \multicolumn{4}{c}{$\lambda$, upper bound} & \multicolumn{4}{c}{$\lambda$, lower bound} \\
\cline{2-9}    \multicolumn{1}{c}{$\text{ARL}_0$} & \multicolumn{1}{c}{0.05} & \multicolumn{1}{c}{0.1} & \multicolumn{1}{c}{0.2} & \multicolumn{1}{c}{0.3} & \multicolumn{1}{c}{0.05} & \multicolumn{1}{c}{0.1} & \multicolumn{1}{c}{0.2} & \multicolumn{1}{c}{0.3} \\
    \hline
    50    & 14    & 11    & 10    & 10    & 5     & 5     & 5     & 5 \\
    100   & 15    & 12    & 11    & 13    & 7     & 7     & 6     & 5 \\
    200   & 16    & 13    & 13    & 16    & 9     & 8     & 7     & 6 \\
    300   & 17    & 14    & 14    & 18    & 10    & 9     & 7     & 6 \\
    370   & 17    & 15    & 15    & 19    & 11    & 9     & 7     & 6 \\
    400   & 18    & 15    & 15    & 20    & 11    & 9     & 7     & 6 \\
    500   & 18    & 15    & 16    & 22    & 12    & 10    & 8     & 7 \\
    1000  & 23    & 21    & 36    & 73    & 18    & 16    & NA    & NA \\
    \hline
    \end{tabular}

    \label{tab:c2s3t5}
\end{table}
\chapter{Detecting a Change in Location}\label{ch:c3}

\section{Using Sequential Normal Scores to Detect a Change in Location}\label{sec:c3s1}
A Shewhart chart of sequences of standard normal random variables is often matched against upper and lower bounds, such as $+3$ for an upper bound and $-3$ for a lower bound. If the graph exceeds either bound it sends a {\em signal} that the process generating the graph may be {\em out of control}, since the probability is very small (0.0027) of a standard normal random variable exceeding those bounds. The number of observations between two consecutive signals is called a {\em run length} (RL), and the {\em average run length} (ARL) is often used as a proxy for the probability of exceeding those bounds. The ARL for independent observations is the reciprocal of the exceedance probability, so the ARL for 3-sigma bounds of normal random variables is $370 = 1/0.0027$.

\begin{myexample}\label{ex:c3s1e1}
As an example, consider the data set used in Example 8.2 on page 323 of \citet{qiu_2014}. Actually Example 8.2 assumes a reference sample of 1000 {\em in control} observations is available, against which 30 batches of size 5 are compared. We do not have access to the 1000 {\em in control} observations, but they are not needed to illustrate this method. So our data set consists only of 30 batches of observations where each batch has 5 observations. \citet{qiu_2014} describes these as a {\em Phase II batch data with a batch size of $m=5$} and compares them with the reference sample of $1\,000$ observations to illustrate a nonparametric test for location proposed by \citet{chakraborti_etal_2009}. The data are given in Table \ref{tab:c3t1}.

\begin{table}[t!]
    \caption{The data used by \citet{qiu_2014} in his Example 8.2, and also used in Example 3.1.1 in this section. The data consist of 30 batches with 5 observations in each batch.}
    \centering
    \begin{tabular}{rrrrrr}
    \multicolumn{1}{l}{Batch} &       &       &       &       &  \\
    1     & -0.623 & -1.068 & 0.605 & -0.002 & -0.807 \\
    2     & -0.105 & -0.037 & -0.595 & 1.222 & -0.545 \\
    3     & -0.264 & -0.668 & -0.443 & -0.638 & -0.749 \\
    4     & -0.981 & -0.056 & 0.231 & -0.940 & 1.825 \\
    5     & -0.454 & 1.527 & -0.946 & -0.830 & -0.309 \\
    6     & 1.025 & -0.821 & -1.128 & 0.087 & 1.941 \\
    7     & -0.710 & -0.293 & -0.39 & 1.164 & -0.539 \\
    8     & 0.938 & -0.590 & -0.111 & -1.038 & -1.103 \\
    9     & 0.221 & 0.945 & 0.402 & -1.045 & -0.391 \\
    10    & -0.774 & 0.816 & -0.662 & -0.306 & -0.538 \\
    11    & -0.982 & -0.961 & -1.057 & 2.157 & 2.545 \\
    12    & -0.672 & -0.255 & -0.86 & -0.313 & -0.487 \\
    13    & 3.071 & -1.175 & 0.160  & -0.353 & 0.109 \\
    14    & -0.735 & -0.555 & 1.852 & 0.857 & -0.056 \\
    15    & -0.666 & -0.591 & -1.163 & 0.720  & 3.184 \\
    16    & -0.415 & 0.315 & 1.304 & -0.255 & 9.987 \\
    17    & -0.562 & -0.861 & 1.872 & -1.078 & 1.023 \\
    18    & -0.788 & -1.031 & -1.101 & -0.725 & -0.516 \\
    19    & -1.162 & -0.217 & -0.941 & -0.943 & -1.081 \\
    20    & 1.287 & -1.071 & -0.624 & -0.298 & -0.726 \\
    21    & 1.926 & 4.347 & 1.38  & 1.018 & 1.506 \\
    22    & 1.162 & 1.437 & 1.208 & 1.618 & 1.758 \\
    23    & 1.052 & 2.737 & 3.096 & 1.816 & 1.927 \\
    24    & 2.149 & 2.880  & 1.339 & 1.753 & 1.151 \\
    25    & 6.248 & 1.748 & 3.521 & 1.743 & 2.203 \\
    26    & 2.804 & 0.933 & 3.250  & 2.100 & 1.961 \\
    27    & 3.005 & 0.785 & 3.131 & 2.800 & 1.348 \\
    28    & 1.940  & 2.562 & 1.286 & 1.353 & 1.822 \\
    29    & 1.096 & 2.553 & 1.064 & 2.519 & 1.501 \\
    30    & 1.387 & 1.600   & 1.372 & 4.486 & 1.614 \\
    \hline
    \end{tabular}
    \label{tab:c3t1}
\end{table}
The sample medians of the batches are graphed in Figure \ref{fig:c3s1f1} as part of the examination of the data.
\begin{figure}[t!]
    \centering
    \includegraphics[width=0.75\textwidth,angle=-90]{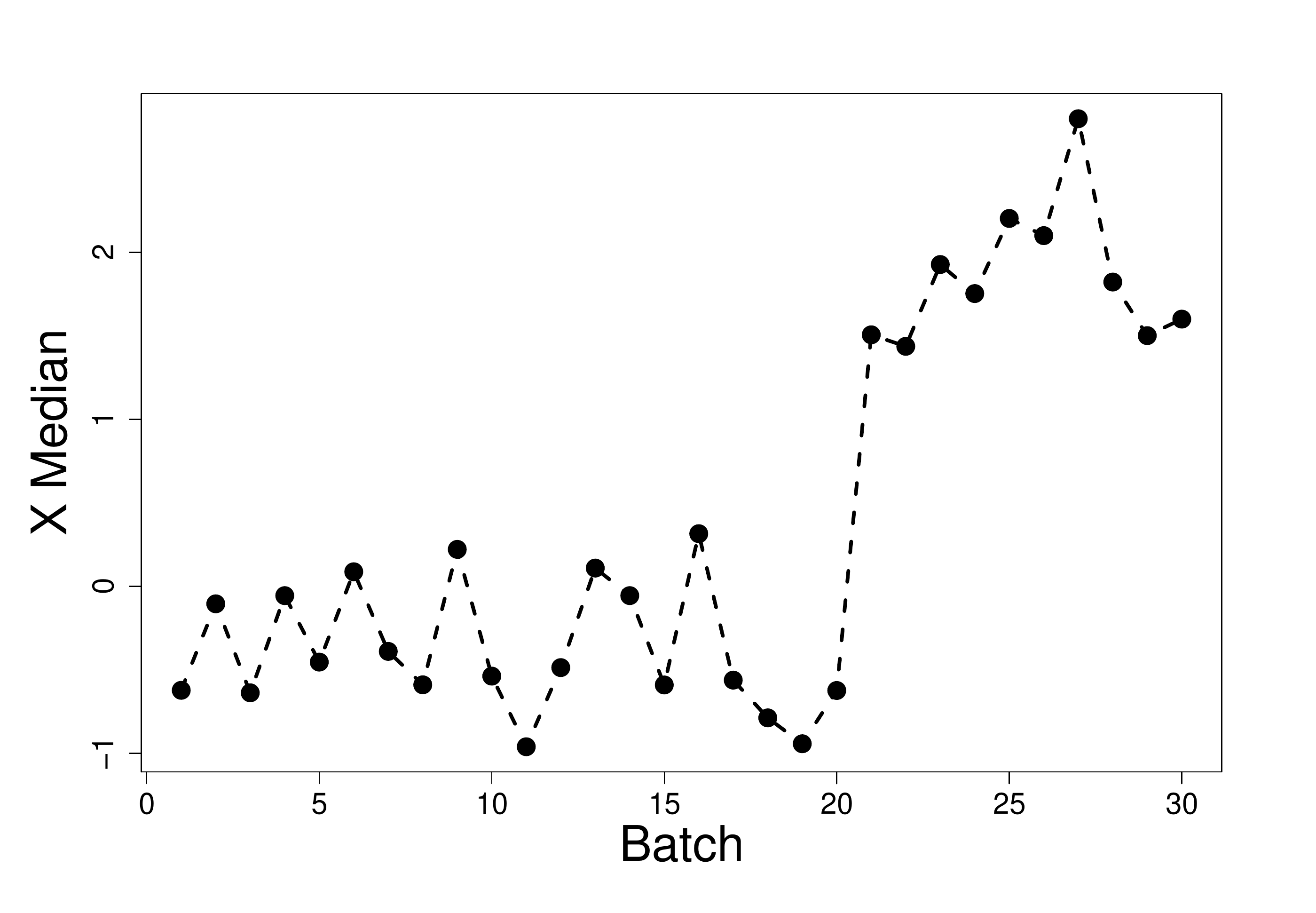}
    \caption{The medians of the observations in the 30 batches in Table 3.1.}
    \label{fig:c3s1f1}
\end{figure}
The data appear to have a negative location parameter until batch 21, at which point it appears to increase. The distribution function of the data is unknown.

Because the reference sample of 1,000 observations is not available, we will illustrate the SNS method in its self-starting mode. The data are ranked as described in Section \ref{sec:c2s1}. The first batch is ranked among itself, and the observations in subsequent batches are ranked against all observations in previous batches. Note that sequential ranks are unchanged when subsequent observations are obtained. The sequential ranks for the data in Table 3.1 are given in Table 3.2.
\begin{table}[t!]
    \caption{The sequential ranks computed on the data in Table \ref{tab:c3t1}. The column on the right is the sample size used in computing the sequential ranks in that row.}
    \centering
    \begin{tabular}{rrrrrrr}
    \multicolumn{1}{c}{Sample} & \multicolumn{5}{c}{Sequential Ranks}  & \multicolumn{1}{c}{N} \\
    1     & 3     & 1     & 5     & 4     & 2     & 5 \\
    2     & 4     & 4     & 4     & 6     & 4     & 6 \\
    3     & 6     & 3     & 6     & 3     & 3     & 11 \\
    4     & 2     & 12    & 14    & 2     & 16    & 16 \\
    5     & 11    & 20    & 3     & 4     & 12    & 21 \\
    6     & 23    & 6     & 1     & 21    & 26    & 26 \\
    7     & 10    & 18    & 17    & 27    & 15    & 31 \\
    8     & 30    & 15    & 23    & 3     & 2     & 36 \\
    9     & 32    & 35    & 33    & 4     & 22    & 41 \\
    10    & 12    & 38    & 15    & 26    & 21    & 46 \\
    11    & 6     & 7     & 4     & 51    & 51    & 51 \\
    12    & 18    & 35    & 12    & 31    & 27    & 56 \\
    13    & 61    & 1     & 46    & 34    & 46    & 61 \\
    14    & 19    & 27    & 62    & 55    & 44    & 66 \\
    15    & 23    & 27    & 2     & 57    & 71    & 71 \\
    16    & 38    & 58    & 68    & 46    & 76    & 76 \\
    17    & 31    & 14    & 75    & 5     & 68    & 81 \\
    18    & 20    & 10    & 5     & 23    & 38    & 86 \\
    19    & 3     & 57    & 16    & 16    & 6     & 91 \\
    20    & 85    & 9     & 37    & 57    & 30    & 96 \\
    21    & 95    & 100   & 91    & 85    & 91    & 101 \\
    22    & 88    & 93    & 89    & 95    & 95    & 106 \\
    23    & 88    & 107   & 108   & 100   & 104   & 111 \\
    24    & 108   & 111   & 95    & 100   & 89    & 116 \\
    25    & 120   & 102   & 119   & 102   & 113   & 121 \\
    26    & 118   & 83    & 122   & 113   & 113   & 126 \\
    27    & 123   & 81    & 125   & 121   & 98    & 131 \\
    28    & 115   & 122   & 96    & 100   & 110   & 136 \\
    29    & 91    & 126   & 91    & 125   & 104   & 141 \\
    30    & 105   & 109   & 104   & 144   & 109   & 146 \\
\end{tabular}

    \label{tab:c3t2}
\end{table}

The sequential normal scores are computed by taking the inverse of the standard normal distribution on their $\text{rankits} = (\text{rank} – 0.5)/N_i$ . The sequential normal scores for the data in Table 3.1 are given in Table \ref{tab:c3t3}.

\begin{table}[t!]
    \caption{The sequential normal scores SNS computed on the data in Table \ref{tab:c3t1}, using the sequential ranks in Table \ref{tab:c3t2}. The column on the right is the charting statistic $Z_i$, the sum of the SNS in that row, divided by the square root of the batch size 5.}
    \centering
    \begin{tabular}{crrrrrr}
    Batch & \multicolumn{5}{c}{Sequential normal scores (SNS)} & \multicolumn{1}{c}{$Z(\text{SNS})$} \\
    1     & 0.000 & -1.282 & 1.282 & 0.524 & -0.524 & 0.000 \\
    2     & 0.210 & 0.210 & 0.210 & 1.383 & 0.210 & 0.995 \\
    3     & 0.000 & -0.748 & 0.000 & -0.748 & -0.748 & -1.003 \\
    4     & -1.318 & 0.579 & 1.010 & -1.318 & 1.863 & 0.365 \\
    5     & 0.000 & 1.465 & -1.180 & -0.967 & 0.120 & -0.251 \\
    6     & 1.105 & -0.801 & -2.070 & 0.801 & 2.070 & 0.494 \\
    7     & -0.506 & 0.162 & 0.081 & 1.057 & -0.081 & 0.319 \\
    8     & 0.913 & -0.246 & 0.319 & -1.480 & -1.732 & -0.995 \\
    9     & 0.733 & 1.000 & 0.816 & -1.370 & 0.061 & 0.555 \\
    10    & -0.674 & 0.897 & -0.481 & 0.137 & -0.137 & -0.116 \\
    11    & -1.238 & -1.139 & -1.486 & 2.334 & 2.334 & 0.360 \\
    12    & -0.489 & 0.295 & -0.823 & 0.112 & -0.067 & -0.434 \\
    13    & 2.400 & -2.400 & 0.662 & 0.124 & 0.662 & 0.647 \\
    14    & -0.582 & -0.249 & 1.489 & 0.938 & 0.410 & 0.897 \\
    15    & -0.476 & -0.323 & -2.031 & 0.827 & 2.455 & 0.202 \\
    16    & -0.016 & 0.695 & 1.217 & 0.250 & 2.479 & 2.068 \\
    17    & -0.315 & -0.967 & 1.403 & -1.593 & 0.967 & -0.226 \\
    18    & -0.750 & -1.224 & -1.623 & -0.638 & -0.161 & -1.966 \\
    19    & -1.919 & 0.308 & -0.953 & -0.953 & -1.551 & -2.267 \\
    20    & 1.176 & -1.350 & -0.305 & 0.224 & -0.504 & -0.339 \\
    21    & 1.519 & 2.174 & 1.259 & 0.981 & 1.259 & 3.217 \\
    22    & 0.936 & 1.139 & 0.974 & 1.235 & 1.235 & 2.468 \\
    23    & 0.800 & 1.744 & 1.859 & 1.261 & 1.494 & 3.202 \\
    24    & 1.452 & 1.670 & 0.895 & 1.070 & 0.716 & 2.595 \\
    25    & 2.245 & 0.990 & 2.040 & 0.990 & 1.474 & 3.461 \\
    26    & 1.495 & 0.398 & 1.803 & 1.242 & 1.242 & 2.764 \\
    27    & 1.515 & 0.291 & 1.649 & 1.404 & 0.657 & 2.467 \\
    28    & 1.002 & 1.245 & 0.531 & 0.618 & 0.860 & 1.903 \\
    29    & 0.363 & 1.227 & 0.363 & 1.190 & 0.625 & 1.685 \\
    30    & 0.570 & 0.653 & 0.550 & 2.117 & 0.653 & 2.032 \\
    \end{tabular}

    \label{tab:c3t3}
\end{table}

In the right-hand column of Table \ref{tab:c3t3} is the graphing statistic $Z_i = \sum_j Z_{i,j}/\sqrt{5}$, which is also approximately standard normal in distribution. The graph of $Z_i$ is given in Figure 3.2. The first signal of a change in location parameter appears with sample 21, when $Z_i$ exceeds 3.0 in absolute value.

\begin{figure}[t!]
    \centering
    \includegraphics[width=0.75\textwidth,angle=-90]{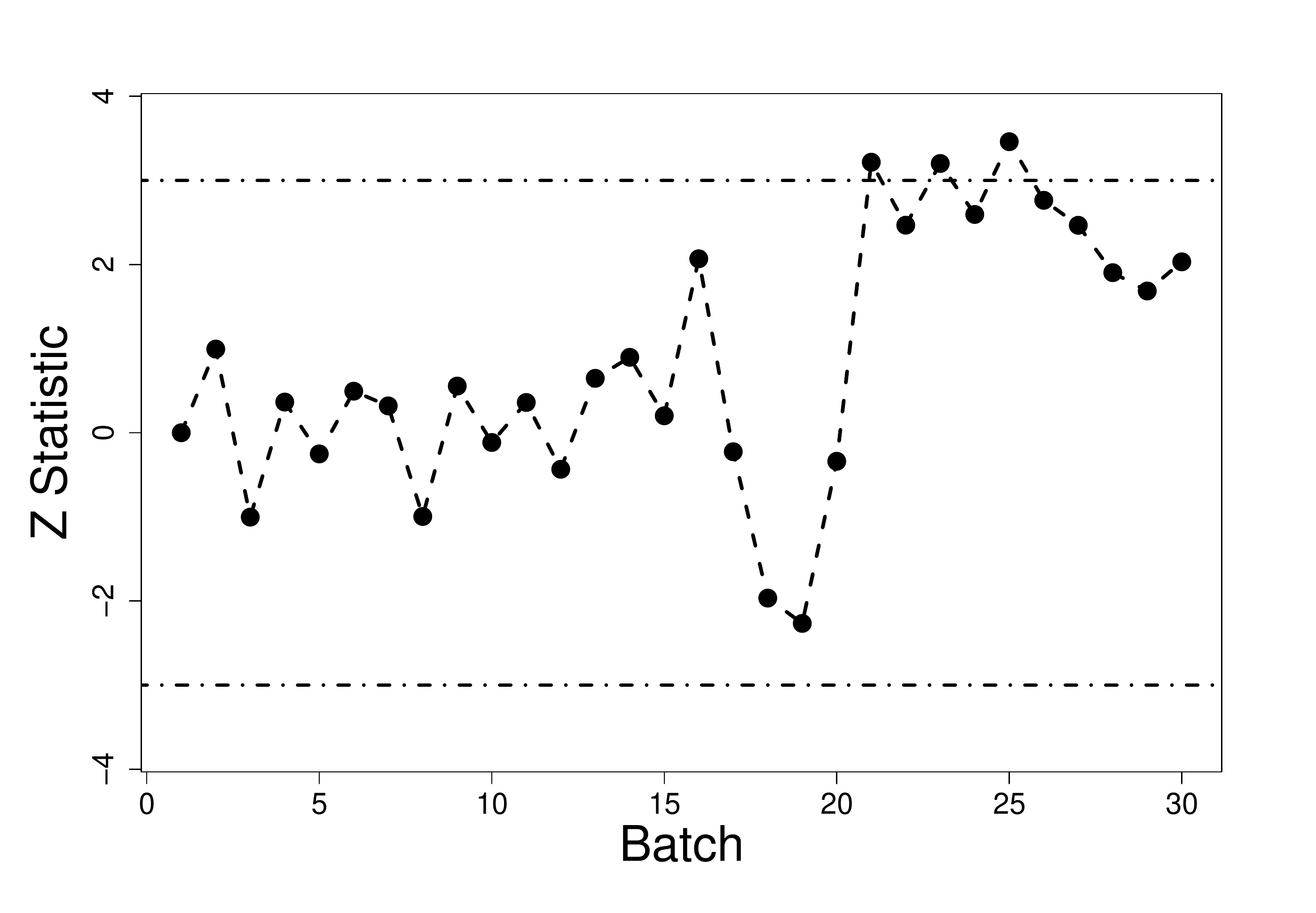}
    \caption{The graphing statistic Z for the SNS in Table 3.3 of the data in Table 3.1, showing a change in location with batch 21.}
    \label{fig:c3s1f2}
\end{figure}

This example illustrates the effect of mixing {\em out of control} batches with {\em in control} batches. Batch 22 is compared with the first 21 batches, which includes one batch that has been declared out-of-control. Normally one would compare each batch with only the batches that were in control, so normally the analyst would stop adding batches to the reference observations, and compare future batches with only the first 20 batches. This is easily done with sequential normal scores. The new charting statistic remains the same as the original one until sample 22, at which it changes because it is being compared with only the first 20 batches. The result is illustrated in Figure \ref{fig:c3s1f3}, which shows the increased power obtained by not mixing out-of-control batches with in-control batches.

\begin{figure}[t!]
    \centering
    \includegraphics[width=0.75\textwidth,angle=-90]{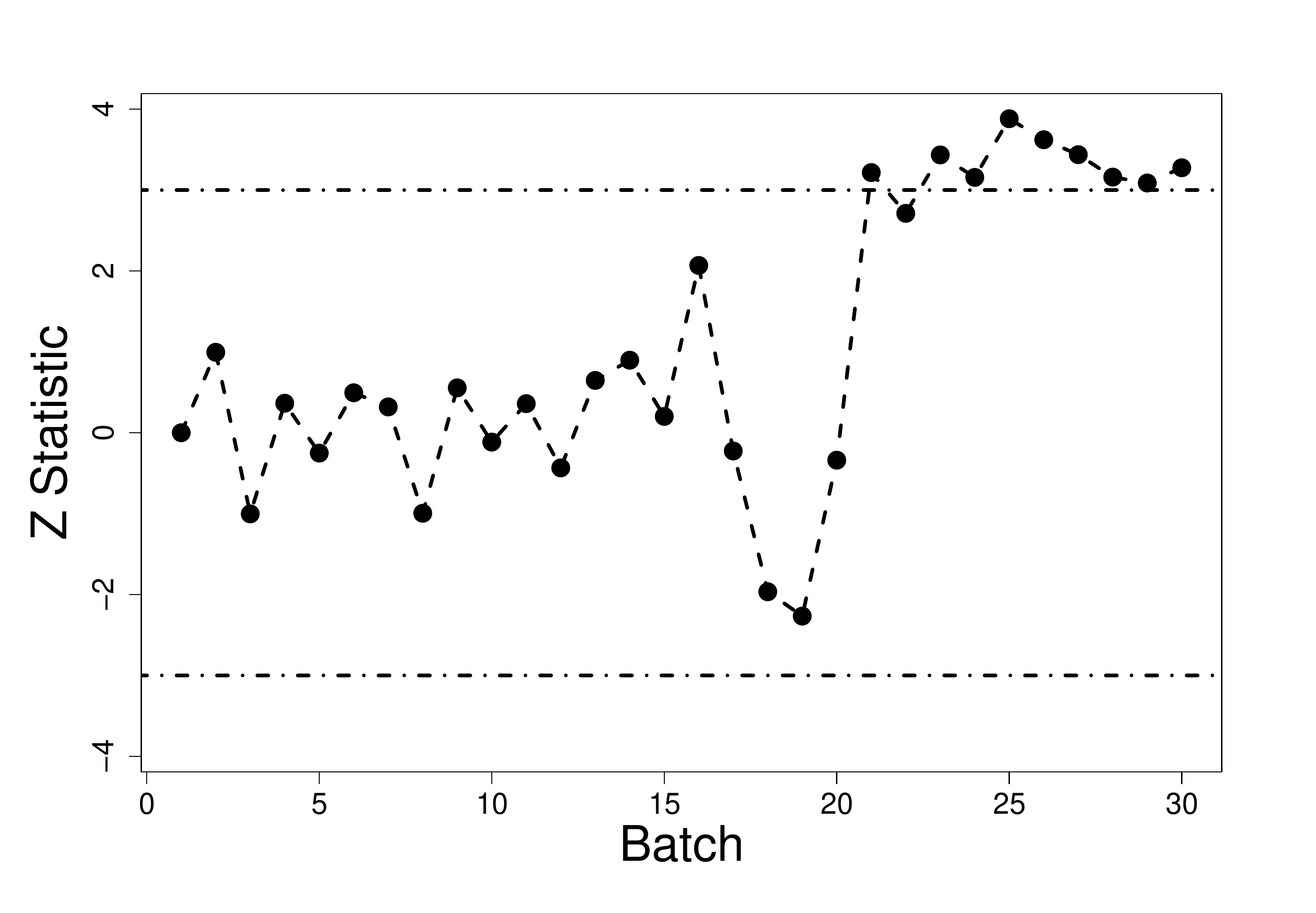}
    \caption{The graphing statistic $Z_i$ for the SNS of the data in Table 3.1, using only the first 20 batches as a reference sample.}
    \label{fig:c3s1f3}
\end{figure}

For comparison purposes the two graphs can be compared with Figure 8.3(b) on page 325 in \citet{qiu_2014}. That graph illustrates a nonparametric method called the {\em distribution-free precedence} (DFP) method proposed by \citet{chakraborti_etal_2004}. It uses the same data used in this example, plus 1,000 observations in a reference set that is not used in this example. That method exceeded the equivalent of our 3-sigma limit on the 23$^\text{rd}$ batch, and 5 of the last 8 batches. So there is good agreement between the DFP method and the method illustrated in this example, even though the DFP method used a 1,000 observation reference set, and this method did not.
\end{myexample}

\section{CUSUM Variation to Detect a Change in Location}\label{sec:c3s2}

In Example \ref{ex:c3s1e1}, individual batches were analyzed in a case where the raw data showed a rather extreme change in location. Example \ref{ex:c3s2e1} looks at a case where the change in location is more subtle, lending itself to the CUSUM approach discussed in Section \ref{sec:c2s2}, which accumulates values as it progresses in time.

\begin{myexample}\label{ex:c3s2e1}
To illustrate this procedure we will consider 30 batches of size $n=5$, from Example 8.4 of \citet{qiu_2014}. The data appear in Table \ref{tab:c3s2t1}, along with the median of each batch.

\begin{table}[t!]
    \caption{The data used in Example 8.4 in \citet{qiu_2014}, and also used in the example is this section. The data consist of 30 batches of observations with 5 observations in each batch. The right hand column contains the median of each batch.}
    \centering
    \begin{tabular}{crrrrrr}
    Batch & \multicolumn{5}{c}{Observations}      & \multicolumn{1}{c}{Median} \\
    1     & -0.393 & -0.685 & 0.360 & 0.148 & 0.867 & 0.148 \\
    2     & -0.552 & -0.462 & -0.979 & -0.580 & -0.008 & -0.552 \\
    3     & -0.908 & -0.489 & 1.081 & -0.850 & -0.577 & -0.577 \\
    4     & -0.150 & -0.397 & 0.393 & -0.009 & 0.647 & -0.009 \\
    5     & -0.059 & -1.095 & 0.078 & -0.273 & 0.772 & -0.059 \\
    6     & -1.097 & 0.574 & -0.721 & 0.219 & -1.119 & -0.721 \\
    7     & -0.548 & -0.250 & -0.945 & 2.411 & 0.309 & -0.250 \\
    8     & -0.604 & 1.609 & -0.688 & -1.089 & -0.901 & -0.688 \\
    9     & -0.151 & -0.355 & 0.887 & -0.688 & 0.521 & -0.151 \\
    10    & -0.683 & -0.311 & 0.303 & -0.557 & -0.344 & -0.344 \\
    11    & 0.597 & 1.670 & -0.754 & -0.122 & 0.662 & 0.597 \\
    12    & 0.309 & -0.937 & 0.724 & 0.016 & 0.310 & 0.309 \\
    13    & -0.334 & -0.912 & -0.302 & -0.981 & 0.939 & -0.334 \\
    14    & -1.002 & -0.105 & -0.892 & -1.193 & 0.553 & -0.892 \\
    15    & 0.228 & 0.580 & -1.040 & -0.509 & -0.754 & -0.509 \\
    16    & 1.557 & -0.323 & -0.158 & 0.901 & -1.193 & -0.158 \\
    17    & -0.407 & -0.239 & 1.006 & -0.912 & -1.093 & -0.407 \\
    18    & -0.902 & 1.091 & -0.956 & 0.647 & 0.182 & 0.182 \\
    19    & -0.866 & 2.270 & -0.722 & -0.009 & 0.802 & -0.009 \\
    20    & -0.436 & 0.057 & -0.797 & 3.871 & -0.921 & -0.436 \\
    21    & 0.322 & 0.784 & 0.359 & 0.687 & 0.789 & 0.687 \\
    22    & 1.565 & 1.102 & 1.468 & 1.607 & 2.692 & 1.565 \\
    23    & 0.215 & 0.365 & 1.163 & 0.130 & 1.947 & 0.365 \\
    24    & 2.689 & 2.182 & 0.203 & 0.641 & 0.277 & 0.641 \\
    25    & 1.810 & 2.726 & -0.141 & 0.070 & 6.091 & 1.810 \\
    26    & 2.808 & 0.766 & 0.328 & 3.424 & 0.969 & 0.969 \\
    27    & 3.909 & 0.056 & 3.154 & 1.238 & 0.952 & 1.238 \\
    28    & 0.490 & 0.749 & 0.362 & 0.325 & -0.037 & 0.362 \\
    29    & 0.665 & 1.592 & 0.548 & -0.052 & 4.201 & 0.665 \\
    30    & 0.304 & 3.285 & 0.630 & 1.034 & -0.063 & 0.630 \\
    \end{tabular}
    \label{tab:c3s2t1}
\end{table}

An exploratory graph of the batch medians is given in Figure \ref{fig:c3s2f1}. It shows an increase in the batch median values over time, possibly at batch 11, but more likely around batch 21 or 22.

\begin{figure}[t!]
    \centering
    \includegraphics[width=0.75\textwidth,angle=-90]{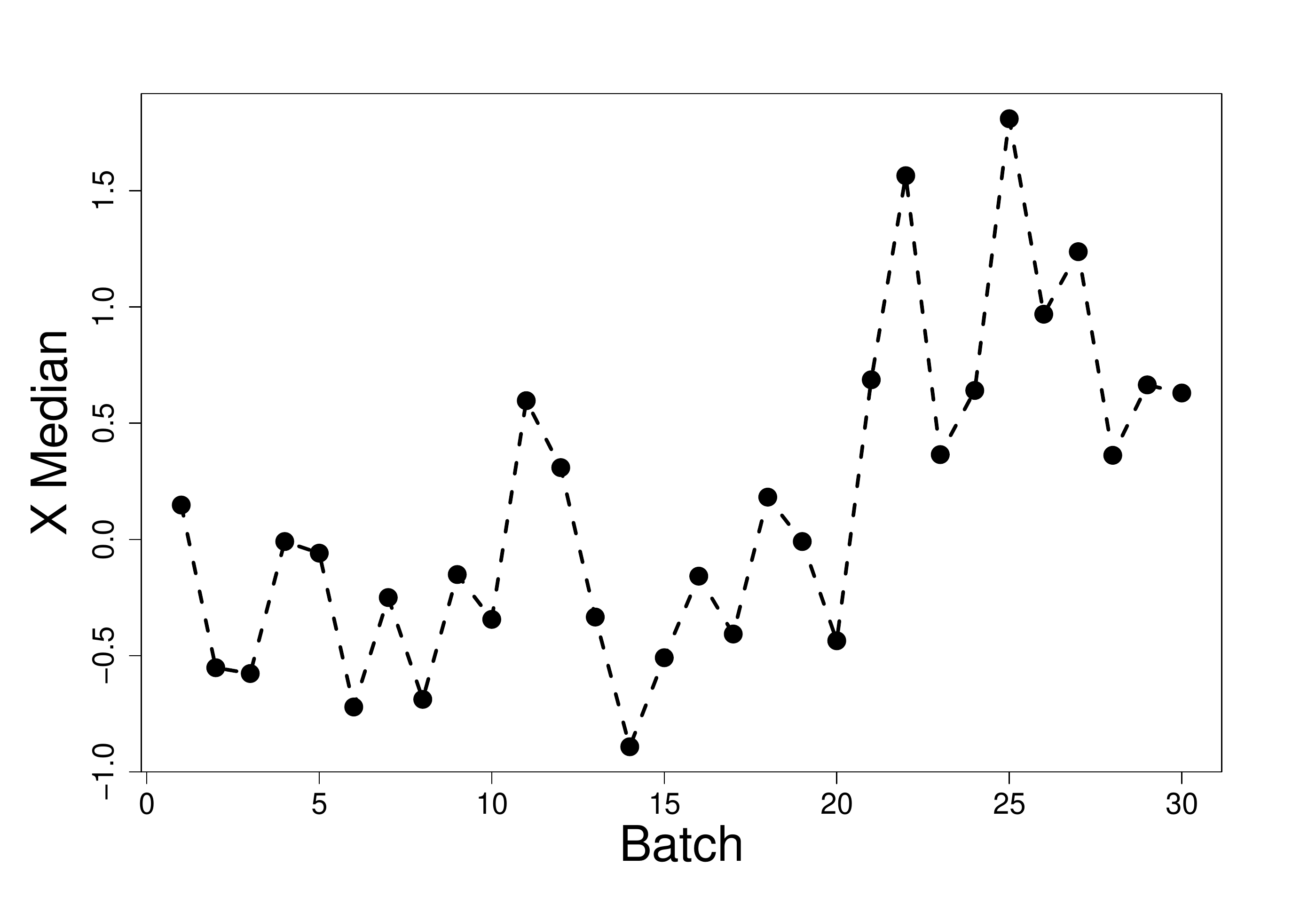}
    \caption{The medians for the 30 batches of data in Table 3.4.}
    \label{fig:c3s2f1}
\end{figure}
In example 8.4 of \citet{qiu_2014} there were also 100 reference values preceding the data in Table 3.4. Then each batch was ranked relative to the 100 reference values, and the batch rank sum was used as a test statistic. The CUSUM approach was used, with $k=0.5$ and $\pm4.389$ as the bounds to achieve an $\text{ARL} = 500$. The first signal was obtained at batch 23. We will compare this with our method, which applies the same CUSUM parameters to sequential normal scores.

We do not have access to the 100 reference values used by \citet{qiu_2014}, so we are using the self-starting method outlined in Section \ref{sec:c2s1}, and illustrated in detail in the previous section. First sequential ranks are obtained, then rankits, and finally the inverse of the standard normal distribution function is applied to get sequential normal scores SNS. A graph of $Z_i=\sum_jZ_{i,j}/\sqrt{5}$, given in Figure \ref{fig:c3s2f2}, shows the same behavior as the graph in Figure 3.4 of the batch medians.

\begin{figure}[t!]
    \centering
    \includegraphics[width=0.75\textwidth,angle=-90]{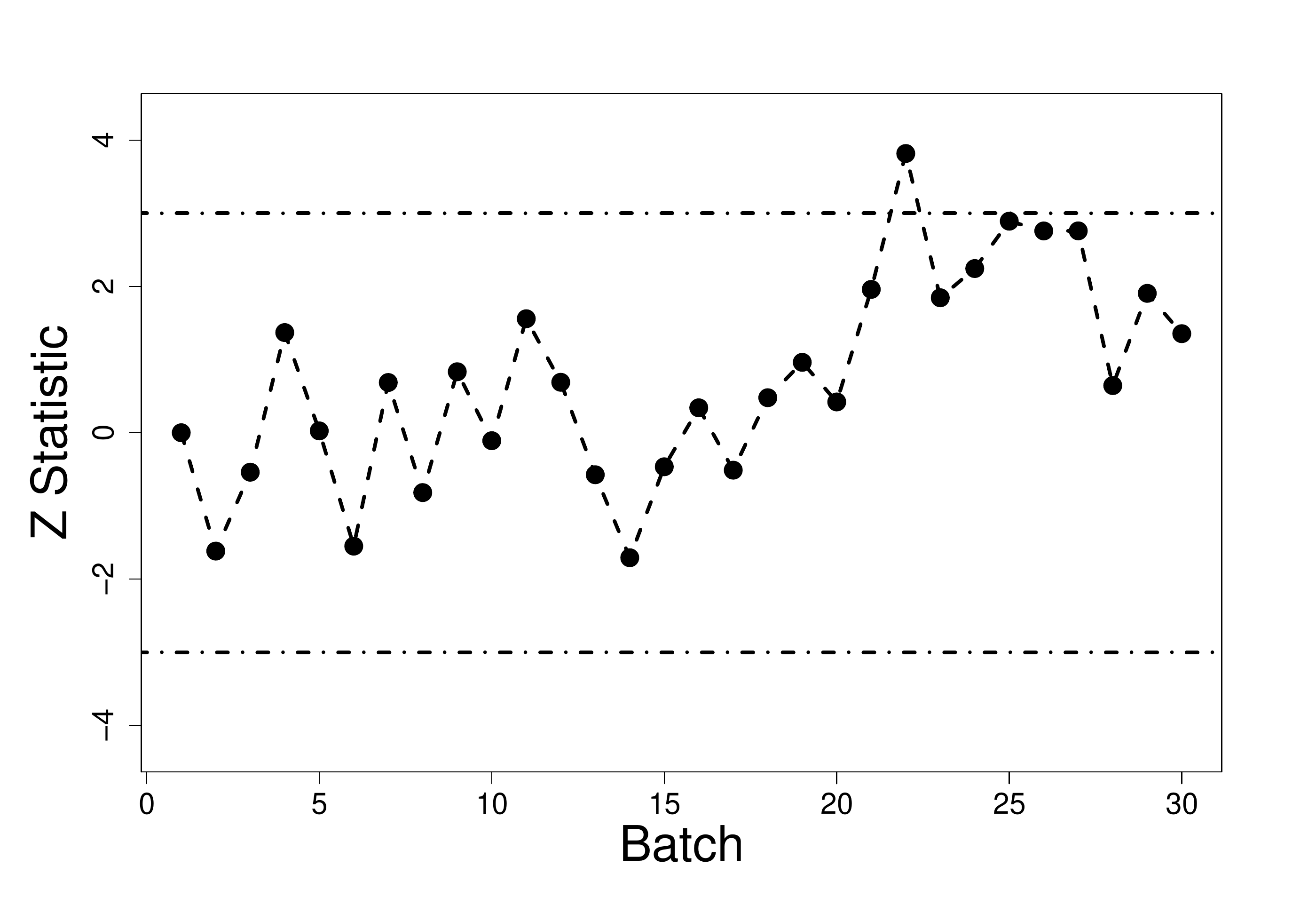}
    \caption{The SNS charting statistic $Z_i$ for the 30 batches in Table 3.4, showing the signal with batch 22.}
    \label{fig:c3s2f2}
\end{figure}

Even though Figure \ref{fig:c3s2f2} of individual batch statistics shows sample 22 exceeds the upper bound 3.090 for $\text{ARL}=500$, we will proceed to illustrate the CUSUM procedure. The CUSUM equations of Section \ref{sec:c2s2} are applied to these $Z$ values, with the results shown in Figure \ref{fig:c3s2f3}. The first exceedance of the upper bound 4.389 appears with the 22$^\text{nd}$ sample, and all batches thereafter. Actually the CUSUM of 5.16 associated with the 22$^\text{nd}$ sample exceeds the upper bound 5.071 for $ARL = 1000$.

\begin{figure}[t!]
    \centering
    \includegraphics[width=0.75\textwidth,angle=-90]{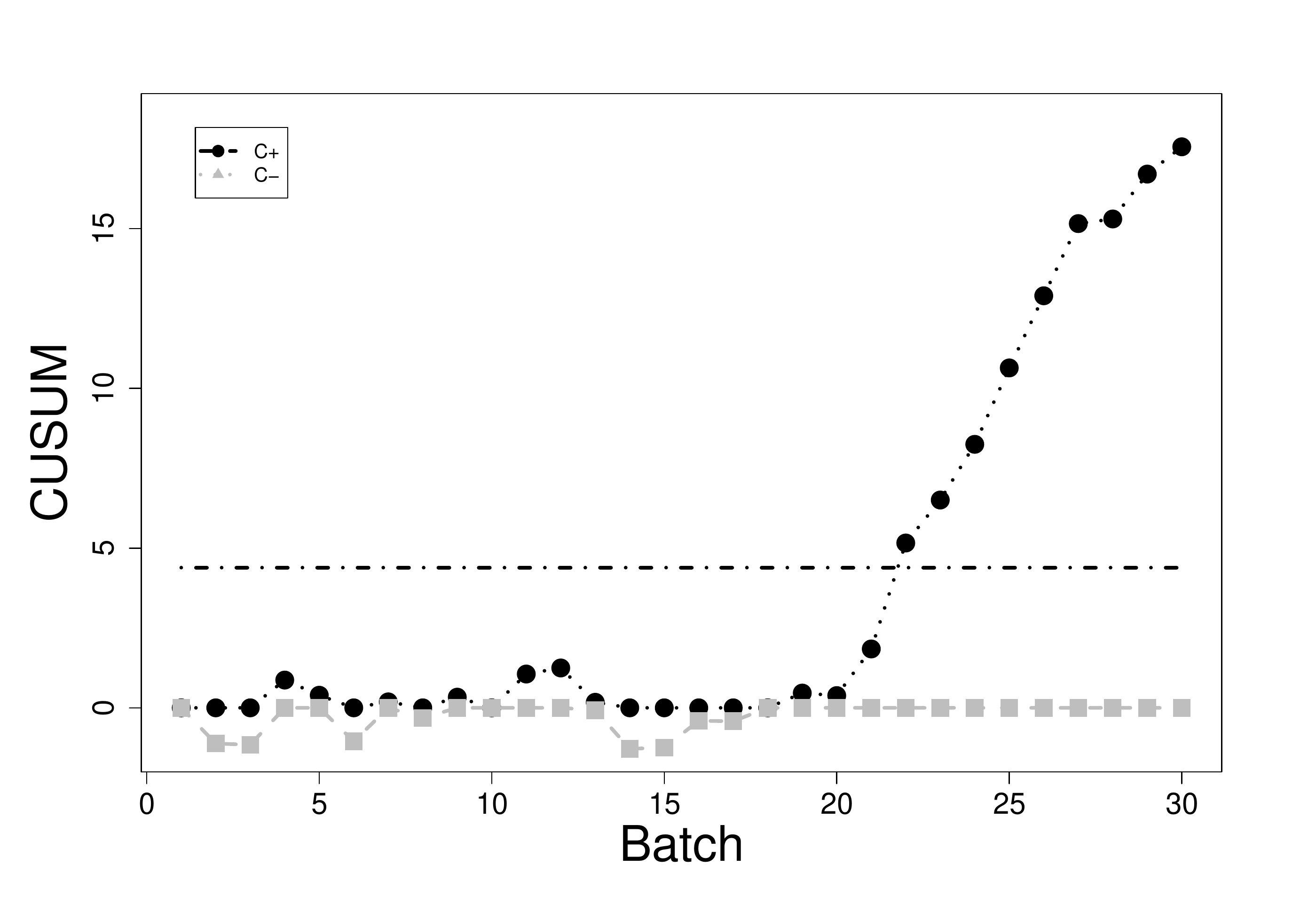}
    \caption{The positive and negative CUSUMs of the SNS in Example 3.2.1, showing a signal with batch 22.}
    \label{fig:c3s2f3}
\end{figure}

Figure \ref{fig:c3s2f3} may be compared with a nonparametric procedure introduced by \citet{li_etal_2010}, graphed in \citet{qiu_2014} as Figure 8.5 on page 329. That procedure used a reference sample of 100, not used in our example, followed by the 30 batches of size 5 that we used. Each subsequent batch of size $n=5$ was compared with the reference sample using a Mann-Whitney test. This involves ranking the new observations among all 105 observations (the reference sample plus one batch) each time a new batch is obtained, and plotting the sum of the ranks of the batch. The first batch to exceed the upper bound is the 23rd batch, and all batches thereafter. Even though the two methods have distinct differences, the results are similar.
\end{myexample}

\section{EWMA Variation to Detect a Change in Location}\label{sec:c3s3}

\begin{myexample}\label{ex:c3s3e1}
To illustrate this procedure we will consider the same 30 batches of size $n=5$ in Table 3.4 that we used to illustrate the CUSUM procedure, from Example 8.4 of \citet{qiu_2014}. The batch sum of SNS is divided by the square root of 5 as in the previous section, and that value $Z$ is used as a charting statistic. The graph of this statistic is given in Figure 3.5 in the previous section. However, the charting statistic is now converted to an EWMA with $\lambda = 0.1$, $E_0 = 0$, and $E_i = (0.1)Z_i + (0.9)E_{i-1}$ as described in Section \ref{sec:c2s3}.

We are choosing $\lambda=0.1$ and $\rho=2.814$ to achieve an $\text{ARL}=500$. for independent standard normal random variables. This matches the choice of values in Figure 8.7 on page 334 of \citet{qiu_2014}. The upper and lower limits are 0.646 and -0.646 respectively from our Table 2.4. The graph of EWMA in Figure \ref{fig:c3s3f1} for SNS shows out-of-control values begin with batch 23 and continue for the rest of the batches, almost the same as in Example 8.4 which uses a different procedure.

\begin{figure}[t!]
    \centering
    \includegraphics[width=0.75\textwidth,angle=-90]{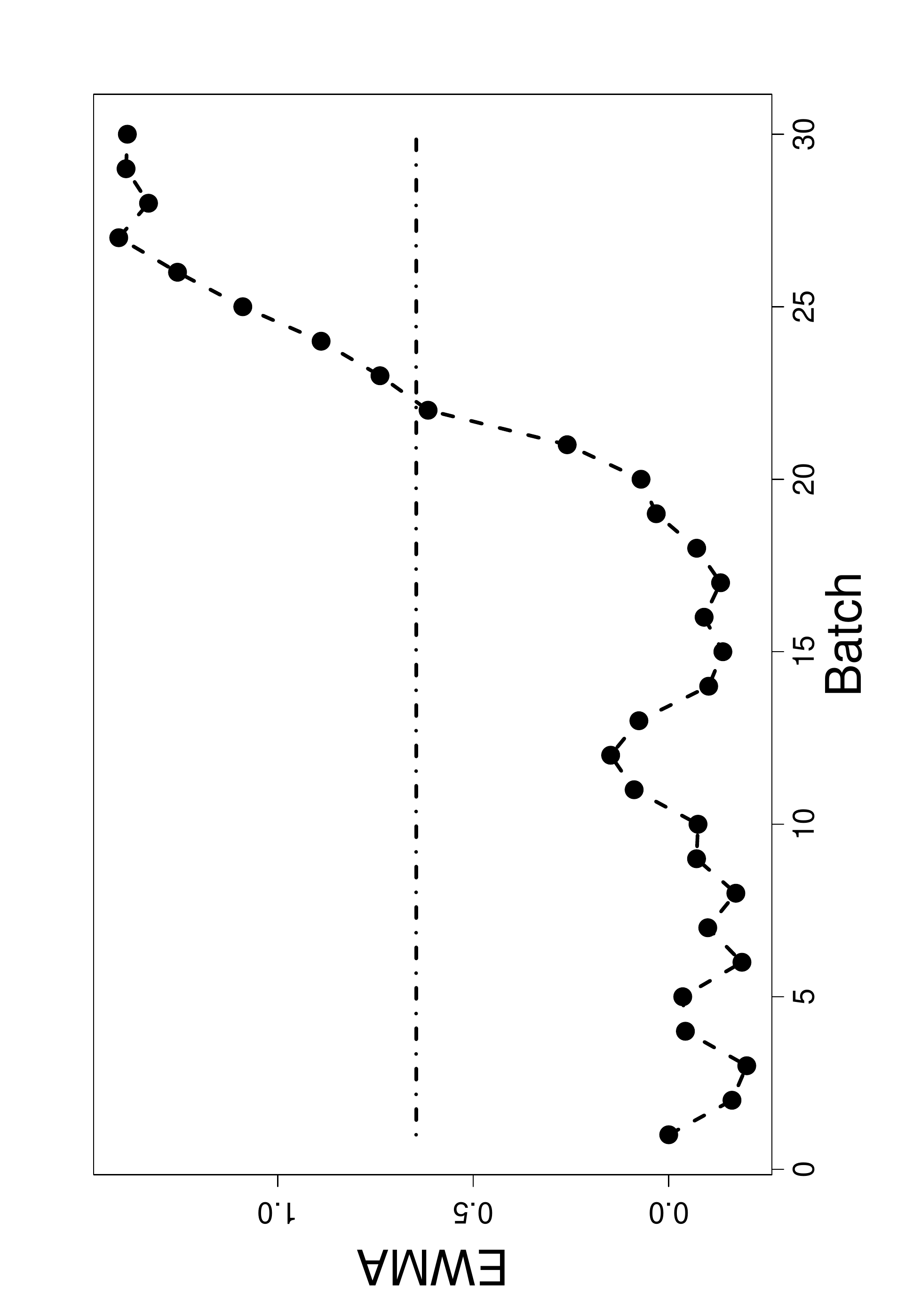}
    \caption{The exponentially weighted moving average of the SNS for the 30 batches, showing a signal with batch 23.}
    \label{fig:c3s3f1}
\end{figure}

In Example 8.4, you may recall from the CUSUM example given previously, the batches are compared individually with a reference sample of 100 observations, using a Mann-Whitney 2-sample rank test, to get a test statistic for each batch. Those are then converted to CUSUM and EWMA bounds that are tailored for the mean and standard deviation of that test statistic. Their EWMA graph for the data in Example 8.4 is shown in their Figure 8.7 (p.334). It shows the first signal on batch 24, and signals on all subsequent batches. Our procedure is self-starting with the 30 batches, without using a reference sample, and each batch is compared with the previous batches using sequential normal scores. Their procedure uses only one batch at a time and compares it with 100 reference values that we did not use. Again, as with the CUSUMs, the end result is similar even though the methods are different.
\end{myexample}

\section{Detecting a Change Point in Location}\label{sec:c3s4}

In a sequence of observations, one primary interest is to determine if there has been a change in location or scale. The previous examples illustrated how sequential normal scores can be used to develop efficient and accurate methods for detecting such changes. Once it has been determined that there has been a change in location, it is often of interest to estimate at which time point the change occurred. These methods are called {\em change-point detection} (CPD) methods.

Sequential normal scores in a sequence start with a mean of zero and a standard deviation of about 1.0. At some point the mean of the original sequence may change, and at a later point that change is detected, by one of many methods including several described in previous sections, such as CUSUM or EWMA. At this time the average of several recent SNS values can be compared with the average of the previous values using the 2-sample t-statistic for normal random variables, to see if the sample means are significantly different.

The 2-sample t-statistic for comparing sample means simplifies in the case of sequential normal scores, because the population standard deviation is 1.0. Under the null hypothesis of no change the standard deviation remains at 1.0. Only the means of the two sub-samples vary in the statistic. Thus the $t$-statistic for a sample of size $n$ vs. a sample of size $m$ becomes
\begin{align}
    T = (\bar{Y}-\bar{X})/\sqrt{1/n+1/m},
\end{align}
where $\bar{X}$ is the average of the SNS of the first group and $\bar{Y}$ is the average of the SNS of the most recent observations up to and including the signal. The sample sizes are the total number of observations that went into each average. If the batch sizes are equal, it is equivalent for this method to use the number of batches that went into each average. The variances are replaced by their population values under the null hypothesis of identical distributions, which is 1.0.

After a signal is received, using perhaps CUSUM or EWMA or some other method, the SNS of the sequence already observed is partitioned into two samples, and the statistic $T$ is computed. The partition that gives the maximum value of $T$ defines our estimate of where the change in location took place. This is the SNS version of the maximum likelihood estimate for the change point in normal random variables. The SNS version takes advantage of the fact that if there is no change in the distribution, the population standard deviation is 1.0, to reduce the sampling variation as much as possible. In the case of batches, the charting statistics $Z$ are used in calculating the sample means.

\begin{myexample}\label{ex:c3s4e1}
In the previous two examples the same 30 batches of size $n=5$ were used to detect a change in location. Our CUSUM detected a change at 22, our EWMA detected a change at 23, and two procedures used in \citet{qiu_2014} detected a change at 23 and a change at 24, respectively. We used all three possibilities (22, 23, and 24) to estimate the change point. All three choices had the same maximum point, estimating the change point at batch 21 as shown in Figure 3.8, which seems to agree with the original data, as shown in Figure 3.4.

\begin{figure}[t!]
    \centering
    \includegraphics[width=0.75\textwidth,angle=-90]{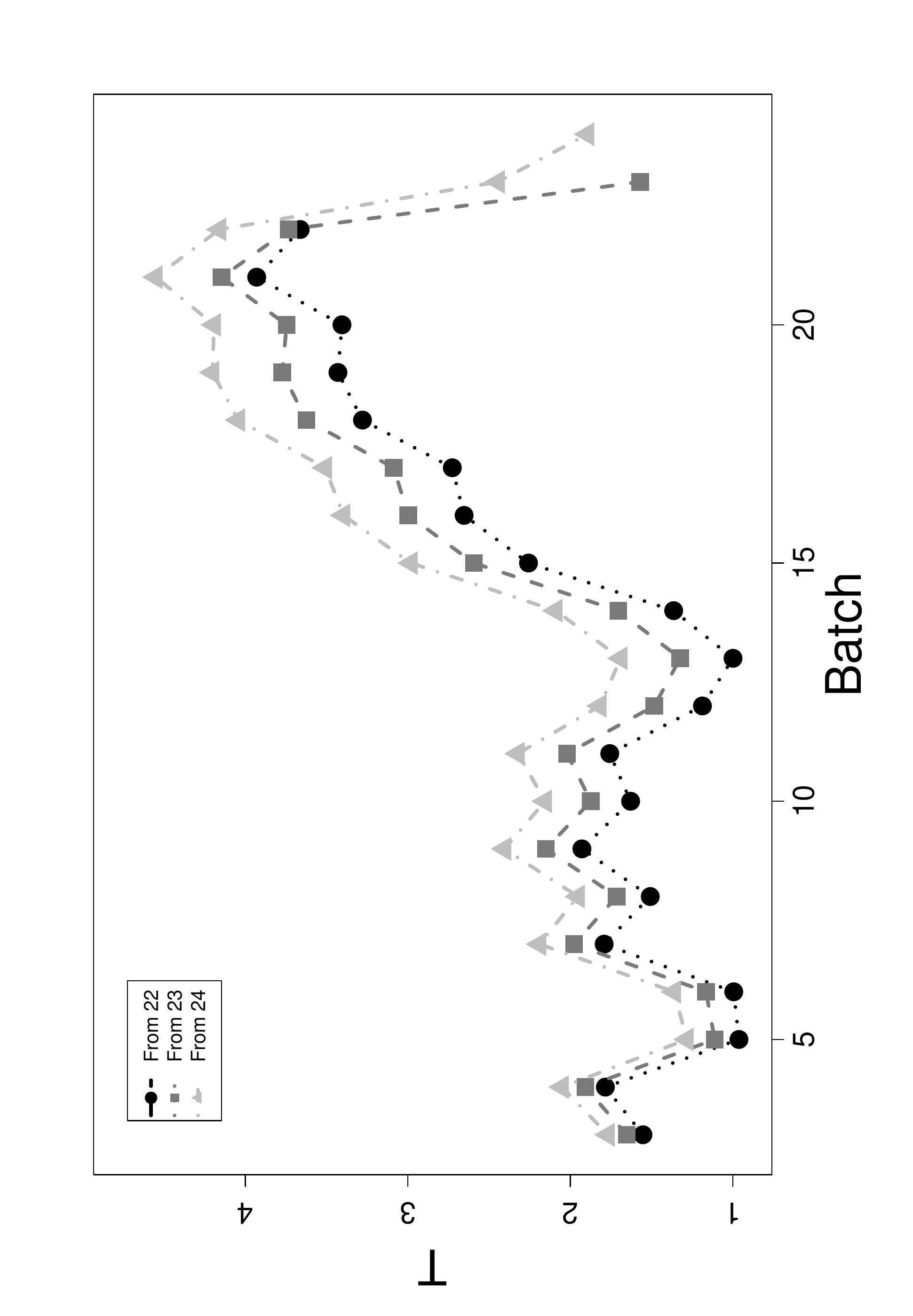}
    \caption{Plots of the change point statistics T given in equation 3.4.1 for SNS in Example 3.4.1, using different batches as the signal. The maximum in all three cases is at batch 21, which is the estimate for the point of change.}
    \label{fig:c3s4f1}
\end{figure}

\end{myexample}

\section{Sequential Normal Scores with a Reference Data Set (Phase 1)}\label{sec:c3s5}

In Example 8.2 of \citet{qiu_2014} a reference sample of 1000 observation was available to illustrate the distribution-free precedence test of \citet{chakraborti_etal_2004}, but it was not available to illustrate the SNS test, so the SNS test used only the 30 batches of size $n=5$ for its comparison. Similarly, in Example 8.4 a reference sample of 100 observations was available to illustrate a nonparametric rank sum test of \citet{li_etal_2010} on 30 subsequent batches of n=5, but the reference sample was not available for the SNS test, so the SNS test was used only on the 30 subsequent batches. Even though the self-starting mode of the SNS tests compared favorably with the other tests in both situations, we will present an illustration of how to use the SNS test when a reference set is available.

It is not necessary to monitor the Phase 1 data. That defeats the purpose of having a reference set. A reference set, by definition, contains only data that are in control, even though some of the observations may not appear to the analyst to fit the analyst’s concept of what an in control set of data should look like. The expert in the field declares the data set to be a representative set, against which other data may be compared. That is why it is important to consider nonparametric methods, because new observations are compared with the reference set in nonparametric methods. Parametric methods are more likely to compare new observations with an artificial mathematical model whose parameters may have been adjusted to fit the reference set in some way.

With that in mind it is clear that a reference set, or set of Phase 1 data, is treated as batch 1 in methods based on sequential normal scores. The observations in batch 1 could be ranked among themselves and converted to normal scores, but as noted earlier that is not necessary, because the charting statistic for batch 1 is always $Z_1 = 0$. Then observations in batch 2 are ranked relative to the observations in batch 1 to get their SNS in the usual way. If there is only one observation in batch 2, that SNS value is $Z_2$. If there are several observations in batch 2 their sum of SNS values is divided by the square root of the size of batch 2 to get $Z_2$, in the usual way as described in Section 1.1. And so on, the procedure is repeated for each observation or batch.

The CUSUM or EWMA is computed on the $Z$’s, no matter if their batch sizes are equal or unequal. The size of the reference sample may be large, but it results in one and only one $Z_1$ and it will be zero, using sequential normal scores.

To estimate the change point, the average of the Z values in the first $k$ batches  is subtracted from the average of the Z values in the rest of the batches to get the numerator of the test statistic. When the batch sizes are equal the computations are simple, as pointed out in the previous section. When the batch sizes are unequal, such as in the presence of a reference sample, the average of several batches is a weighted average of the charting statistics.

For example, if the first $k$ batches have respective sample sizes $N_1, N_2, \ldots, N_k$ and respective charting statistics $Z_1, Z_2, \ldots, Z_k$ then the weighted average of those batches is $\Sigma_i(Z_i\sqrt{N_i})/\Sigma_i(N_i)$ and the sample size, used in the denominator of the test statistic, is $\Sigma_i(N_i)$.

\begin{myexample}
In Example 8.7 on page 339 of \citet{qiu_2014} the first 100 observations obtained from a production process for phase II monitoring contain 14 observations that are in control, the first 14 observations. They could be ranked among themselves, but it is not necessary, because we know that $Z_1 = 0$, and that is all we need to know. However the subsequent observations are ranked against all previous observations, and that includes the observations in the reference sample. We don't actually use the ranks of the observations in Phase 1, just the observations themselves, for subsequent sequential ranks. We don’t repeat the 100 observations here, but we show the results.
\begin{figure}[t!]
    \centering
    \includegraphics[width=0.75\textwidth,angle=-90]{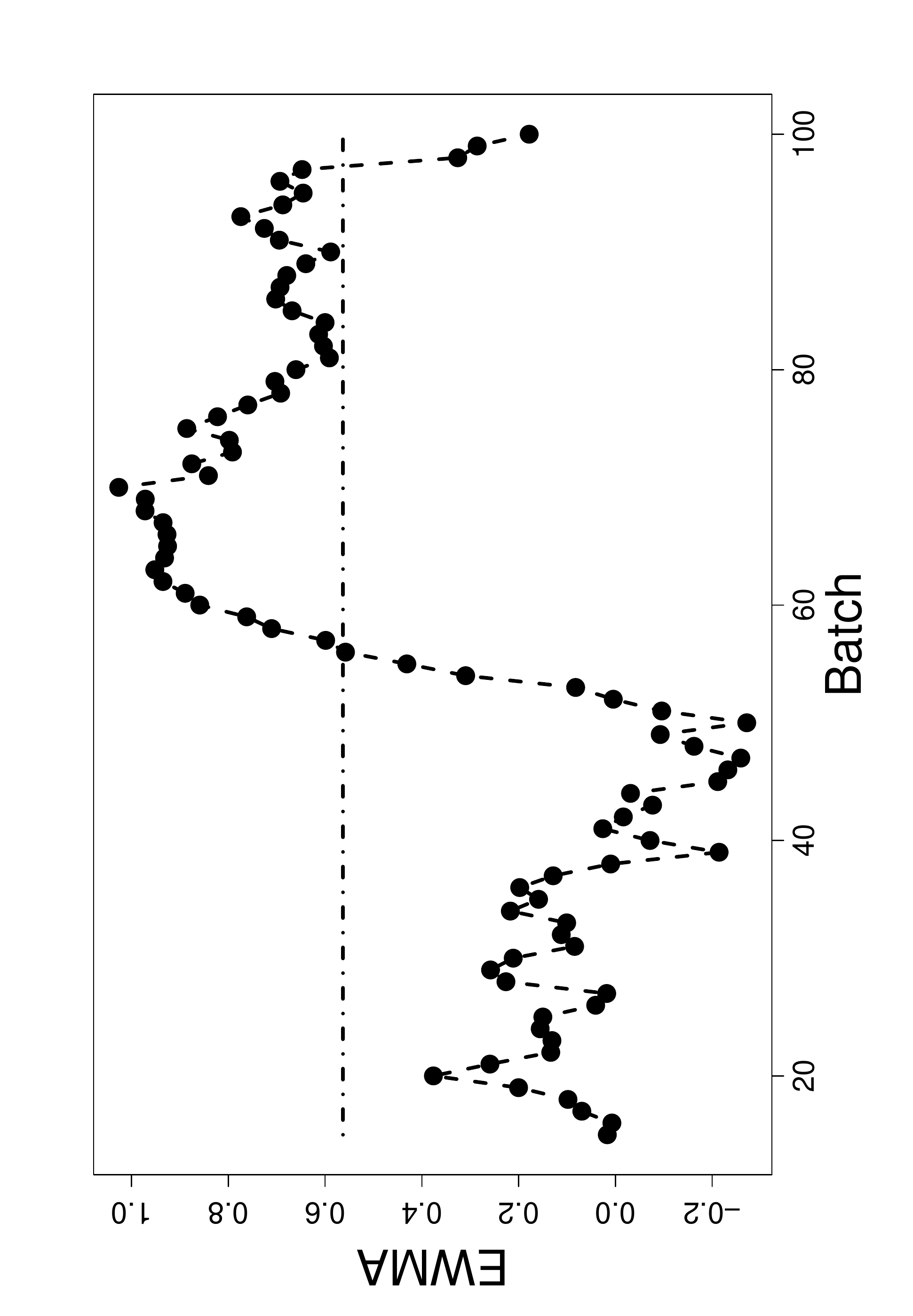}
    \caption{The EWMA computed on the SNS of 100 observations in Example 3.5.1, where the first 14 observations are assumed to be in control. The EWMA sends a signal of change of location with observation 57.}
    \label{fig:c3s5f1}
\end{figure}
The EWMA with $\lambda=0.1$ is shown in Figure \ref{fig:c3s5f1}. It crosses the upper limit 0.563 for $\text{ARL}=200$ on the 57$^\text{th}$ observation. To estimate the change point the test statistic used in the previous section is used with the modification for unequal batch sizes described above. The first batch size is 14 and the subsequent batch sizes are 1 each. The test statistic appears in Figure \ref{fig:c3s5f2}. The maximum value of the test statistic occurs on observation 51. That represents the SNS estimate of the point at which the location parameter changes in the observed sequence. The method used in Example 8.7 of \citet{qiu_2014} was proposed by \citet{hawkins_etal_2010}, and estimated the change point at observation 55. Since the actual change point is unknown all we can say is there is good agreement in the two estimates.
\begin{figure}[t!]
    \centering
    \includegraphics[width=0.75\textwidth,angle=-90]{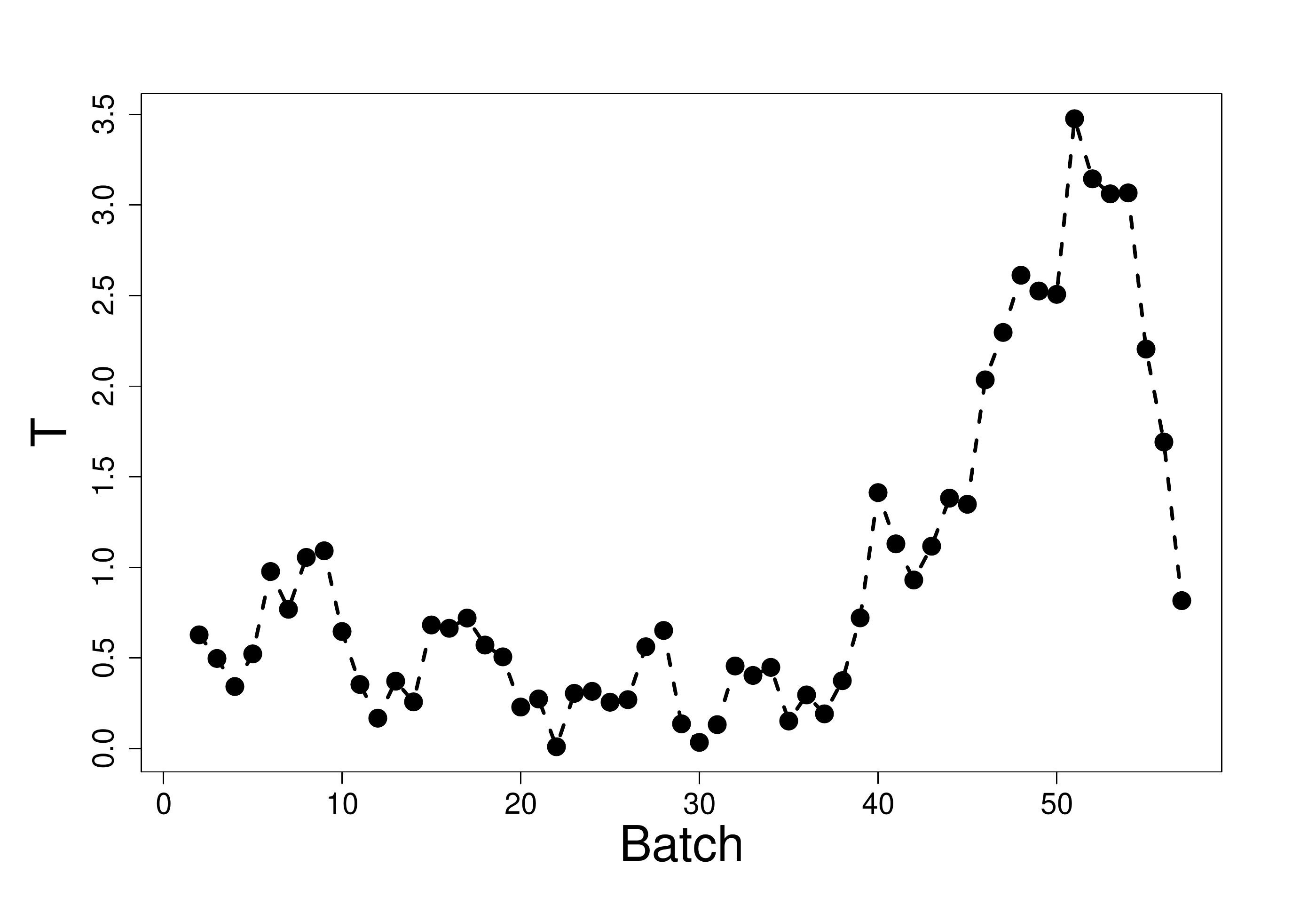}
    \caption{The change-point statistic T using SNS in Example 3.5.1, based on a change signaled by EWMA at observation 57.This shows a maximum at observation 51, which is our estimate of the change point.}
    \label{fig:c3s5f2}
\end{figure}
\end{myexample}
\chapter{Detecting a Change in Scale}\label{ch:c4}

Our attention has been directed thus far to detecting a change in location. Some situations demand the ability to detect a change in scale. A sequence with a change in location will tend to exceed either the upper bound too frequently, or the lower bound too frequently, but not both. A graph of sequential normal scores indicates a change in scale if the SNS frequently exceed the bounds in both directions. This suggests using the squares of the SNS as a monitoring statistic, because, with a mean of zero, the squared value will be large if the original SNS is either large or small.

The square of a standard normal random variable has a chi-squared distribution with 1 degree of freedom, with mean 1.0 and variance 2.0. Therefore SNS$^2$ will be approximately chi-squared in distribution, with 1 degree of freedom. The sum of a batch of $k$ squared independent SNS values will have approximately a chi-squared distribution with $k$ degrees of freedom, and thus will have a mean of $k$ and a variance of $2k$. First we will illustrate the SNS$^2$ procedure on a sequence of individual observations (batches of size $n=1$), and we will use the EWMA to detect changes in the spread of the observations. The EWMA will tend to be relatively large when the spread (variance) increases. A method for estimating the change point is also illustrated. Later we will illustrate these methods on batches of data.

\section{Using SNS$^2$ to Detect a Change in Scale, along with EWMA, and Change Point Estimation}\label{sec:c4s1}

Chapter 3 discussed how sequential normal scores can be used with parametric methods designed for normal random variables to detect changes in location. The most popular measure of location for a random variable $X$ is the mean, defined as the expected value of $X$, or $E(X) = \mu$. The most popular measure of spread is the standard deviation $\sigma$, which is the square root of the variance. The variance is also an expected value, $E(X-\mu)^2 = \sigma^2$.

Thus tests for spread follow the steps used in tests for location, but with SNS$^2$ used instead of SNS. The bounds for Shewhart charts need to be adjusted from assuming standard normal random variables, to assuming chi-squared random variables with 1 degree of freedom. Shewhart charts for individual observations of SNS$^2$ will have an $\alpha$-level upper bound, for detecting an increase in spread, equal to the ($1-\alpha$) quantile of a chi-squared random variable with 1 degree of freedom. A lower $\alpha$-level bound for SNS$^2$, for detecting a decrease in spread, is equal to the $\alpha$ quantile of a chi-squared random variable with 1 degree of freedom.
A similar adjustment is needed for EWMA for SNS$^2$. Table 2.7 in Section 2.3 provides the new bounds for EWMA that apply to SNS$^2$. If a change in variance is detected, the point at which the variance changes may be estimated as before. An example will clarify the steps involved.

\begin{myexample}
Example 6.5 on page 246 of \citet{qiu_2014} presents 30 observations to illustrate a parametric procedure introduced by \citet{hawkins_etal_2005} for detecting an increase in variance. The first 9 observations are assumed to be a reference sample, which we will consider to be batch 1 with 9 observations. The remaining 21 observations are assumed to be individual values, which we take as batches of size 1. The data are given in the second column of Table \ref{tab:c4s1t1}, and graphed in Figure \ref{fig:c4s1f1}.

\begin{table}[t!]
    \caption{The data used in Example 6.5 of \citet{qiu_2014} and also used in the example in this section. The first 9 observations are assumed to be a reference sample, which we will consider to be batch 1 with 9 observations. The remaining 21 observations are assumed to be individual values, which we take as batches of size 1. The data are given in the second column of the table. Column 3 contains the sequential normal scores SNS, using the first 9 observations as batch 1. Column 4 contains the squared values of SNS. Column 5 shows the EWMA for chi-squared random variables $SNS^2$, one observation per batch, using $\lambda = 0.1$. The first significant value using control limits calibrated for an in-control ARL of 200 is in boldface, at observation 29. The last column shows the computed values of the statistic T using equation (4.1.1) to estimate the change point. The maximum value is in boldface, indicating observation 19 as the estimated change point.}
    \centering
    \begin{tabular}{crrrrr}
    Obs. & \multicolumn{1}{c}{X} & \multicolumn{1}{c}{SNS} & \multicolumn{1}{c}{SNS$^2$} & \multicolumn{1}{c}{EWMA(0.1)} & \multicolumn{1}{c}{T} \\
    1     & -0.502 & -0.589 & 0.347 &       &  \\
    2     & 0.132 & 0.282 & 0.080 &       &  \\
    3     & -0.079 & -0.282 & 0.080 &       &  \\
    4     & 0.887 & 1.593 & 2.538 &       &  \\
    5     & 0.117 & 0.000 & 0.000 &       &  \\
    6     & 0.319 & 0.589 & 0.347 &       &  \\
    7     & -0.582 & -0.967 & 0.936 &       &  \\
    8     & 0.715 & 0.967 & 0.936 &       &  \\
    9     & -0.825 & -1.593 & 2.538 & 1.000 &  \\
    10    & -0.360 & -0.385 & 0.148 & 0.915 & 1.243 \\
    11    & 0.090 & 0.000 & 0.000 & 0.823 & 1.543 \\
    12    & 0.096 & 0.105 & 0.011 & 0.742 & 1.878 \\
    13    & -0.202 & -0.396 & 0.157 & 0.684 & 2.208 \\
    14    & 0.740 & 1.242 & 1.542 & 0.769 & 2.503 \\
    15    & 0.123 & 0.341 & 0.116 & 0.704 & 2.442 \\
    16    & -0.029 & -0.237 & 0.056 & 0.639 & 2.767 \\
    17    & -0.389 & -0.821 & 0.674 & 0.643 & 3.123 \\
    18    & 0.511 & 0.862 & 0.742 & 0.653 & 3.334 \\
    19    & -0.914 & -1.938 & 3.756 & 0.963 & \textbf{3.550} \\
    20    & 2.310 & 1.960 & 3.841 & 1.251 & 2.960 \\
    21    & -0.876 & -1.465 & 2.147 & 1.340 & 2.336 \\
    22    & 1.528 & 1.489 & 2.219 & 1.428 & 2.185 \\
    23    & 0.524 & 0.709 & 0.503 & 1.336 & 2.016 \\
    24    & 1.547 & 1.534 & 2.354 & 1.438 & 2.406 \\
    25    & -1.629 & -2.054 & 4.218 & 1.716 & 2.232 \\
    26    & -0.877 & -1.304 & 1.700 & 1.714 & 1.355 \\
    27    & -1.440 & -1.593 & 2.538 & 1.796 & 1.385 \\
    28    & 0.462 & 0.514 & 0.264 & 1.643 & 1.051 \\
    29    & -2.315 & -2.114 & 4.471 & \textbf{1.926} & 2.243 \\
    30    & 0.494 & 0.573 & 0.328 & 1.766 &  \\
    \end{tabular}

    \label{tab:c4s1t1}
\end{table}
\begin{figure}[t!]
    \centering
    \includegraphics[width=0.75\textwidth,angle=-90]{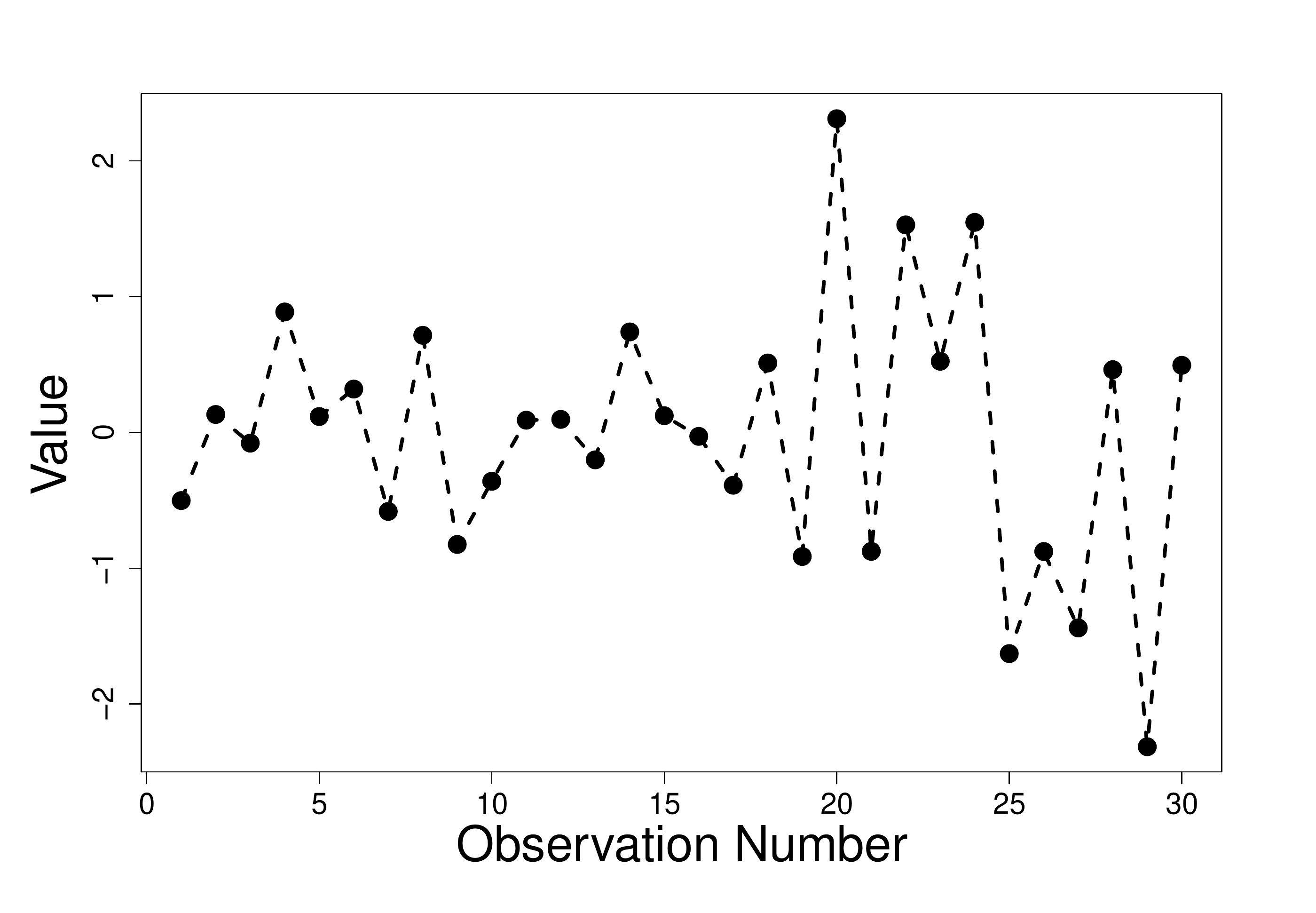}
    \caption{A graph of the original 30 observations in Table 4.1, showing a possible change in spread.}
    \label{fig:c4s1f1}
\end{figure}
The first 9 observations are considered a reference sample in \citet{qiu_2014} so we rank them among themselves and convert them to sequential normal scores. Note that their average is always exactly zero, and their standard deviation approximately 1.0. The remaining 21 observations are ranked relative to the previous observations in the sequence, including the observation being ranked, and converted to sequential normal scores. They appear in the third column in the above table, and their squared values appear in the fourth column. The graphs of the original data and their sequential normal scores are given in Figures \ref{fig:c4s1f1} and \ref{fig:c4s1f2}.
\begin{figure}[t!]
    \centering
    \includegraphics[width=0.75\textwidth,angle=-90]{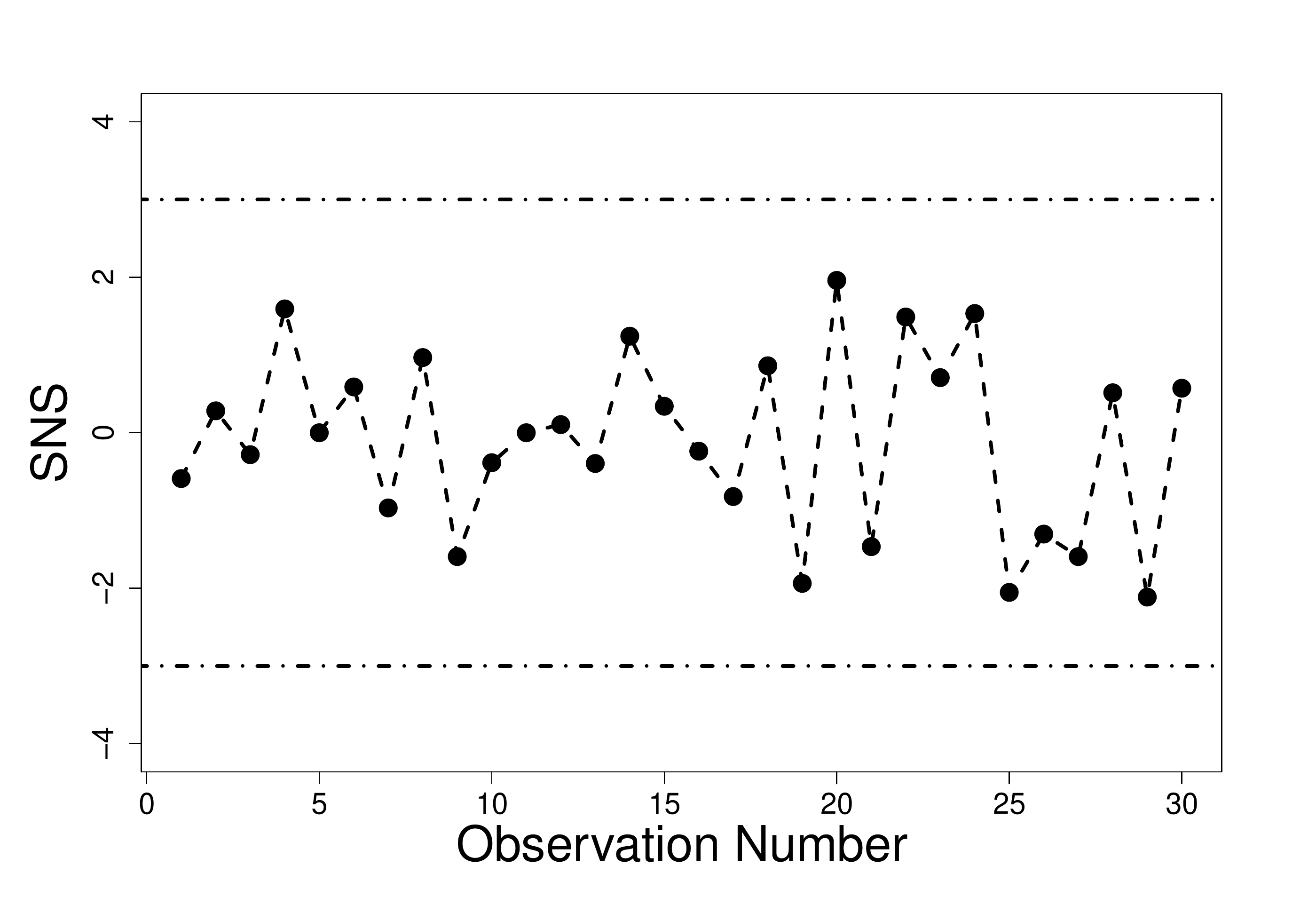}
    \caption{The sequential normal scores in Table 4.1, using the first 9 observations as the reference sample, so the first 9 observations are ranked among themselves. The remaining 21 observations are ranked relative to all the observations that preceded them in the series. No observation signals a change (exceeding $\pm2.807$, at p=0.005).}
    \label{fig:c4s1f2}
\end{figure}
It is clear from both graphs that the spread increases somewhere near observation 19 and continues through the end of the sequence. The two graphs appear similar, with a correlation coefficient of 0.957. The increase in spread results in observations that tend to be greater than or less than the earlier observations, so the square of the SNS is appropriate. They are graphed in Figure \ref{fig:c4s1f3}.
\begin{figure}[t!]
    \centering
    \includegraphics[width=0.75\textwidth,angle=-90]{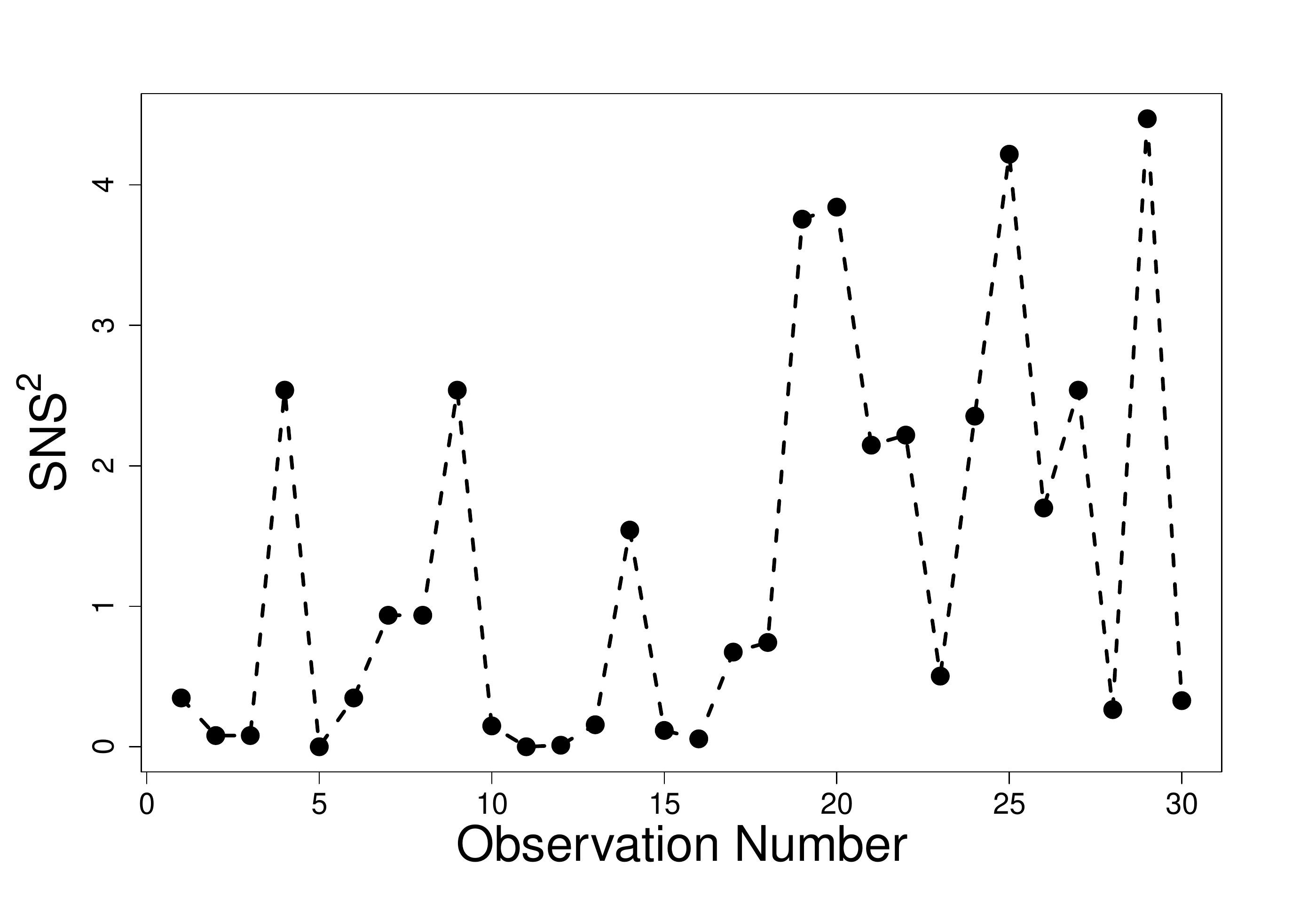}
    \caption{The squares of the SNS values in Figure \ref{fig:c4s1f2}, given in Table 4.1, showing how a change in spread in SNS becomes a change in location in SNS$^2$. No observation signals a change (greater than $7.88=2.807^2$).}
    \label{fig:c4s1f3}
\end{figure}

We will examine the SNS$^2$ values for a change in location. There are too few observations to detect statistical significance individually, so a EWMA is appropriate. The bounds given in our Table 2.7 for chi-squared random variables show that for $\lambda = 0.1$ an upper bound of 1.842 is appropriate for an $\text{ARL}=200$, which matches the target ARL in \citet{qiu_2014} Example 6.5.

The SNS$^2$ of a single observation is approximately a chi-squared with 1 degree of freedom. The EWMA for chi-squared random variables with 1 degree of freedom begins with observation 10 (SNS$^2$=0.148) the first single observation. (By ranking the first batch among themselves, their ranks are no longer independent, and the sum of the SNS is always zero, so they cannot be considered in the EWMA.) Therefore $E_9 = 1$, $E_{10} = \lambda(0.148)+(1-\lambda)E_9$, and so on. For $\lambda = 0.1$, $E_{10} = 0.915$, as given in the table. See Figure \ref{fig:c4s1f4} for a graph of the EWMA.

\begin{figure}[t!]
    \centering
    \includegraphics[width=0.75\textwidth,angle=-90]{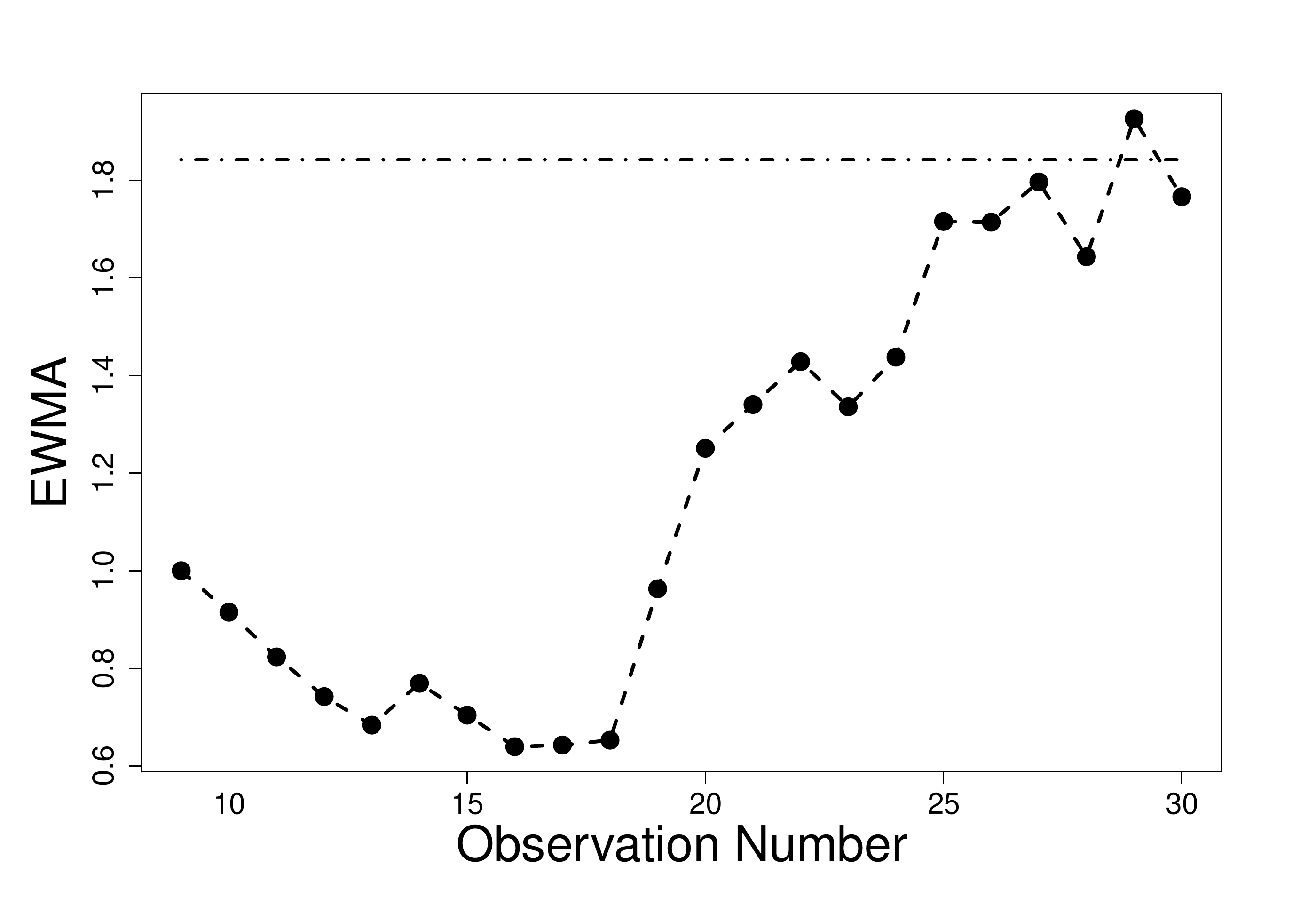}
    \caption{The EWMA computed on the SNS$^2$ in Figure 4.3 and Table 4.1, beginning with observation 10. This shows a signal at observation 29.}
    \label{fig:c4s1f4}
\end{figure}
The EWMA exceeds its upper bound 1.842 on observation 29. Therefore the change in variance is statistically significant. It remains to estimate the change point.

The estimated point of change is found as before, in Section \ref{sec:c3s4}. The $t$-statistic is computed on the two partitions of the SNS$^2$ values from observation 10 to observation 29 as shown in Table~4.1. The population variance is 2 in this case, so the statistic is given as follows
\begin{align}
    T = (\bar{Y}-\bar{X})/\sqrt{2/n+2/m}.
\end{align}
The observations begin with observation 2, but the $T$ statistic is meaningless until after the reference sample, because the order of the observations in the reference sample is meaningless. The maximum value of $T$ occurs with observation 19, as seen in Figure \ref{fig:c4s1f5}, so that is our estimate of the change point.
\begin{figure}[t!]
    \centering
    \includegraphics[width=0.75\textwidth,angle=-90]{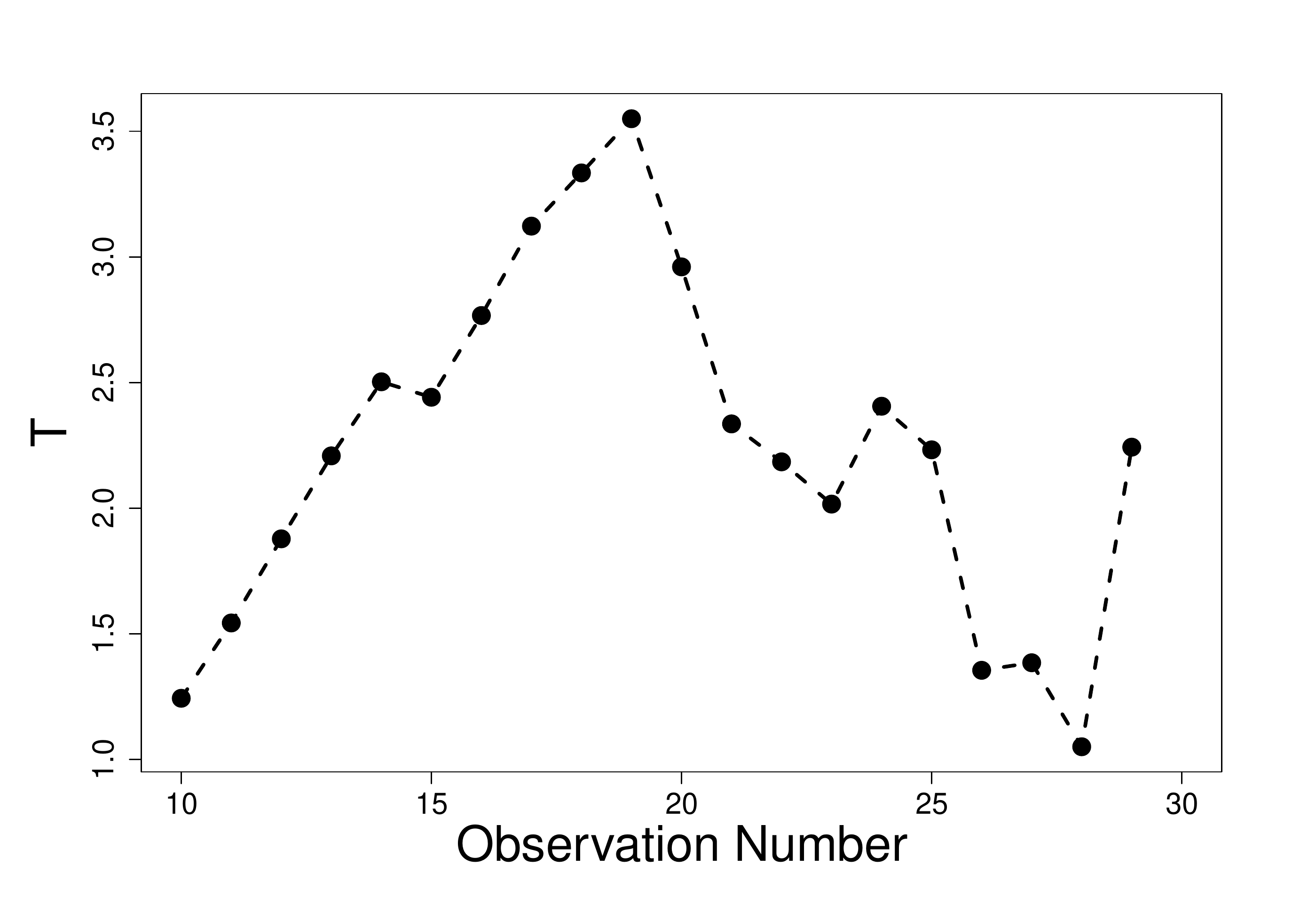}
    \caption{The change-point statistic computed on the SNS$^2$ values in Figure 4.3 and in Table 4.1, assuming a signal at observation 29. This shows a maximum at observation 19, which is our estimate of the change point in spread.}
    \label{fig:c4s1f5}
\end{figure}
The estimate of the change point in Example 6.5 on page 246 of \citet{qiu_2014} is observation 22 using the parametric procedure introduced by \citet{hawkins_etal_2005}. The data in this example were generated as $N(0,1)$ for observations 1 through 20, and $N(0,2^2)$ for observations 21 through 30. Thus the actual change point was 21 for this data set. Both methods produce good estimates of the change point.
\end{myexample}

Our example used single observations, which furnished a statistic SNS$^2$ with a chi-squared distribution with 1 degree of freedom. If batches of $m$ observations each are obtained the sum of the SNS$^2$ for that batch is approximately chi-squared with $m$ degrees of freedom. Thus the charting statistic, and the statistic used for EWMA, is approximately chi-squared with $m$ degrees of freedom. The procedure derived on page 204 in \citet{qiu_2014} can be modified to fit batches of independent SNS$^2$ values as follows.

Let $Y_i$ be the sum of the $m$ independent SNS$^2$ values in the $i^\text{th}$ batch. Then, except for batch 1 with the reference sample, $Y_2, Y_3, \ldots$, are independent chi-squared random variables with $m$ degrees of freedom. Following the reasoning used in equation (5.20) on p. 204 of \cite{qiu_2014} the EWMA is defined as $E_1 = 1$, and $E_i = \lambda Y_i /m + (1-\lambda)E_{i-1}$ for $i>1$. The upper and lower bounds depend on $\rho_U$ and $\rho_L$ which are given in Section \ref{sec:c2s3}, Table 2.6, copied from \citet{qiu_2014}, for selected values of $\lambda$ and ARL. Specifically the steady state upper bound $U$ and the steady state lower bound $L$ are
\begin{align}
    U  =  1 + \rho_U \sqrt {2\lambda/(2-\lambda)m} & & \text{ and }\\
    L  =  1 – \rho_L \sqrt{2\lambda/(2-\lambda)m}.
\end{align}
A signal is obtained when the EWMA is above $U$ or below $L$.

\section{Detecting a Change in Both Location and Scale (SNS$^2$ Method)}\label{sec:c4s2}

Suppose the amount of detergent in a box is supposed to weigh 16 ounces. If the production process strays below that standard the consumer is the loser, and if the production process strays above that standard the producer loses product. A quality control chart needs to be sensitive to changes in both directions. However, if the variability increases, it becomes more difficult to monitor the average accurately. Therefore it is advisable to monitor the process for an increase in spread as well as a change in location in either direction.

\begin{myexample}
Our example in this section follows Example 4.9 on page 152 of \citet{qiu_2014}. In that example a parametric method (assuming normality) is developed for monitoring both location and spread. The method involves combining two tests. One test is designed to be optimal for detecting a shift in mean of 1.0, from $N(0,1)$ to $N(1,1)$ that has an ARL of 400. A second test is designed to be optimal for detecting a shift in standard deviation of 1.0 from $N(0,1)$ to $N(0,2^2)$, also with an ARL of 400. If either test exceeds its control limit the series is declared out of control, with a combined ARL of 200.

That procedure is used on two data sets, the first of which is given in Table 4.2. The first data set has 10 batches of size 5 from $N(0,1)$ followed by 10 batches of size 5 from $N(1,1)$. The first test in \citet{qiu_2014}, for change in location, detects the change at the 11$^\text{th}$ batch, and thus the joint monitoring scheme signals at this point also. The second test, for spread, fails to signal a change in spread, as expected, because there is no change in spread. The batch means $\bar{X}$ are plotted in Figure \ref{fig:c4s2f1}. The shift in location is evident from this graph, shifting on batch 11 and lasting through batch 20. No change in spread is apparent.

\begin{table}[t!]
    \caption{The first of two data sets used in Example 4.9 of \citet{qiu_2014}. The data set consists of 20 batches of 5 observations each. The first 10 batches are $N(0,1)$ and the second 10 batches are $N(1,1)$. Column 7 is the average of the batch, and column 8 is the sum of the squared values of SNS for each batch, which is compared with the chi-squared distribution with 5 degrees of freedom.}
    \centering
    \begin{tabular}{rrrrrccc}
    \multicolumn{5}{c}{First data set}    & Batch & $\bar{X}$ & Z for SNS$^2$ \\
    0.019 & -0.184 & -1.371 & -0.599 & 0.295 & 1     & -0.368 & 3.835 \\
    0.390 & -1.208 & -0.364 & -1.627 & -0.256 & 2     & -0.613 & 4.369 \\
    1.102 & 0.756 & -0.238 & 0.987 & 0.741 & 3     & 0.670 & 11.486 \\
    0.089 & -0.955 & -0.195 & 0.926 & 0.483 & 4     & 0.070 & 2.021 \\
    -0.596 & -2.185 & -0.675 & -2.119 & -1.265 & 5     & -1.368 & 10.272 \\
    -0.374 & -0.688 & -0.872 & -0.102 & -0.254 & 6     & -0.458 & 0.743 \\
    -1.854 & -0.078 & 0.969 & 0.185 & -1.380 & 7     & -0.432 & 5.925 \\
    -1.436 & 0.362 & -1.759 & -0.325 & -0.652 & 8     & -0.762 & 3.689 \\
    1.087 & -0.763 & -0.829 & 0.834 & -0.968 & 9     & -0.128 & 5.669 \\
    -0.029 & 0.233 & -0.301 & -0.678 & 0.655 & 10    & -0.024 & 1.483 \\
    0.599 & 0.665 & 2.368 & 3.138 & 1.506 & 11    & 1.655 & \textbf{18.070} \\
    1.786 & 0.098 & 1.533 & 0.354 & 1.291 & 12    & 1.012 & 17.079 \\
    -0.238 & 0.544 & 0.170 & 1.340 & 2.066 & 13    & 0.776 & 11.973 \\
    2.216 & 1.736 & 0.519 & 1.563 & -0.246 & 14    & 1.157 & 17.144 \\
    1.381 & -0.430 & -0.048 & 0.781 & -0.490 & 15 & 0.239 & 6.944 \\
    2.173 & -0.480 & 0.570 & -0.052 & 2.523 & 16    & 0.947 & 11.851 \\
    1.593 & 0.777 & 1.713 & 1.717 & 1.440 & 17    & 1.448 & 23.082 \\
    1.159 & 1.660 & 3.221 & -0.184 & 0.926 & 18    & 1.356 & 18.228 \\
    0.584 & 0.809 & 1.070 & 2.155 & 1.595 & 19    & 1.242 & 15.721 \\
    -0.420 & -0.607 & 1.893 & 1.148 & 2.227 & 20    & 0.848 & 16.468 \\
    \end{tabular}

    \label{tab:c4s2t1}
\end{table}

\begin{figure}[t!]
    \centering
    \includegraphics[width=0.75\textwidth,angle=-90]{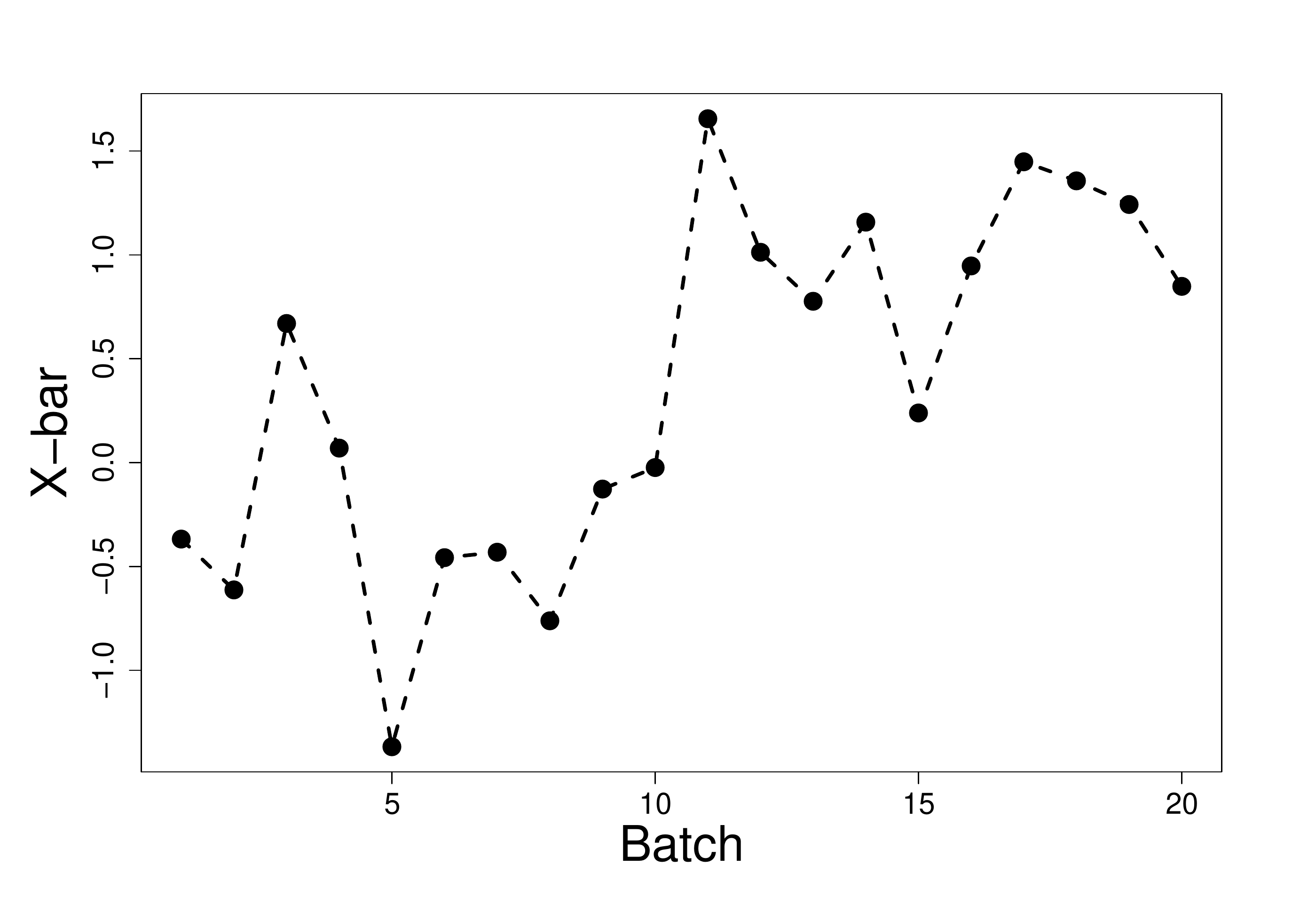}
    \caption{The batch averages for the data in the 20 batches of the first data set, as given in Table 4.2, showing an apparent change in location.}
    \label{fig:c4s2f1}
\end{figure}
The right hand column in Table \ref{tab:c4s2t1} is obtained as follows.

1. Each observation in batch 1 is ranked relative to the other observations in batch 1, converted to a rankit, and then to a sequential normal score SNS using the inverse of the standard normal distribution, as described in Section \ref{sec:c1s1}. The SNS is squared and summed over the five observations in batch 1 to get the graphing statistic $Z$ for batch 1. That value of $Z$ is a constant depending only on the batch size.

2. Each observation in subsequent batches is ranked relative to all of the observations in the preceding batches, converted to a rankit, and then to a SNS using the inverse of the standard normal distribution, as described in Section \ref{sec:c1s1}. The SNS is squared and summed over the five observations in that batch to get $Z$ for that batch. That $Z$ is approximately chi-squared with 5 degrees of freedom. It is given in the far right column of Table 4.2.

3. When the value of $Z$ exceeds the upper 0.005 quantile (for ARL = 200) of a chi-squared random variable with 5 degrees of freedom, which is 16.7, the reference set is frozen. That is, batch 11 exceeds the value 16.7, so the observations in batch 12 and all subsequent batches are ranked relative to the first 10 batches, as batch 11 was, because a change point has been observed.

The values of $Z$ for the first data set are shown in Figure \ref{fig:c4s2f2}. It is evident in the graph that the values of (SNS)$^2$ have increased from its theoretical mean of 5 as a result of the change in location of the original observations. The squared SNS test statistic, with an upper bound only, is sensitive to a change in location in either direction, and to an increase in spread.

\begin{figure}[t!]
    \centering
    \includegraphics[width=0.75\textwidth,angle=-90]{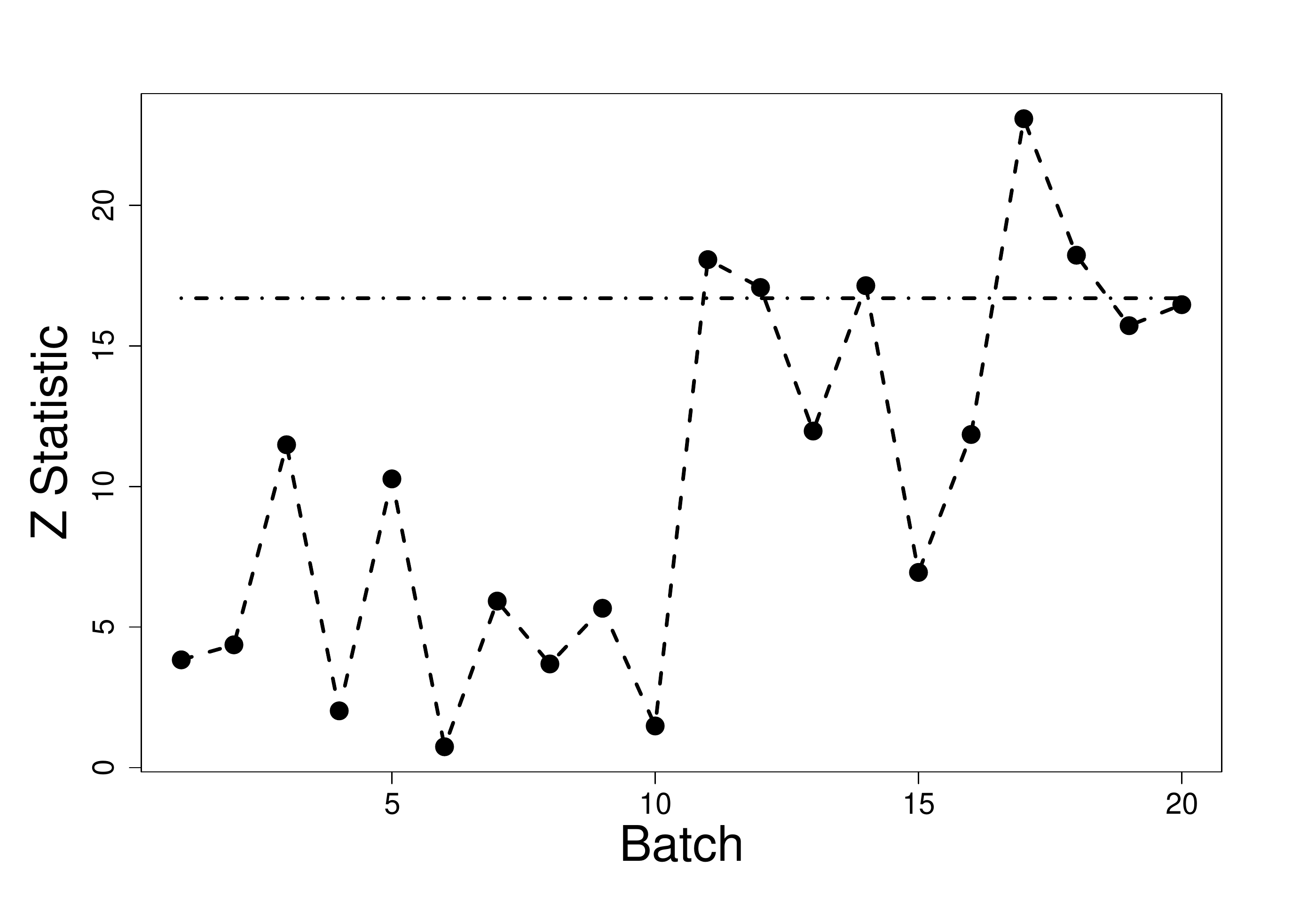}
    \caption{The charting statistic for the batch sum of the SNS$^2$ in the 20 batches of the first data set, given as Z in Table 4.2. A signal of change is sent with batch 11.}
    \label{fig:c4s2f2}
\end{figure}

The second data set also has 10 batches of size 5 from $N(0,1)$, but in this case they are followed by 10 batches of size 5 from $N(0,2^2)$, indicating a change in standard deviation from 1.0 to 2.0. As expected, the first test in \citet{qiu_2014} fails to detect a change in mean, but the second test signals a change in spread on the 12$^\text{th}$ batch, thus signaling a change on the 12$^\text{th}$ batch with the joint monitoring scheme.

The second data set is given in Table \ref{tab:c4s2t2}.

\begin{table}[t!]
    \caption{The second of two data sets used in Example 4.9 of \citet{qiu_2014}. The data set consists of 20 batches of 5 observations each. The first 10 batches are $N(0,1)$ and the second 10 batches are $N(0,2^2)$. Column 7 is the average of the batch, and column 8 is the sum of the squared values of SNS for each batch, which is compared with the chi-squared distribution with 5 degrees of freedom.}
    \centering
    \begin{tabular}{rrrrrcrr}
    \multicolumn{5}{c}{Second data set}   & Batch & \multicolumn{1}{c}{$\bar{X}$} & \multicolumn{1}{c}{Z for SNS$^2$} \\
    -0.762 & 0.419 & -1.040 & 0.712 & -0.633 & 1     & -0.261 & 3.835 \\
    0.563 & 0.661 & -1.658 & 1.028 & 1.128 & 2     & 0.344 & 6.648 \\
    -1.280 & 1.129 & -0.464 & -0.316 & 0.924 & 3     & -0.001 & 4.726 \\
    0.077 & 1.040 & 0.742 & 1.256 & 0.951 & 4     & 0.813 & 5.434 \\
    -0.481 & 0.203 & -0.032 & -1.196 & 0.624 & 5     & -0.176 & 1.990 \\
    -0.915 & 0.249 & -1.063 & -0.364 & -1.207 & 6     & -0.660 & 4.011 \\
    1.429 & 0.633 & -1.997 & -0.682 & -0.460 & 7     & -0.215 & 9.811 \\
    -0.983 & 0.495 & 0.726 & 0.667 & 0.955 & 8     & 0.372 & 2.254 \\
    -1.675 & -1.205 & -1.963 & 1.471 & 0.372 & 9     & -0.600 & 12.999 \\
    1.066 & 0.531 & 0.102 & 1.338 & 0.087 & 10    & 0.625 & 4.022 \\
    -0.782 & -0.500 & 2.310 & -1.729 & -1.733 & 11    & -0.487 & 11.660 \\
    -4.642 & 1.218 & 2.300 & -2.399 & -3.160 & 12    & -1.337 & \textbf{22.225} \\
    1.306 & -1.099 & 1.042 & -1.399 & -0.878 & 13    & -0.205 & 5.659 \\
    -1.355 & 1.918 & -2.936 & 0.368 & -2.870 & 14    & -0.975 & 16.386 \\
    -2.275 & -0.829 & 0.288 & 2.124 & -1.142 & 15    & -0.367 & 10.554 \\
    2.554 & 0.457 & -0.618 & 1.920 & 1.098 & 16    & 1.082 & 10.798 \\
    0.851 & 1.287 & -2.721 & -0.397 & 1.239 & 17    & 0.052 & 9.802 \\
    4.136 & -0.611 & 0.562 & 1.383 & 0.093 & 18    & 1.113 & 8.244 \\
    0.226 & 1.991 & -1.362 & -2.554 & -2.937 & 19    & -0.927 & 16.378 \\
    -0.627 & -3.407 & -2.701 & -2.204 & -2.199 & 20    & -2.228 & 22.633 \\
    \end{tabular}

    \label{tab:c4s2t2}
\end{table}

A graph of the values of $\bar{X}$, the batch means, is given in Figure \ref{fig:c4s2f3}. The erratic behavior of the later batches is a result of the increase in standard deviation of the generated observations.

\begin{figure}[t!]
    \centering
    \includegraphics[width=0.75\textwidth,angle=-90]{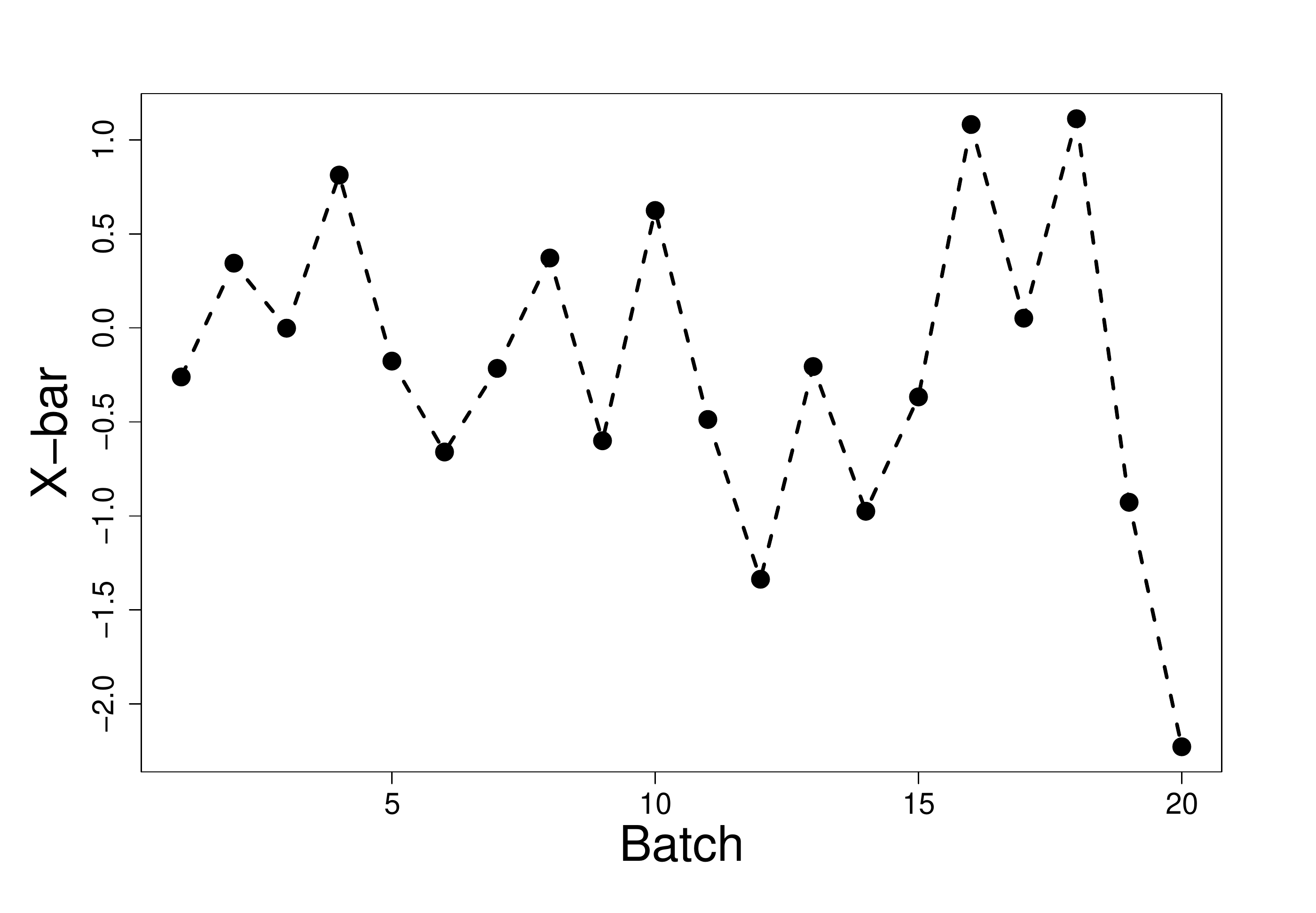}
    \caption{The batch averages for the data in the second data set, as given in Table 4.3, showing a possible change in spread.}
    \label{fig:c4s2f3}
\end{figure}

The values of the charting statistic $Z$ are found exactly the same as they were found in the first data set, with one slight change. The statistic $Z$ exceeds its upper bound 16.7 (for $\text{ARL}=200$) on batch 12, so the reference set is frozen at batch 11, instead of at batch 10 as in the first data set. All observations in batches 13 through 20 are ranked relative to only the first 11 batches, as were the observations in batch 12. A graph of the values of $Z$ for the second data set is given in Figure \ref{fig:c4s2f4}. Again, it is evident from the graph that the mean of $Z$ increases from its theoretical mean of 5 as the spread of the original data increases. Thus a single nonparametric control chart of squared sequential normal scores is able to detect changes in either location (both directions) or scale (increases only).

\begin{figure}[t!]
    \centering
    \includegraphics[width=0.75\textwidth,angle=-90]{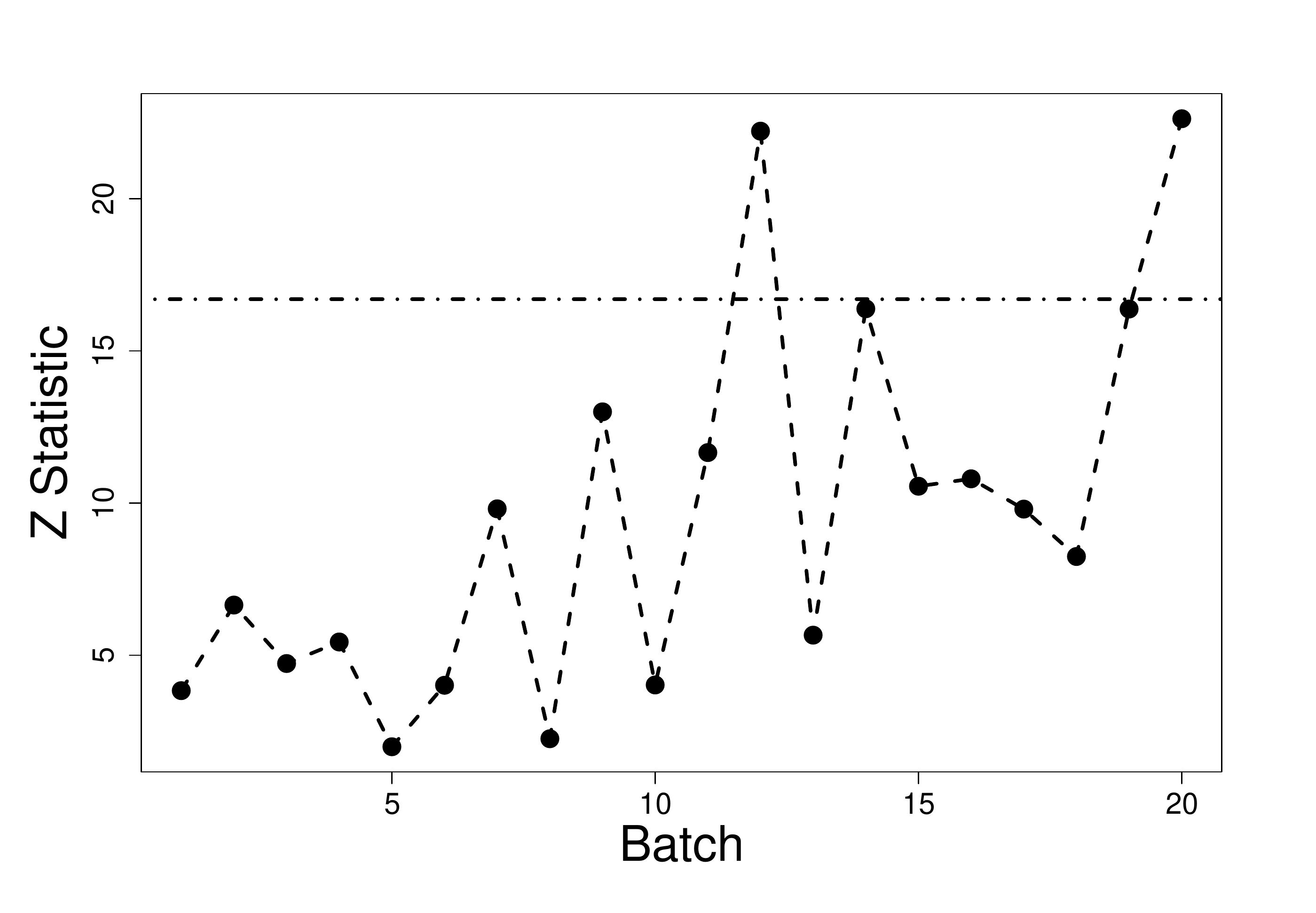}
    \caption{The sum of the SNS$^2$ in each batch for the data in the second data set, given as Z in Table 4.3, showing how a possible change in spread shows up as a possible change in location.}
    \label{fig:c4s2f4}
\end{figure}

The single graph of SNS$^2$ replaces two graphs for the first data set, the CUSUM on the batch means $\bar{X}$ ($k=0.5$, $h=0.881$) illustrated in Figure 4.10(b) on page 153 of \citet{qiu_2014} for the first data set, and the CUSUM on the standardized batch means $\bar{X}^2$ (because the data start with mean 0 and standard deviation 1) illustrated in Figure 4.10(c) using $k=1.848$ and $h=2.533$. The second graph of SNS$^2$ replaces the two graphs for the second data set, illustrating the same two CUSUMs, in Figures 4.10(e) and 4.10(f). Keep in mind that the SNS methods are nonparametric, while Figure 4.10 in \citet{qiu_2014} illustrates a parametric procedure designed to be optimal for the given shifts in location and scale.
\end{myexample}

Several nonparametric procedures have been proposed for detecting changes in either the location or spread or both. Most of them involve two control charts, such as was suggested in the above Example 4.9 \citet{qiu_2014} for a parametric test. Some of the more recent suggestions involve combining the two test statistics into one test statistic. It will be interesting to compare their power with the squared SNS procedure. The advantage of combining the two test statistics is that in cases where a slight change in location might not be detectable by itself, a slight change in spread at the same time might be sufficient to combine to signal an event.

\begin{mycomment}
A control chart using SNS instead of the squared SNS would be just as effective for detecting a change in location, but would be less effective in detecting a change in spread, because a change in spread could exceed both the upper bound and the lower bound. By squaring the SNS both types of change would be reflected in large values, exceeding the upper bound only. If it is desired to identify whether a change in SNS$^2$ is due to a change in location or to a change in scale, one simply analyzes a graph of SNS to see if a change in location is statistically significant. If it is not, then the change in SNS$^2$ is likely due to a change in scale, or to a combination of location and scale.
\end{mycomment}

\begin{mycomment}
These squared SNS charts with only an upper bound will not be effective in detecting a decrease in variance. Usually quality control is focused on detecting an increase in variance, but if a situation calls for signaling a decrease in spread, the squared SNS statistic can be used with a lower bound rather than an upper bound, if the location parameter remains unchanged. In order to make the test independent of the change in location, the batch data can first be “aligned” by subtracting the batch sample mean before finding the squared SNS, but such an adjustment removes the nonparametric nature of the method. At best it becomes only “robust” to the underlying distribution.
\end{mycomment}

\section{Detecting a Change in Both Location and Scale (SNS Method)}\label{sec:c4s3}

A second method for adapting the SNS to test for both location and scale involves squaring the data, rather than squaring the sequential normal scores. Before squaring the data, the observations need to be centered around a reasonable estimate  $\theta$  of the location parameter, such as the population median or mean. This location parameter is subtracted from each observation, and the result is squared.

That is, $Y_i = (X_i - \theta )^2$ is now the variable that is converted to a sequential normal score. These new squared observations will reflect any increase or decrease in location, or increase in standard deviation, by increasing in size. Thus they are converted to sequential normal scores in the usual way (see Section 1.1) and compared with the usual bounds in a Shewhart chart, or in a CUSUM or EWMA procedure.

\begin{myexample}
Figures \ref{fig:c4s3f1} and \ref{fig:c4s3f2} graph the batch graphing statistic on the batches of SNS in Example 4.2.1, computed on the $Y$'s. Figure \ref{fig:c4s3f1} is for the first data set, where the mean shifts with batch 11. This new test almost signals a shift at batch 11 with a value of 2.55, but not quite exceeding the upper limit 2.58 for $\text{ARL}=200$. Figure \ref{fig:c4s3f2} is for the second data set, where the spread increases with batch 11. This new test signals a shift at batch 12, when it exceeds the upper limit 2.58. The end result is almost the same as when the squared SNS is used in the previous section. This second method of detecting a change in both location and spread, which uses SNS instead of SNS$^2$, is useful for analyzing multivariate data as we will show in Chapter \ref{chap:c6}.

\begin{figure}[t!]
    \centering
    \includegraphics[width=0.75\textwidth,angle=-90]{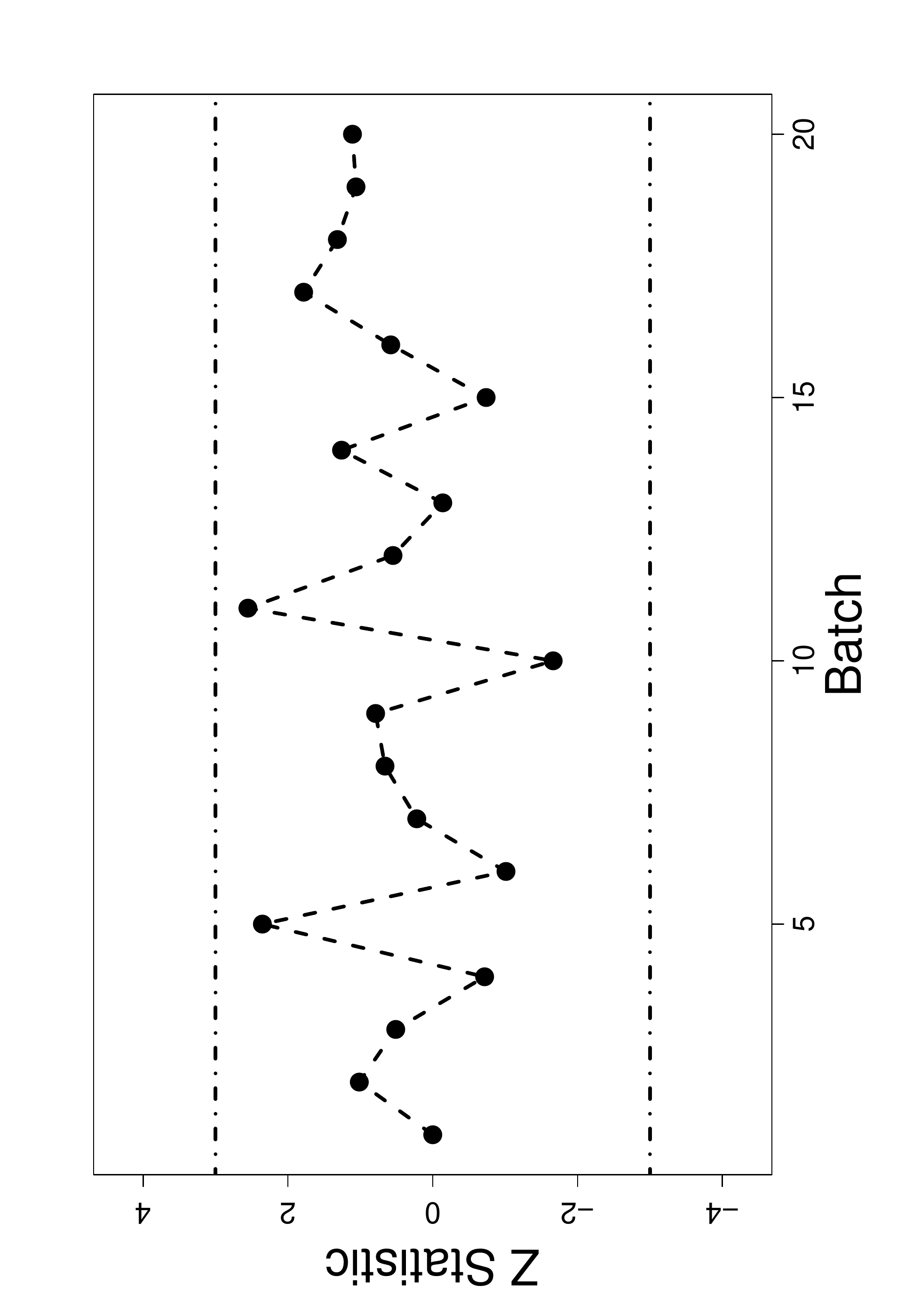}
    \caption{The batch graphing statistics for the batches of SNS values computed on the transformed data in the first data set (Table 4.2), where the transformation is $Y_i = (X_i - \theta)^2$, and where  $\theta$  is a location parameter, equal to 0 in this example. Batches 11-20 use only batches 1-10 as a reference set.}
    \label{fig:c4s3f1}
\end{figure}

\begin{figure}[t!]
    \centering
    \includegraphics[width=0.75\textwidth,angle=-90]{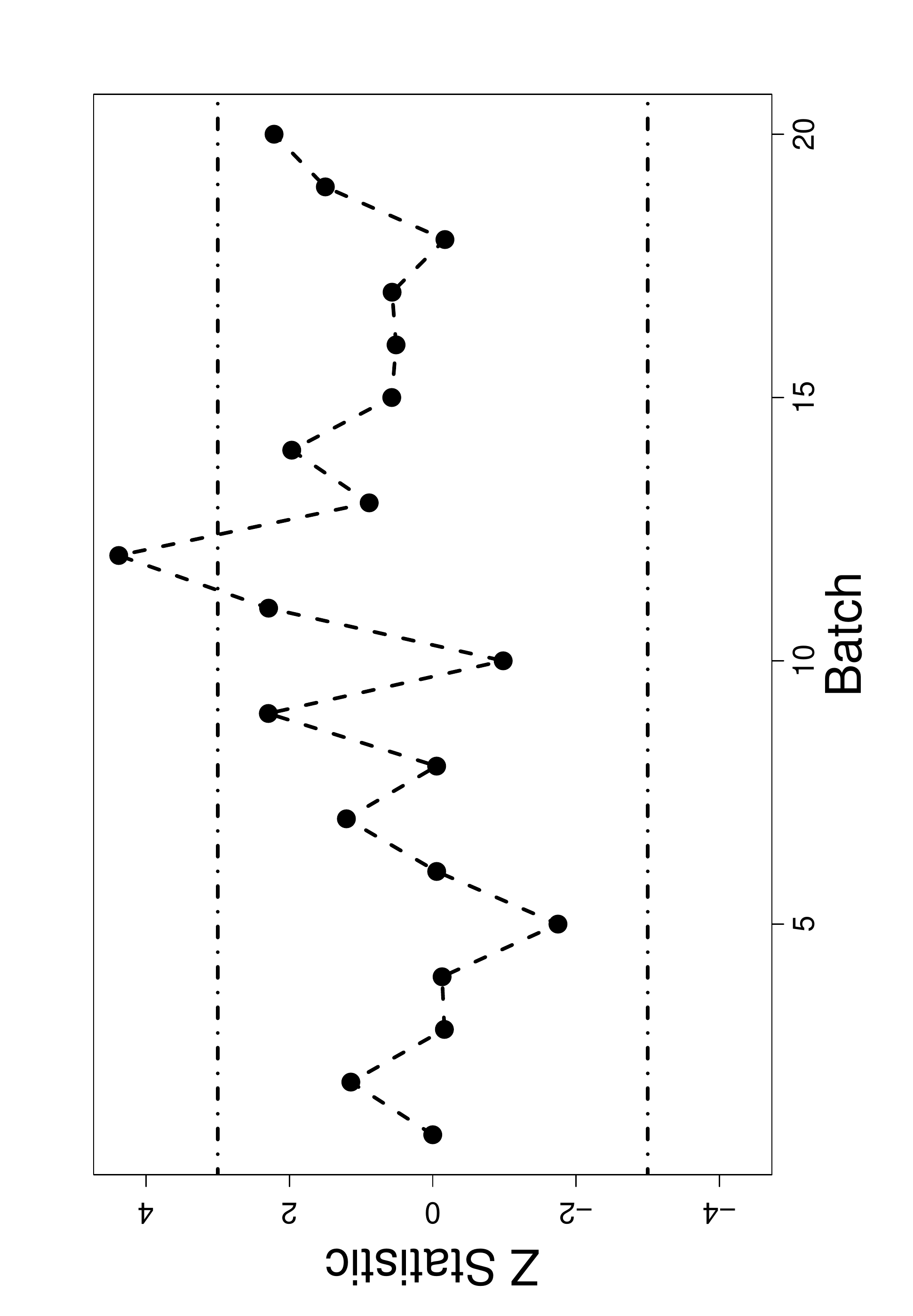}
    \caption{The batch graphing statistics for the batches of SNS values computed on the transformed data in the second data set (Table 4.3), where the transformation is $Y_i = (X_i - \theta)^2$, and where  $\theta$  is a location parameter, equal to 0 in this example. The graph sends a signal at batch 12. Batches 12--20 use only batches 1--11 as a reference set.}
    \label{fig:c4s3f2}
\end{figure}
\end{myexample}
\chapter{Conditional Sequential Normal Scores Given a Quantile $\theta$}\label{ch:c5}

\section{Using Conditional Sequential Normal Scores to Detect a Change in Location}\label{sec:c5s1}
As explained in Section \ref{sec:c1s2}, additional knowledge about the unknown distribution function may enhance the sensitivity of the statistical procedure. In particular, knowledge of a quantile can be used to improve the method. Assume a quantile $\theta$ is known, such as the median where $F(\theta)=0.5$. Then we can use conditional sequential normal scores in much the same way as we have demonstrated unconditional sequential normal scores in the previous sections.

\begin{myexample}
As an example, consider the data set used in Example 8.1 on page 319 of \citet{qiu_2014} to detect a change in location. It consists of 30 batches of observations where each batch has 10 observations. These are given in Table \ref{tab:c5s1t1}. The first 20 batches (rows) were generated from a $t_3$ distribution, and the last 10 batches were generated by $t_3+1$. Both distributions are symmetric, although that is not a necessary assumption for this method. The given quantile is the mean (and the median) of the first 20 batches, which is zero. In Qiu's example, the Signed Ranks statistic is found for each batch, independent of the other batches, as suggested by \citet{bakir_2004}. For this comparison we have converted the signed ranks statistic to its standard normal form by subtracting its mean and dividing by its standard deviation. It is reported as NSR in Table \ref{tab:c5s1t1}.

\begin{sidewaystable}[t!]
ll
    \caption{The data set used in Example 8.1 of \citet{qiu_2014}, and in Example 5.1.1 of this section. It consists of 30 batches of 10 observations in each batch. The column NSR contains the standard normal approximation to the signed ranks statistic used by \citet{qiu_2014}, and the column $Z$ contains the sum of the conditional sequential normal scores for each batch, divided by the square root of the batch size 10.}
    \centering
    \begin{tabular}{crrrrrrrrrrrr}
    Batch & \multicolumn{10}{c}{Observations COPIED FROM P. QIU'S FILES}                  & \multicolumn{1}{c}{NSR} & \multicolumn{1}{c}{$Z$} \\
    1     & -0.577 & -0.633 & 1.249 & 0.270 & -1.032 & -0.308 & 0.428 & 0.794 & -0.533 & 1.222 & 0.255 & 0.000 \\
    2     & -0.067 & -1.057 & -0.181 & 1.434 & 0.042 & -0.328 & 1.043 & 1.435 & 3.155 & 3.205 & 1.274 & 1.766 \\
    3     & 0.124 & 0.021 & -0.573 & 0.302 & 0.042 & 1.060 & -1.090 & 0.898 & 0.239 & -0.775 & 0.459 & -0.659 \\
    4     & -0.525 & 0.466 & 0.350 & -0.288 & -0.285 & -0.025 & -2.620 & -1.901 & -0.544 & -0.187 & -1.682 & -1.522 \\
    5     & -1.232 & 0.076 & -2.098 & -0.155 & 0.944 & 0.318 & 2.459 & 0.378 & 1.112 & -0.135 & 0.561 & 0.301 \\
    6     & 0.510 & -0.844 & -0.353 & 0.012 & 0.487 & -0.419 & -0.974 & 2.179 & -2.751 & -1.760 & -0.866 & -1.295 \\
    7     & -0.730 & -0.104 & -0.509 & 0.140 & 0.020 & -1.155 & -0.711 & -0.924 & -2.765 & -0.292 & -2.395 & -2.196 \\
    8     & -0.417 & 5.786 & 0.713 & 0.788 & -0.005 & 2.550 & 0.298 & -0.338 & 0.052 & 0.560 & 1.784 & 1.982 \\
    9     & 1.282 & 0.710 & 0.519 & 0.258 & -0.387 & 1.150 & 0.601 & 1.625 & -0.061 & -0.380 & 1.988 & 1.653 \\
    10    & 0.925 & -0.090 & -3.007 & 1.581 & -0.713 & 1.065 & -0.258 & -0.333 & 0.191 & 3.873 & 0.561 & 0.464 \\
    11    & 0.827 & -0.999 & -0.228 & -2.269 & 1.955 & 0.728 & -1.032 & 3.002 & 1.194 & 0.068 & 0.561 & 0.508 \\
    12    & 1.598 & 0.338 & -0.501 & 0.745 & 0.517 & 0.828 & -2.418 & 0.996 & 0.739 & 3.882 & 1.682 & 1.633 \\
    13    & -1.479 & -0.031 & -0.890 & -1.173 & 0.884 & -1.686 & 1.450 & 0.483 & 0.702 & 0.085 & -0.561 & -0.590 \\
    14    & -1.262 & 1.477 & 5.203 & 0.206 & -0.987 & -1.030 & 6.325 & 0.120 & 1.029 & 1.301 & 1.376 & 1.731 \\
    15    & -0.059 & -0.176 & 0.587 & 1.977 & -0.336 & 0.965 & 2.177 & -0.281 & 0.469 & 2.773 & 1.784 & 1.699 \\
    16    & -0.179 & 0.690 & 1.441 & -0.765 & 0.750 & -0.658 & 0.402 & 0.191 & -0.705 & 0.371 & 0.561 & 0.214 \\
    17    & 0.801 & -0.292 & -0.303 & -0.866 & -0.377 & 2.428 & -10.94 & -0.009 & 1.103 & -0.779 & -0.459 & -0.810 \\
    18    & 0.333 & -0.845 & 1.758 & -1.007 & -10.15 & -4.631 & -5.097 & 0.166 & 0.345 & 0.037 & -1.070 & -2.184 \\
    19    & 1.252 & -3.496 & 0.108 & 0.191 & -0.917 & 1.583 & 0.344 & -0.228 & 1.760 & -3.460 & 0.051 & -0.205 \\
    20    & 2.359 & 1.430 & 2.043 & -0.629 & -1.963 & -1.748 & 0.464 & -0.738 & -0.614 & -1.239 & -0.153 & -0.477 \\
    21    & 0.078 & 3.303 & 0.523 & 1.114 & 0.980 & 3.034 & 3.861 & 2.772 & -0.593 & 0.760 & 2.497 & \bf{3.124} \\
    22    & 1.374 & 2.265 & 1.161 & 1.315 & 2.175 & 1.306 & -1.495 & 1.257 & 2.221 & 0.602 & 2.090 & 2.850 \\
    23    & 0.294 & 0.090 & 0.349 & -0.576 & -0.447 & 2.712 & -0.487 & 5.125 & 1.048 & 4.445 & 1.274 & 1.904 \\
    24    & 1.164 & 0.862 & 1.883 & 0.737 & 1.981 & 0.939 & 2.121 & 1.255 & 1.911 & -0.253 & 2.701 & 3.017 \\
    25    & 1.286 & 1.086 & 0.433 & -2.654 & 1.915 & 0.807 & 2.594 & 3.136 & 0.795 & 1.843 & 1.886 & 2.655 \\
    26    & 0.694 & 0.913 & -0.925 & 1.773 & -0.793 & 1.130 & 1.604 & 0.200 & 1.852 & 2.871 & 1.988 & 2.071 \\
    27    & -1.227 & 3.695 & -2.125 & 1.431 & 1.494 & 1.616 & 2.169 & 2.309 & -2.388 & 0.914 & 1.070 & 1.608 \\
    28    & 2.083 & -3.002 & 0.989 & 2.195 & 1.522 & -1.948 & 0.097 & -1.190 & 1.599 & 1.996 & 0.866 & 1.110 \\
    29    & 1.198 & 1.377 & -2.482 & 0.899 & 1.348 & -0.493 & 1.843 & 1.501 & 0.820 & 2.099 & 1.682 & 2.077 \\
    30    & 1.731 & 1.860 & 0.397 & 1.789 & 4.945 & 2.189 & 6.197 & 1.939 & 0.333 & 0.984 & 2.803 & 4.129 \\
    \end{tabular}
    \label{tab:c5s1t1}
\end{sidewaystable}
The graphing sequential normal score statistic is reported as $Z$ in the far right column of Table \ref{tab:c5s1t1}. It is found as follows.

1. The conditional ranks for observations in each batch are found as described in Section \ref{sec:c1s2}. As an example, the conditional ranks for the first five batches are given in Table \ref{tab:c5s1t2}.

\begin{table}[t!]
  \caption{The conditional ranks for the first five batches in Table \ref{tab:c5s1t1}. The first batch is ranked within itself, as described in Section \ref{sec:c1s2}, given the $\text{median} = 0$. Subsequent batches are ranked relative to only the previous batches. The number of observations used in the rankings are given in the last two columns, one for observations less than or equal to 0, and the other for observations
greater than 0.}
    \centering
    \begin{tabular}{crrrrrrrrrrrr}
    Batch & \multicolumn{10}{c}{CONDITIONAL RANKS}                                        & \multicolumn{1}{c}{$N^{-}$} & \multicolumn{1}{c}{$N^{+}$} \\
    1     & 3     & 2     & 5     & 1     & 1     & 5     & 2     & 3     & 4     & 4     & 5     & 5 \\
    2     & 6     & 1     & 6     & 6     & 1     & 5     & 4     & 6     & 6     & 6     & 6     & 6 \\
    3     & 2     & 1     & 5     & 3     & 1     & 6     & 1     & 5     & 2     & 3     & 10    & 12 \\
    4     & 9     & 9     & 8     & 11    & 11    & 13    & 1     & 1     & 8     & 11    & 13    & 19 \\
    5     & 3     & 4     & 2     & 19    & 13    & 8     & 19    & 9     & 15    & 19    & 21    & 21 \\
    \end{tabular}
    
    \label{tab:c5s1t2}
\end{table}

2. The conditional ranks are converted to conditional rankits as described in Section \ref{sec:c1s2}, and then to unconditional rankits $P_{i,j}$. For example the unconditional rankits are given in Table \ref{tab:c5s1t3} for the first five
batches.
\begin{table}[t!]
    \caption{The unconditional rankits (see equation 1.2.1) for the first five batches in Table \ref{tab:c5s1t1}, using the conditional ranks in Table \ref{tab:c5s1t2}, and the equations found in Section \ref{sec:c1s2}.}
    \centering
    \begin{tabular}{crrrrrrrrrr}
    Batch & \multicolumn{10}{c}{UNCONDITIONAL RANKITS $P_{i,j}$} \\
    1     & 0.250 & 0.150 & 0.950 & 0.550 & 0.050 & 0.450 & 0.650 & 0.750 & 0.350 & 0.850 \\
    2     & 0.458 & 0.042 & 0.458 & 0.958 & 0.542 & 0.375 & 0.792 & 0.958 & 0.958 & 0.958 \\
    3     & 0.563 & 0.521 & 0.225 & 0.604 & 0.521 & 0.729 & 0.025 & 0.688 & 0.563 & 0.125 \\
    4     & 0.327 & 0.724 & 0.697 & 0.404 & 0.404 & 0.481 & 0.019 & 0.019 & 0.288 & 0.404 \\
    5     & 0.060 & 0.583 & 0.036 & 0.440 & 0.798 & 0.679 & 0.940 & 0.702 & 0.845 & 0.440 \\
    \end{tabular}
     
    \label{tab:c5s1t3}
\end{table}

3. The unconditional rankits are converted to conditional sequential normal scores by transforming
them using the inverse of the standard normal distribution function. The conditional SNS for the first five batches are given in Table \ref{tab:c5s1t4}. The sum of the conditional SNS in each batch is divided by the square root of 10 (the batch size) to obtain a graphing statistic that is approximately standard normal in distribution, listed as Z in Tables 5.1 and 5.4. This statistic will not always be zero for batch 1, so the actual computations in batch 1 are needed here.

\begin{table}[t!]
    \caption{The conditional sequential normal scores for the first five batches in Table \ref{tab:c5s1t1}, using the inverse of the standard normal distribution function on the unconditional rankits in Table \ref{tab:c5s1t3}. The Z column on the far right is the sum of the SNS in that batch, divided by the square root of the batch size
10.}
    \centering

    \begin{tabular}{crrrrrrrrrrr}
    Batch & \multicolumn{10}{c}{CONDITIONAL SNS (BATCH DATA, GIVEN MEDIAN = 0)}  & \multicolumn{1}{c}{$Z$} \\
    1     & -0.674 & -1.036 & 1.645 & 0.126 & -1.645 & -0.126 & 0.385 & 0.674 & -0.385 & 1.036 & 0.000 \\
    2     & -0.105 & -1.732 & -0.105 & 1.732 & 0.105 & -0.319 & 0.812 & 1.732 & 1.732 & 1.732 & 1.766 \\
    3     & 0.157 & 0.052 & -0.755 & 0.264 & 0.052 & 0.610 & -1.960 & 0.489 & 0.157 & -1.150 & -0.659 \\
    4     & -0.448 & 0.594 & 0.517 & -0.243 & -0.243 & -0.048 & -2.070 & -2.070 & -0.558 & -0.243 & -1.522 \\
    5     & -1.559 & 0.210 & -1.803 & -0.150 & 0.833 & 0.464 & 1.559 & 0.531 & 1.016 & -0.150 & 0.301 \\
    \end{tabular}
    
    \label{tab:c5s1t4}
\end{table}

4. Once the sequence is declared “out of control” the reference sample is frozen. In this example the Z statistic exceeded its 3-sigma limit with batch 21, so all the observations in batches 22 to 30 were ranked relative to only the first 20 batches, as was batch 21. 

The objective is to see how the shift of one unit upwards affects the graphing statistic. The result is given in Figure \ref{fig:c5s1f1}. The sequential normal scores are shown by the  squares. The  circles represent the alternative method proposed by \citet{bakir_2004} which uses the Wilcoxon Signed Ranks test on each batch separately, comparing the observations with the proposed median $\theta = 0$. 

\begin{figure}[t!]
    \centering
    \includegraphics[width=0.75\textwidth,angle=-90]{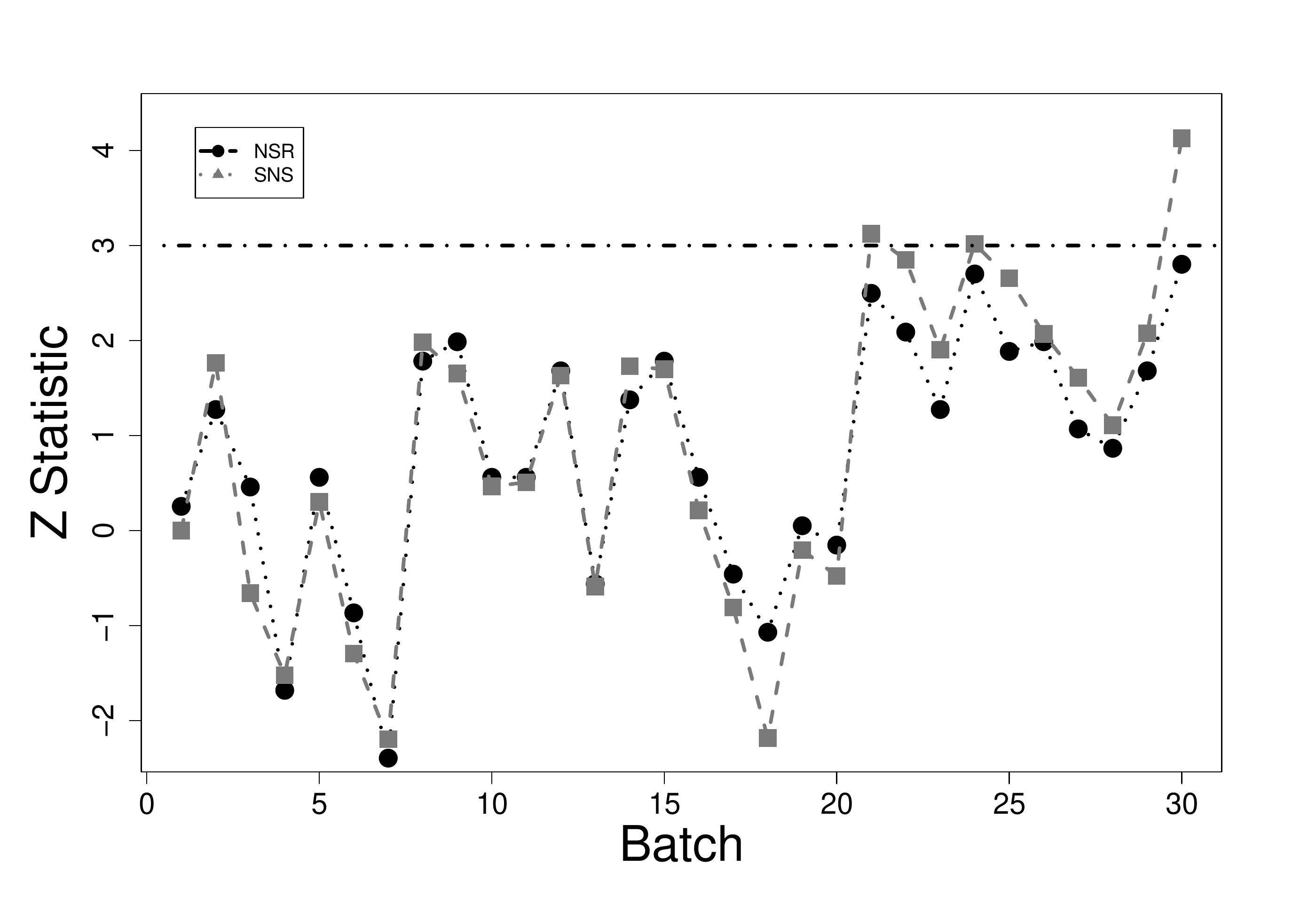}
    \caption{The summary statistics Z for 30 batches of size 10, using a nonparametric signed ranks statistic NSR, and the nonparametric SNS method. The SNS method sends a signal (greater than 3) on batch 21, the actual point of change, while the NSR method does not send a comparable signal until batch 30 ($p<0.00135$ using exact probabilities for signed ranks).}
    \label{fig:c5s1f1}
\end{figure}

Bakir’s test assumes the unknown distribution function is symmetric about zero, which it is in this example, but symmetry is not a necessary assumption for the Sequential Normal Scores test. Another difference between the two tests is that Bakir’s test does not use the information contained in the previous batches, while the SNS test ranks each batch relative to all of the previous batches prior to the first signal. Both tests successfully detect the shift in the 21$^\text{st}$ batch at $p=0.005$, although only the SNS test exceeds the upper 3-sigma boundary, usually assumed in Shewhart charts like this one. When the exact probabilities are used in Bakir’s test the final batch exceeds the equivalent of the 3-sigma limit.

\begin{mycomment}
The batches used in this discussion and in the example were all the same size. This is not necessary. The adjustment for unequal sizes is obvious, because each rankit depends on the actual number of observations involved in the ranking, which varies from one batch to another. 
\end{mycomment}

\begin{mycomment}
The method described here is self-starting, which means it does not depend on a known series of observations that are “in control” before starting. The adjustment for prior data is simple. The first “batch” may contain a large number of observation, which includes all the “in control” observations, but is treated only as the first batch. An obvious adjustment for the batch size needs to be made.
\end{mycomment}

\begin{mycomment}
In practice, once a batch exceeds the upper or lower control limit, it is no longer considered part of the reference sample. That means that in this example batch 21 is considered “out of control” in reference to the first 20 batches. Then batch 22 and subsequent batches are ranked relative only to the first 20 batches. This improves the sensitivity of the test because “out-of-control” batches are not being mixed in with known “in-control” batches.
\end{mycomment}
\end{myexample}

\section{CUSUM Variation to Detect a Change in Location}\label{sec:c5s2}

\begin{myexample}
It is a simple matter to apply the CUSUM variation to conditional sequential normal scores, just as it was applied to regular SNS in Section \ref{sec:c3s2}. To illustrate, consider the data set used in Example 8.3 on page 326 of \citet{qiu_2014}, and given in Table \ref{tab:c5s2t1}. It consists of 30 batches of observations where each batch has 6 observations. It is assumed that the data comes from a production process with a symmetric distribution with median zero, although the symmetry assumption is not needed for this SNS method. All we will assume is that the median is known, and in this case it is zero.

\begin{table}[t!]
    \caption{The data set used in Example 8.3 of \citet{qiu_2014} and in Example 5.2.1 in this section. The data consists of 30 batches of 6 observations in each batch. The columns on the right represent the graphing statistics Z of the signed ranks statistic NSR and the sequential normal scores statistic SNS, and the respective CUSUM statistics. The SNS statistics for batches 21-30 were compared only with the first 20 batches because a signal was obtained with the CUSUM at batch 21.}
    \centering
    \begin{tabular}{crrrrrrrrrr}
    Batch & \multicolumn{6}{c}{COPIED FROM P. QIU'S FILES} & \multicolumn{1}{l}{$Z_{\text{NSR}}$} & \multicolumn{1}{l}{$Z_{\text{SNS}}$} & \multicolumn{1}{l}{$C^+_{\text{NSR}}$} & \multicolumn{1}{l}{$C^+_{\text{SNS}}$} \\
    1     & 0.015 & 0.020 & -0.041 & 2.936 & 0.583 & 0.330 & 1.572 & 1.303 & 0.734 & 0.465 \\
    2     & -0.765 & 0.066 & 0.603 & 0.319 & -0.612 & 0.585 & -0.105 & 0.448 & 0.000 & 0.074 \\
    3     & 0.696 & 0.468 & -0.785 & -0.669 & 2.052 & 0.163 & 0.524 & 0.681 & 0.000 & 0.000 \\
    4     & -0.185 & 0.257 & 0.065 & 0.906 & -0.786 & 1.392 & 0.943 & 0.523 & 0.105 & 0.000 \\
    5     & 0.565 & -0.504 & -0.445 & -0.316 & 1.138 & 0.419 & 0.524 & 0.588 & 0.000 & 0.000 \\
    6     & -0.113 & -0.399 & -0.542 & 0.457 & 2.083 & -0.759 & -0.314 & 0.061 & 0.000 & 0.000 \\
    7     & 0.502 & -1.134 & -0.349 & -1.258 & -0.794 & -0.672 & -1.782 & -2.911 & 0.000 & 0.000 \\
    8     & -0.748 & -0.674 & 2.734 & 0.474 & -0.881 & 0.071 & -0.314 & -0.231 & 0.000 & 0.000 \\
    9     & 1.195 & 0.629 & 0.135 & 1.157 & -0.701 & 1.650 & 1.572 & 1.784 & 0.734 & 0.945 \\
    10    & -0.402 & -0.266 & 1.039 & -0.437 & 1.603 & -0.263 & 0.105 & 0.551 & 0.000 & 0.658 \\
    11    & -0.712 & -0.740 & 0.705 & 0.268 & 0.038 & 0.480 & -0.105 & -0.015 & 0.000 & 0.000 \\
    12    & -0.403 & -2.126 & 0.536 & 0.248 & 0.238 & -0.507 & -0.524 & -0.869 & 0.000 & 0.000 \\
    13    & 0.155 & 0.122 & 0.602 & -0.148 & -0.538 & 0.074 & 0.524 & 0.323 & 0.000 & 0.000 \\
    14    & -1.364 & -1.211 & -0.309 & 0.320 & -0.142 & -0.348 & -1.572 & -1.611 & 0.000 & 0.000 \\
    15    & 0.537 & -0.422 & -0.056 & 1.423 & 0.098 & -0.076 & 0.734 & 0.765 & 0.000 & 0.000 \\
    16    & -0.859 & 0.472 & 0.281 & -0.135 & -0.153 & -0.877 & -0.734 & -0.876 & 0.000 & 0.000 \\
    17    & -1.485 & -0.222 & -0.380 & -0.766 & -0.129 & 0.233 & -1.572 & -1.551 & 0.000 & 0.000 \\
    18    & -0.602 & 0.119 & -0.071 & 0.478 & 0.384 & 0.471 & 0.734 & 0.513 & 0.000 & 0.000 \\
    19    & -0.413 & 0.412 & -0.422 & 0.053 & -0.481 & -0.838 & -1.572 & -1.012 & 0.000 & 0.000 \\
    20    & -0.694 & -0.066 & -0.626 & -0.228 & 0.473 & 0.187 & -0.943 & -0.430 & 0.000 & 0.000 \\
    21    & 1.109 & -0.520 & 1.309 & 0.576 & 1.692 & 1.026 & 1.992 & 2.471 & \bf{1.153} & \bf{1.632} \\
    22    & 1.076 & 0.687 & 1.560 & 0.730 & 0.223 & 1.184 & 2.201 & 2.856 & 2.516 & 3.649 \\
    23    & 0.772 & 1.025 & 1.230 & 1.559 & 1.407 & 0.483 & 2.201 & 3.228 & 3.879 & 6.039 \\
    24    & -1.409 & 1.254 & 1.568 & 1.019 & 0.750 & 1.965 & 1.363 & 2.154 & 4.403 & 7.355 \\
    25    & 0.790 & 1.667 & 0.563 & 2.175 & 0.580 & 0.422 & 2.201 & 3.066 & 5.766 & 9.582 \\
    26    & 0.947 & 1.060 & 0.592 & 1.916 & 0.471 & 0.382 & 2.201 & 2.678 & 7.128 & 11.422 \\
    27    & 1.374 & 1.243 & 1.082 & 0.469 & 1.010 & 0.498 & 2.201 & 2.854 & 8.491 & 13.437 \\
    28    & 2.304 & 1.262 & 1.309 & 0.509 & 0.778 & 0.981 & 2.201 & 3.390 & 9.854 & 15.989 \\
    29    & 1.505 & 0.582 & 0.982 & 2.231 & 1.147 & 0.600 & 2.201 & 3.359 & 11.217 & 18.509 \\
    30    & 1.644 & 1.156 & 0.818 & 1.801 & 1.104 & 0.596 & 2.201 & 3.413 & 12.579 & 21.084 \\
    \end{tabular}
    \label{tab:c5s2t1}
\end{table}

The column labeled $Z_{\text{NSR}}$ is the large-sample standard normal approximation to the signed ranks test statistic used in Qiu's Example 8.3. Next to it is the charting statistic $Z_{\text{SNS}}$ used for the conditional SNS, computed in the same way as described in the previous section. Figure \ref{fig:c5s2f1}  compares the two charting statistics $Z_{\text{NSR}}$ and $Z_{\text{SNS}}$. Both statistics are approximately standard normal, but the NSR statistic reaches its maximum value (when all six observations in a batch are above the median) before the statistic can reach significance on its own. That is why a CUSUM is useful here.

\begin{figure}[t!]
    \centering
    \includegraphics[width=0.75\textwidth,angle=-90]{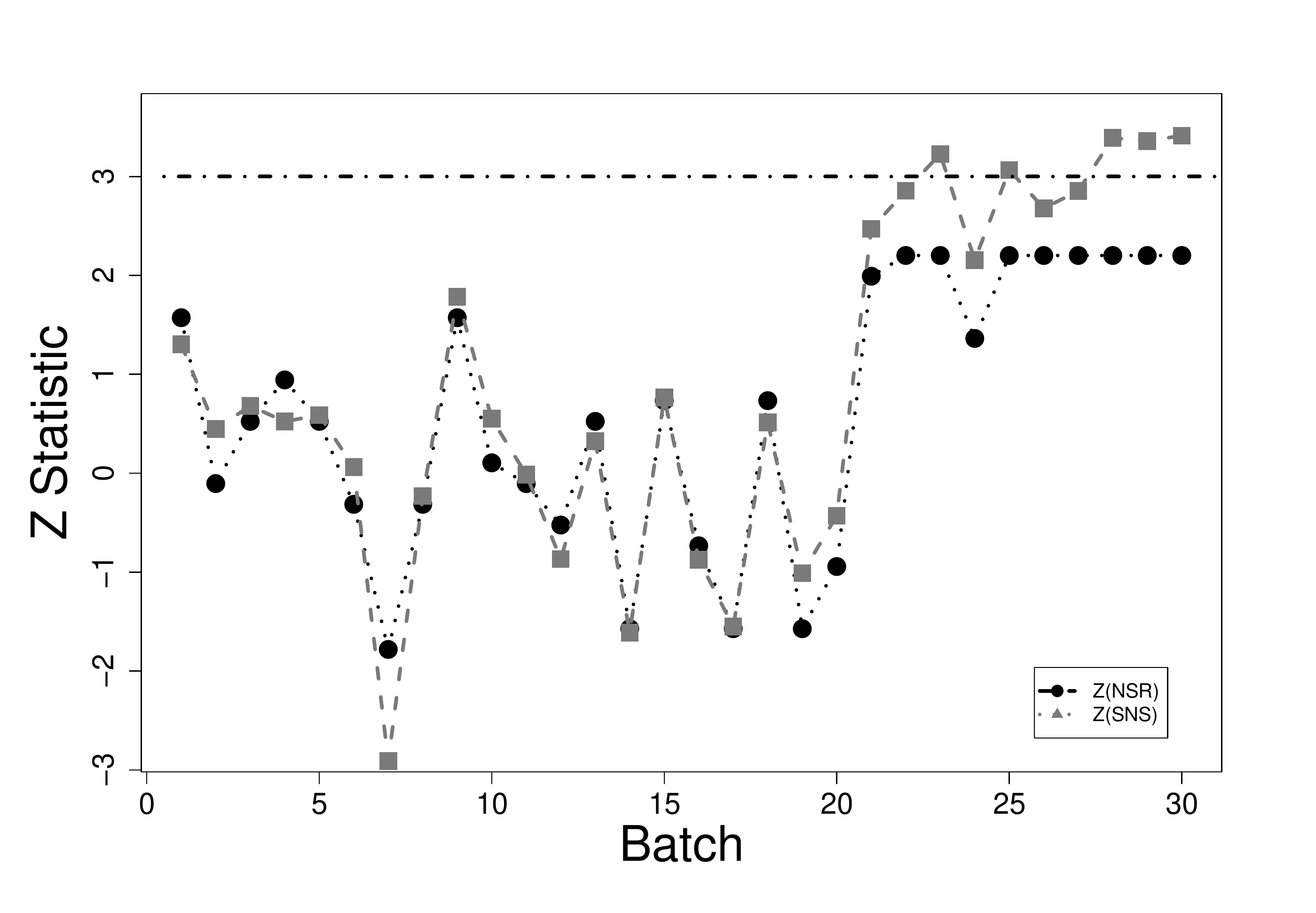}
    \caption{The charting statistics Z for 30 batches of size 6 in Table 5.5, using a nonparametric signed ranks test NSR and a test based on SNS. Although the SNS method signals (greater than 3) at batch 23, the NSR method is unable to achieve that limit.}
    \label{fig:c5s2f1}
\end{figure}

The columns labeled $C^+_{NSR}$ and $C^+_{SNS}$ are the positive CUSUM statistics using $k = 8/\sqrt{91} = 0.8386$ and $h = 10/\sqrt{91} = 1.083$ to match the $k=8$ and $h=10$ used in Qiu’s Example 8.3 for the original signed ranks statistic. Here, $91$ corresponds to the variance of the signed ranks statistic for sample size 6. Both CUSUMs then become $C_0^+=0$ and
\begin{align}
   C_i^+ = \max(0, Z_i + C_{i-1}^+ - 0.8386), i>0
\end{align}
with an upper boundary of 1.083, as discussed in Section \ref{sec:c3s2}. A graph of both CUSUMs is given in Figure \ref{fig:c5s2f2}.

\begin{figure}[t!]
    \centering
    \includegraphics[width=0.75\textwidth,angle=-90]{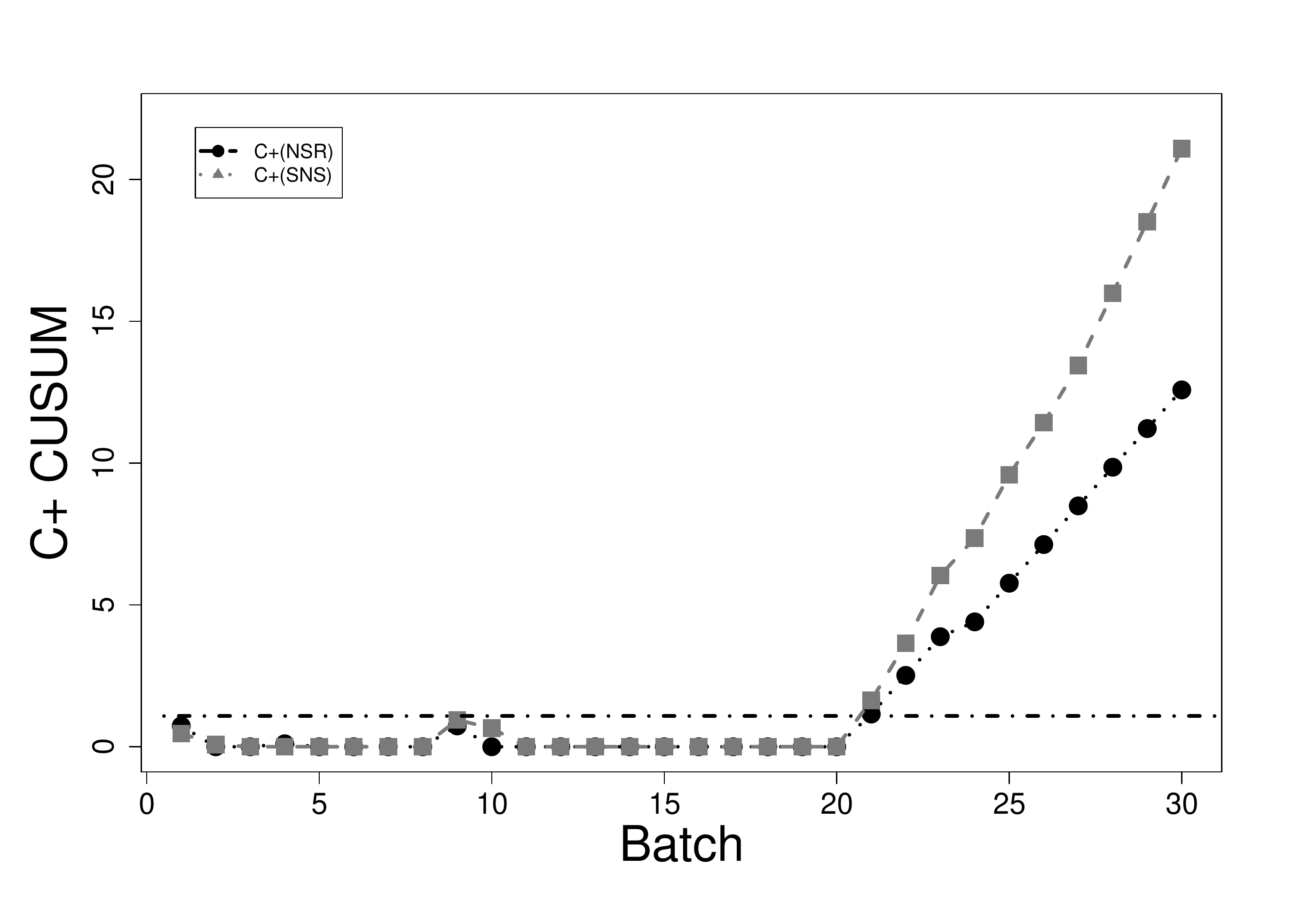}
    \caption{The positive CUSUM calculated on the NSR statistics and on the SNS statistics in Table 5.5. Both methods signal at batch 21, but the SNS method appears to be more sensitive to the change at that point.}
    \label{fig:c5s2f2}
\end{figure}

Both CUSUMs cross the upper boundary 1.083 on batch 21. At that point the reference sample for the conditional sequential normal scores is frozen to consist of only the first 20 batches, and batches 22--30 are ranked relative to only batches 1--20, as was batch 21. The SNS statistic appears to be more sensitive to the shift in location than the NSR statistic, at least for this set of data.
\end{myexample}

\section{EWMA Variation to Detect a Change in Location}\label{sec:c5s3}

Unlike the CUSUM procedure for sequential normal scores which uses two statistics, the EWMA method uses one moving average to detect location changes in both directions as discussed in Section \ref{sec:c2s3}. Because the sequential normal scores charting statistic $Z_i$ has a mean of zero and a standard deviation of 1 the EWMA is given by $E_0 = 0$ and
\begin{align}
   E_i = \lambda Z_i + (1-\lambda)E_{i-1} \text{ for } i>0,
\end{align}
for a “suitably chosen” smoothing constant $\lambda$. The upper and lower control limits $U$ and $L$ depend on $\lambda$ and another parameter $\rho$, which govern the value of the ARL, in the equation $U = \rho\sqrt{\lambda/(2-\lambda)}$, and $L = -U$. See Table \ref{tab:c2s3t1} in Section \ref{sec:c2s3} for upper and lower bounds for EWMA.

\begin{myexample}
As an example, consider the data set used in Example 8.5 on page 332 of \citet{qiu_2014} and given in Table \ref{tab:c5s3t1}. It consists of 30 batches of observations where each batch has 10 observations. It is assumed that the data comes from a production process with a symmetric distribution with median zero, although that symmetry assumption is not needed for this SNS method. All we will assume is that the median (or some other quantile) is known, and in this case it is zero. \citet{qiu_2014} use this data set to illustrate a method suggested by \citet{graham_etal_2011} based on the signed ranks statistic. The column labeled $Z_{\text{SR}}$ is the standardized signed ranks statistic, to be approximately standard normal.

\begin{sidewaystable}[t!]
    \caption{The data set used in Example 8.5 in \citet{qiu_2014}, and in Example 5.3.1 in this section. The data consist of 30 batches of 10 observations each. The last four columns on the right represent the charting statistics for the signed ranks method and for the sequential normal scores method, followed by their respective EWMA using $\lambda = 0.1$ and $\text{ARL} = 370$, resulting in an upper bound of 0.620.}
    \centering
    \begin{tabular}{crrrrrrrrrrrrrr}
    Batch & \multicolumn{10}{c}{Data copied from P. Qiu's webpage for Example 8.5}        & \multicolumn{1}{c}{$Z_{\text{SR}}$} & \multicolumn{1}{c}{$Z_{\text{SNS}}$} & \multicolumn{1}{c}{$E_{\text{SR}}$} & \multicolumn{1}{c}{$E_{\text{SNS}}$} \\
    1     & 0.015 & 0.020 & -0.041 & 2.936 & 0.583 & 0.330 & 0.109 & 0.309 & 0.692 & -0.054 & 2.090 & 1.516 & 0.209 & 0.152 \\
    2     & -0.765 & 0.066 & 0.603 & 0.319 & -0.612 & 0.585 & 0.076 & 0.560 & -0.777 & -0.180 & -0.255 & -0.352 & 0.163 & 0.101 \\
    3     & 0.696 & 0.468 & -0.785 & -0.669 & 2.052 & 0.163 & -0.228 & 0.230 & 0.407 & 1.728 & 1.172 & 1.283 & 0.264 & 0.219 \\
    4     & -0.185 & 0.257 & 0.065 & 0.906 & -0.786 & 1.392 & -2.409 & 0.568 & -0.250 & -0.905 & -0.051 & -0.843 & 0.232 & 0.113 \\
    5     & 0.565 & -0.504 & -0.445 & -0.316 & 1.138 & 0.419 & -0.210 & -0.437 & -0.420 & -0.526 & -0.561 & -0.307 & 0.153 & 0.071 \\
    6     & -0.113 & -0.399 & -0.542 & 0.457 & 2.083 & -0.759 & -0.053 & -0.408 & -0.529 & 0.443 & -0.663 & -0.240 & 0.071 & 0.040 \\
    7     & 0.502 & -1.134 & -0.349 & -1.258 & -0.794 & -0.672 & 0.374 & 0.082 & 0.010 & 0.123 & -1.070 & -1.743 & -0.043 & -0.138 \\
    8     & -0.748 & -0.674 & 2.734 & 0.474 & -0.881 & 0.071 & 1.304 & 0.309 & -0.222 & -0.908 & -0.051 & -0.172 & -0.044 & -0.142 \\
    9     & 1.195 & 0.629 & 0.135 & 1.157 & -0.701 & 1.650 & 0.505 & -0.018 & 0.147 & 0.661 & 1.988 & 2.198 & 0.159 & 0.092 \\
    10    & -0.402 & -0.266 & 1.039 & -0.437 & 1.603 & -0.263 & 0.644 & -0.182 & 0.104 & -1.497 & -0.153 & -0.018 & 0.128 & 0.081 \\
    11    & -0.712 & -0.740 & 0.705 & 0.268 & 0.038 & 0.480 & -1.520 & -0.424 & 0.026 & 0.652 & -0.357 & -0.450 & 0.080 & 0.028 \\
    12    & -0.403 & -2.126 & 0.536 & 0.248 & 0.238 & -0.507 & -0.313 & -0.270 & 0.184 & 0.133 & -0.866 & -0.612 & -0.015 & -0.036 \\
    13    & 0.155 & 0.122 & 0.602 & -0.148 & -0.538 & 0.074 & 0.025 & 0.559 & -0.517 & -0.259 & 0.255 & 0.206 & 0.012 & -0.012 \\
    14    & -1.364 & -1.211 & -0.309 & 0.320 & -0.142 & -0.348 & 0.254 & 0.019 & 0.965 & 0.540 & -0.357 & -0.370 & -0.025 & -0.048 \\
    15    & 0.537 & -0.422 & -0.056 & 1.423 & 0.098 & -0.076 & 0.667 & 1.175 & -0.578 & -0.240 & 0.866 & 1.105 & 0.064 & 0.068 \\
    16    & -0.859 & 0.472 & 0.281 & -0.135 & -0.153 & -0.877 & 0.060 & 0.916 & -0.618 & -0.315 & -0.663 & -0.599 & -0.008 & 0.001 \\
    17    & -1.485 & -0.222 & -0.380 & -0.766 & -0.129 & 0.233 & 0.243 & -0.531 & -0.502 & 1.054 & -1.172 & -1.053 & -0.125 & -0.104 \\
    18    & -0.602 & 0.119 & -0.071 & 0.478 & 0.384 & 0.471 & 0.262 & -0.491 & -0.019 & 0.022 & 0.459 & 0.352 & -0.066 & -0.059 \\
    19    & -0.413 & 0.412 & -0.422 & 0.053 & -0.481 & -0.838 & -0.418 & 1.231 & -0.400 & 0.304 & -1.070 & -0.487 & -0.167 & -0.102 \\
    20    & -0.694 & -0.066 & -0.626 & -0.228 & 0.473 & 0.187 & 0.156 & 0.801 & -0.404 & 0.864 & 0.255 & 0.315 & -0.125 & -0.060 \\
    21    & 1.748 & 1.061 & 0.054 & 1.678 & 2.219 & 2.475 & 0.559 & 1.372 & 0.707 & 2.079 & 2.803 & \bf{5.051} & 0.168 & 0.451 \\
    22    & 0.080 & 1.188 & 1.193 & 0.891 & 0.451 & 3.048 & 1.747 & 2.743 & 1.066 & 1.195 & 2.803 & 4.918 & 0.432 & \bf{0.898} \\
    23    & 1.624 & 0.254 & 1.378 & 1.082 & 1.276 & 0.380 & 1.120 & 0.545 & -0.078 & 0.870 & 2.701 & 3.539 & \bf{0.659} & 1.162 \\
    24    & 3.382 & 0.661 & 1.736 & 0.290 & 0.234 & 1.773 & 0.168 & 2.057 & 0.807 & 0.697 & 2.803 & 4.376 & 0.873 & 1.483 \\
    25    & 1.305 & 2.510 & 0.054 & 2.141 & 0.861 & 2.042 & 1.281 & -0.576 & 1.535 & 0.477 & 2.497 & 4.081 & 1.035 & 1.743 \\
    26    & 1.145 & 0.972 & -0.143 & 2.139 & 1.701 & 1.855 & 1.679 & 0.151 & 1.299 & 0.457 & 2.701 & 4.258 & 1.202 & 1.995 \\
    27    & 1.053 & 0.897 & 4.036 & 1.023 & 0.761 & 1.185 & 1.997 & 0.975 & 1.107 & 0.431 & 2.803 & 4.805 & 1.362 & 2.276 \\
    28    & 0.840 & -1.871 & -1.976 & 0.322 & 0.508 & 0.498 & 0.607 & 1.553 & -0.138 & 0.428 & 0.764 & 0.754 & 1.302 & 2.123 \\
    29    & 2.345 & 0.591 & 0.689 & 0.449 & 1.451 & 1.407 & 1.256 & 0.868 & 1.362 & 0.348 & 2.803 & 4.383 & 1.452 & 2.349 \\
    30    & 0.058 & 1.036 & 0.383 & 2.776 & 0.291 & -0.283 & 0.611 & 0.718 & 1.602 & 0.540 & 2.599 & 3.046 & 1.567 & 2.419 \\
    \end{tabular}
    \label{tab:c5s3t1}
\end{sidewaystable}

The data are converted to conditional sequential normal scores in the usual manner, as described in detail in previous sections. The sum of the SNS in each batch is divided by the square root of the batch size 10 to get the usual charting statistic $Z_{\text{SNS}}$, which is approximately standard normal. A comparison of these two statistics for this data set is given in Figure \ref{fig:c5s3f1}. Although both statistics are approximately standard normal, and both statistics are close together for the first 20 batches, on batch 21 both suddenly increase in size. The charting statistic for signed ranks achieves its mathematical maximum, which is still less than the usual 3-sigma bound, so there is a need for EWMA to help out.

\begin{figure}[t!]
    \centering
    \includegraphics[width=0.75\textwidth,angle=-90]{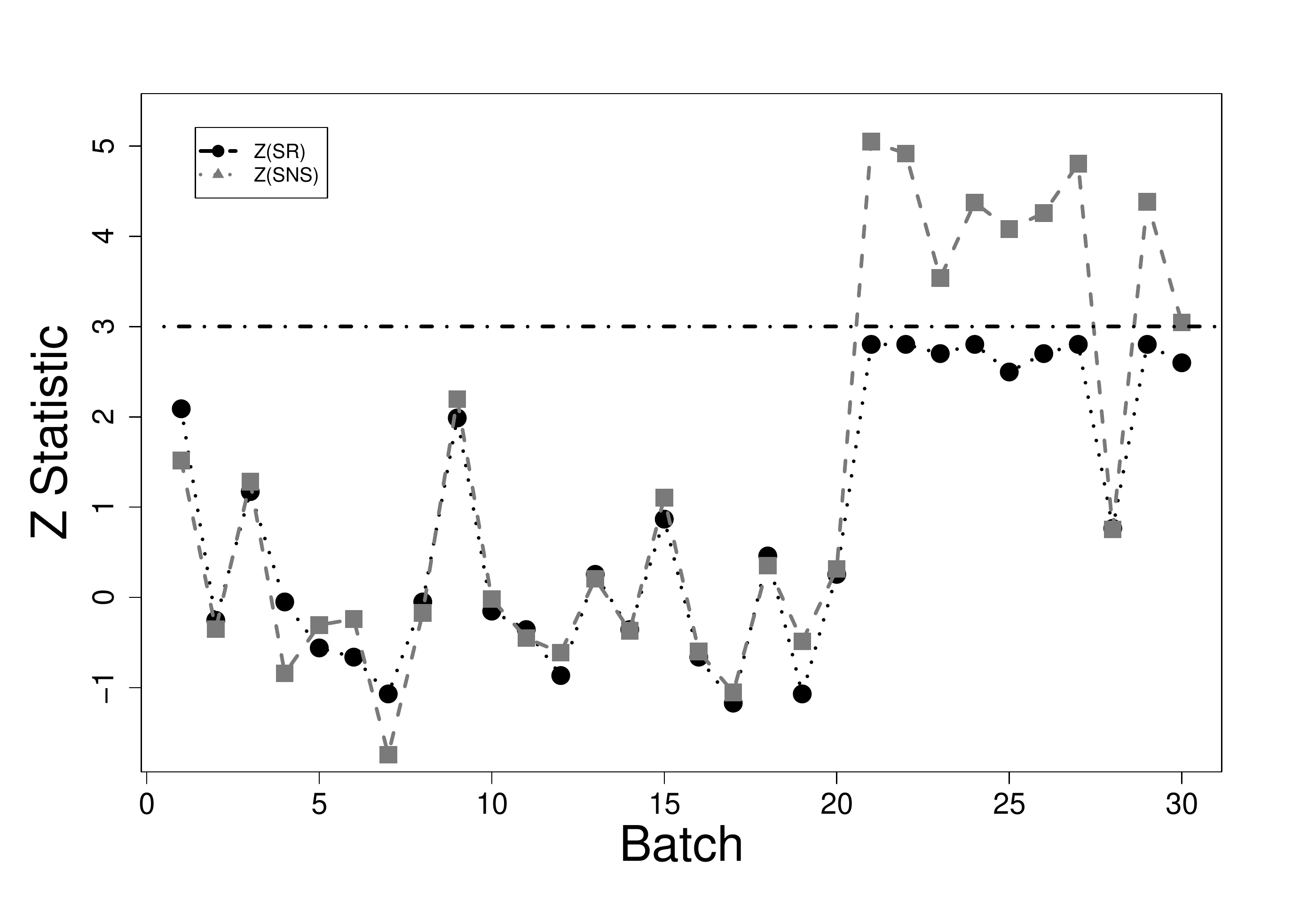}
    \caption{The charting statistics for 30 batches of size 10 given in Table \ref{tab:c5s3t1}, using a nonparametric signed ranks test (SR) and a test based on SNS, to illustrate the EWMA method. Although the SNS method signals (greater than 3) at batch 21, the SR method is unable to achieve that limit.}
    \label{fig:c5s3f1}
\end{figure}

These $Z$ statistics are the usual statistics that are presented to see if there is a shift in location, as was done in a previous section. However now we are interested in achieving greater statistical significance, so we will use the EWMA procedure. To match our example with Example 8.5 in \citet{qiu_2014} we need to use $\lambda = 0.1$ and $\rho = 2.701$, obtained from Table 5.1 on page 188 of Qiu (2014) to match an $\text{ARL} = 370$. This gives an upper limit of 0.620 as also shown in our Table 2.4 previously. The resulting EWMA values are given in the two columns on the far right in Table 5.6. Their graphs appear in Figure \ref{fig:c5s3f2}.

\begin{figure}[t!]
    \centering
    \includegraphics[width=0.75\textwidth,angle=-90]{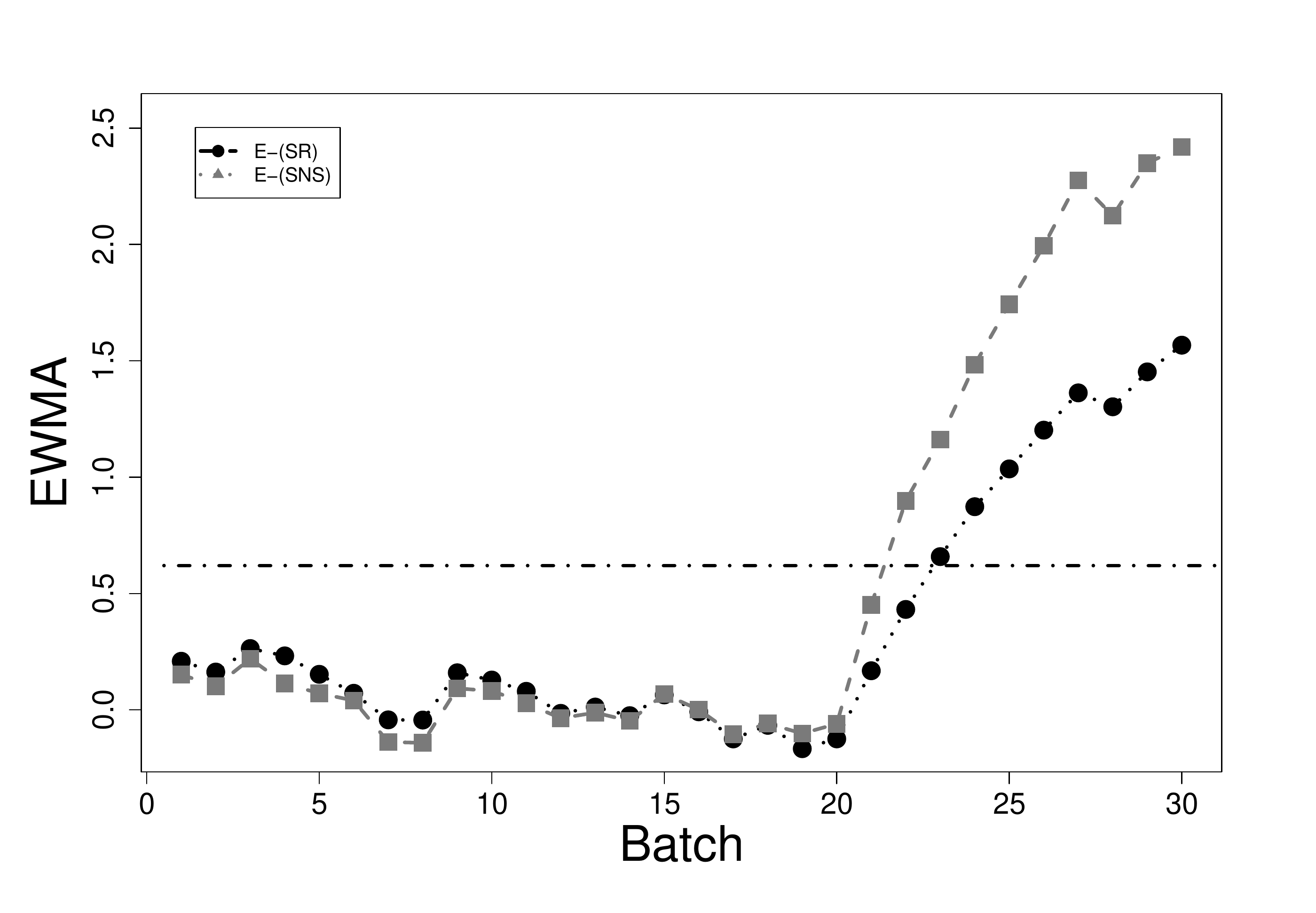}
    \caption{The EWMA calculated on the SR statistics and on the SNS statistics in Table \ref{tab:c5s3t1}, denoted by E-(SR) and E-(SNS) respectively. The EWMA on SR signals on batch 23 (exceeds 0.620) and the EWMA on SNS signals one batch sooner (batch 22).}
    \label{fig:c5s3f2}
\end{figure}
The EWMA for the signed ranks crosses the upper limit with batch 23, and the EWMA for SNS crosses the upper limit one batch sooner, with batch 22. At this point the reference set for SNS is frozen, so batches 23 to 30 are ranked relative to the first 21 batches, as was batch 22. It is clear that the EWMA for SNS is more sensitive to the shift in location than the EWMA for signed ranks, which is limited by its mathematical maximum.
\end{myexample}
\chapter{Multivariate Sequential Normal Scores}\label{chap:c6}
The previous examples all dealt with univariate data, where a single measurement was being monitored to detect a possible change in location or scale. Many times the measurement of quality is a multivariate quantity, where the components in the measurement may be correlated with each other. Most of the best parametric methods are based on some variant of Hotelling’s $T^2$, a multivariate extension of the $t$-test to multivariate observations. Attempts to develop nonparametric methods for this problem usually require large batches, at least 50 for extensions of the sign test and the signed ranks test, plus a known median, and symmetric distributions in the case of the signed ranks test. Attempts to transform the data to a multivariate normal distribution have met with only limited success.
In this paper we consider four variations of how Sequential Normal Scores can be used to develop nonparametric methods for analyzing multivariate data.

Variation 1. If we are interested only in a shift of the mean, then the values of $T^2$ on the raw data can be used to determine if an upper limit, with known probability, has been exceeded, by converting the values of $T^2$ to SNS as illustrated in Chapter 3. If a boundary has been exceeded as judged by SNS, or EWMA, or CUSUM, then the reference set is determined as the observations up to but not including the signal point, for use with additional observations.

Variation 2. If comparisons of the individual components are of interest, such as in multiple comparisons or contrasts, the individual components are transformed into approximately standard normal random variables, as illustrated in Examples 6.1.1 and 6.2.1 for batch data and 6.3.1 for single observations. Then the distribution of $T^2$ may be approximately chi-squared, but the method in Variation 1 above is used to be assured of accurate results.

Variation 3. If it is desired to test for a change in location, or an increase in scale, for one or more components, and if a base vector of location parameters is known or assumed to be known about which the scale measurement is computed, then the method illustrated in Example 6.4.1 may be used.

Variation 4. If it is desired to test for a change in location, or an increase in scale, for one or more components, but a base vector of location parameters is not known, so Variation 3 cannot be used, then the method illustrated in Example 6.4.2 may be used.

\section{Multivariate Sequential Normal Scores to Detect a Change in Location}\label{sec:c6s1}

 The sequential normal scores can transform the marginal distributions to approximate standard normal distributions, but that is no guarantee that the multivariate distribution will be multivariate normal, so that Hotelling’s $T^2$ will have a $\chi^2$ (or $F$) distribution. Therefore a second application of SNS is used to determine the significance of $T^2$. Until the use of the $\chi^2$ distribution as an approximation for the distribution of $T^2$ is justified we will consider the distribution of $T^2$ as unknown, and convert the values of $T^2$ to sequential normal scores for analysis.

Consider a sequence of multivariate observations with $p$ components in each observation. Each component is converted to a sequential normal score in the usual manner, one observation at a time for individual observations, or relative to previous batches of data for batch data. Each component in the multivariate random variable is ranked relative to only that component in previous observations, and converted to a SNS, as detailed in Section \ref{sec:c1s1}. Hotelling’s $T^2$ requires estimating the mean vector for each component, and the covariance matrix for the joint distribution. The covariance matrix for sequential normal scores is the sample correlation matrix because the mean is zero and the variance is close to 1.

\begin{myexample}\label{ex:c6s1e1}
As our first example we will illustrate this method on 1000 observations on bivariate data, used to illustrate two multivariate nonparametric methods in Example 9.1 on page 368 in \citet{qiu_2014}. The data ($X,Y$) consist of 20 batches with 50 observations in each batch. Graphs of these observations are given in Figure 9.3(a) for $X$ and Figure 9.3(b) for $Y$ on page 369 of \citet{qiu_2014}. Unfortunately there are too many observations to make it practical to put into a table for this paper, so we will just present data summaries and direct the interested reader to Peiqua Qiu’s home page where a downloadable data set is available. The batch means are plotted in Figures \ref{fig:c6s1f1} and \ref{fig:c6s1f2}.

\begin{figure}[t!]
    \centering
    \includegraphics[width=0.75\textwidth,angle=-90]{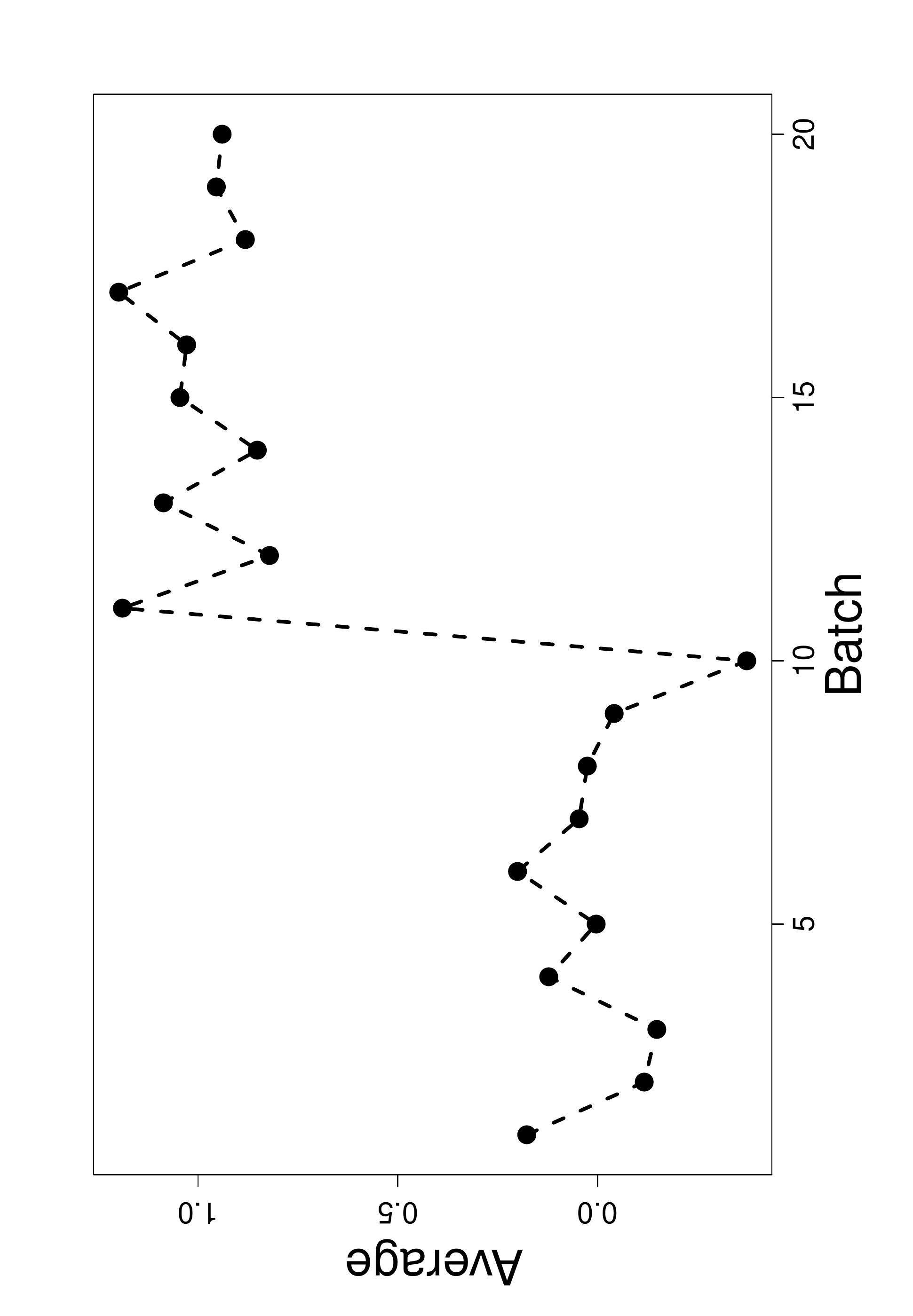}
    \caption{The batch averages for 20 batches of size 50, for the first variable in the bivariate $(X,Y)$ in Example 6.1.1.}
    \label{fig:c6s1f1}
\end{figure}
The $X$ variable appears to be following a Phase 1 distribution until batch 11, at which time the location parameter appears to increase by about 1 unit. The $Y$ variable does not show any apparent change in distribution throughout the 20 batches.
\begin{figure}[t!]
    \centering
    \includegraphics[width=0.75\textwidth,angle=-90]{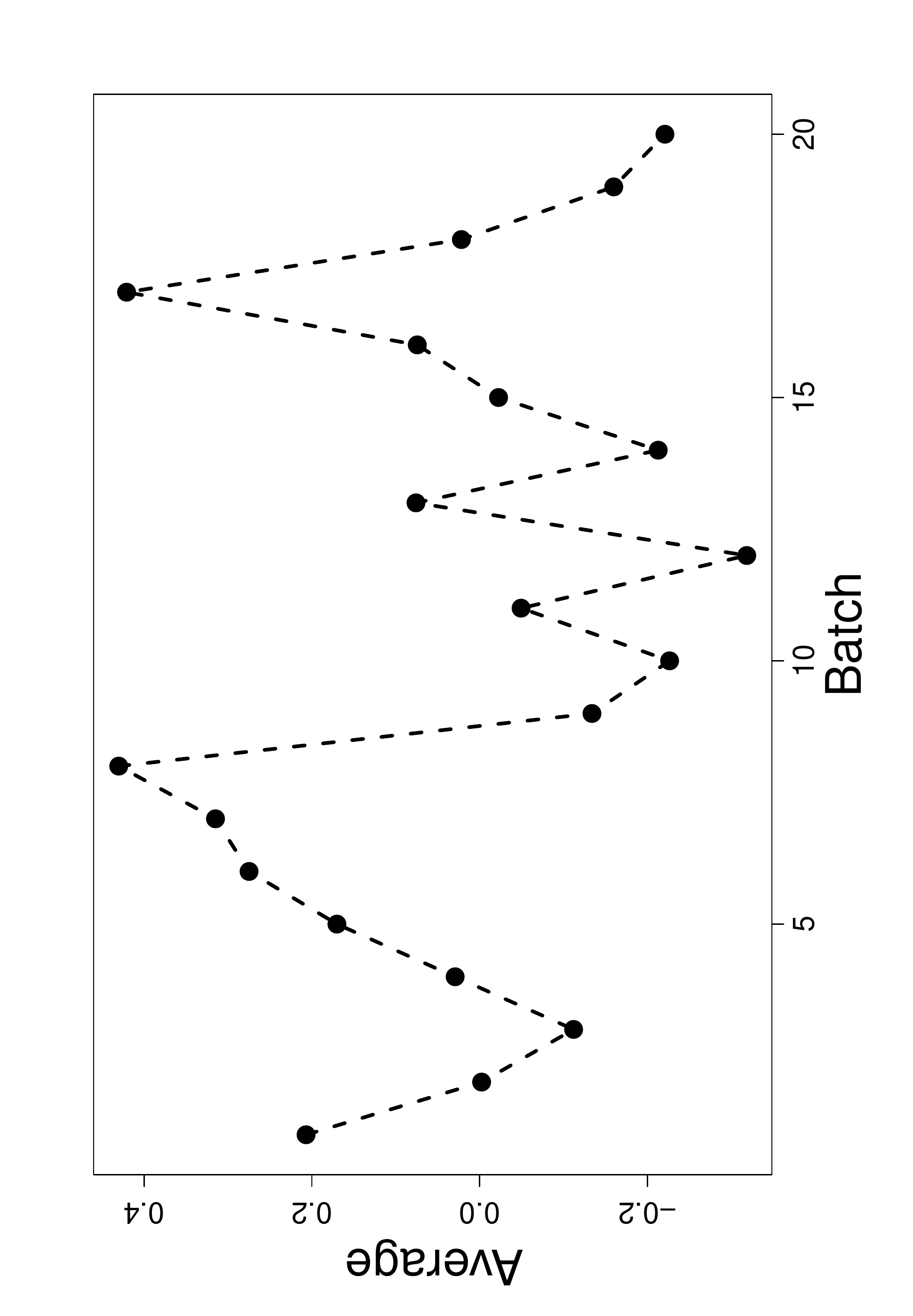}
    \caption{The batch averages for 20 batches of size 50, for the second variable in the bivariate $(X,Y)$ in Example 6.1.1.}
    \label{fig:c6s1f2}
\end{figure}

Conversion to sequential normal scores for each of the two variables is straightforward in the usual manner for batch data. The variables $X$ and $Y$ are correlated. Their cumulative correlation coefficients, computed on the raw data as well as the sequential normal scores as each new batch is obtained, are illustrated in Figure \ref{fig:c6s1f3}. It is clear that the sequential normal scores mimic the raw data in this respect.

\begin{figure}[t!]
    \centering
    \includegraphics[width=0.75\textwidth,angle=-90]{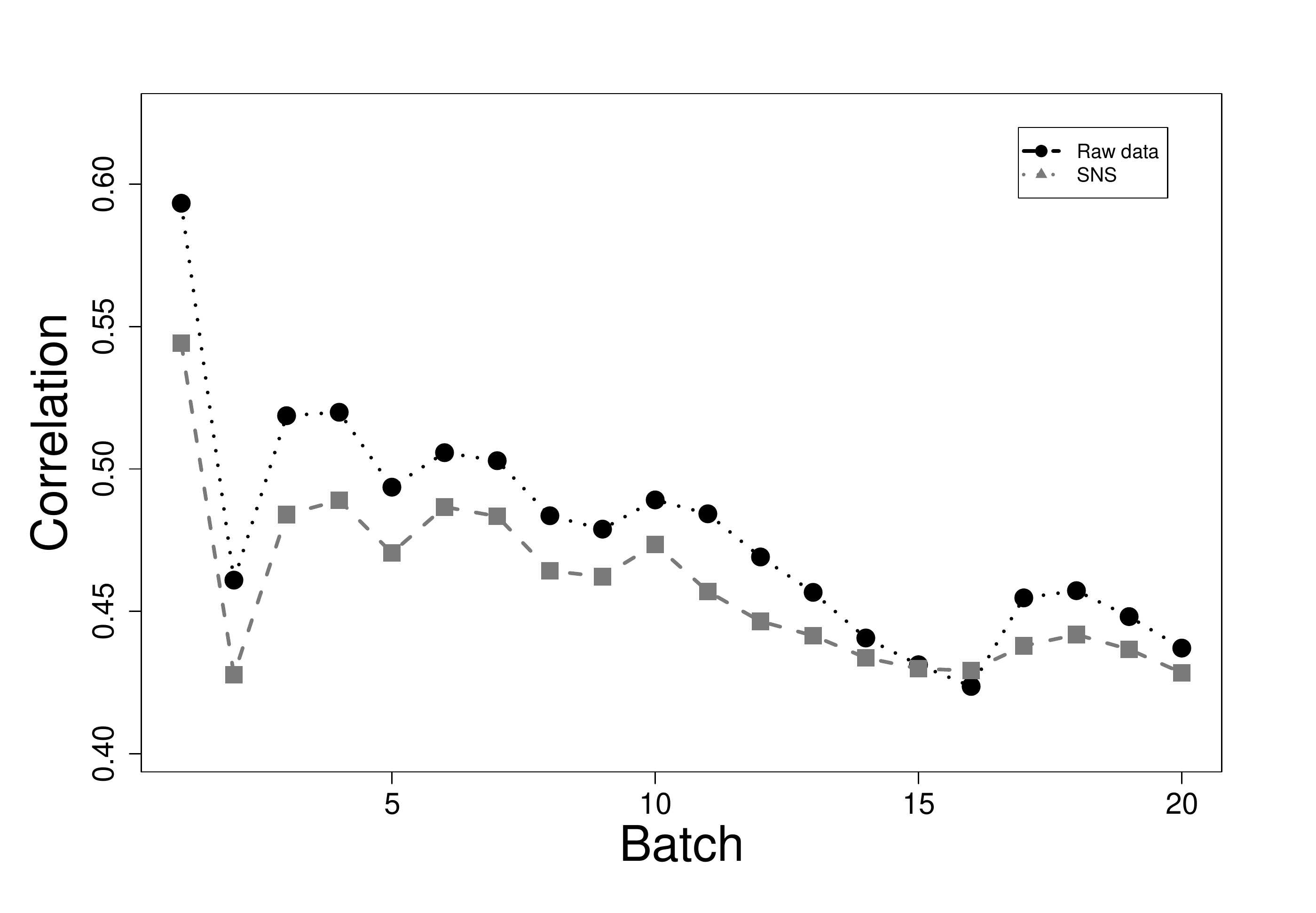}
    \caption{This graph shows the close correlation between the raw data and the sequential normal scores in Example 6.1.1, where the correlation is computed between $X$ and $Y$ for the data up to and including the indicated batch for the raw data, and between the SNS values for $X$ and $Y$.}
    \label{fig:c6s1f3}
\end{figure}

The cumulative correlation coefficients for the SNS are used in the test statistic $T^2$ given as follows,
\begin{align}
T^2 = n(\bar{X}, \bar{Y})(S^2)^{-1}(\bar{X}, \bar{Y})',
\end{align}
where $S^2$ represents the sample correlation matrix for all the batches up to but not including the current batch, and $\bar{X}$ and $\bar{Y}$ are the sample means of the SNS for the current batch. Many computer programs will find the inverse of a matrix, but they are time consuming, so for simulation purposes it is faster to use the following equation for bivariate data:
\begin{align}
    T^2 = 50(\bar{X}^2+\bar{Y}^2-2r\bar{X}\bar{Y})/(1-r^2),
\end{align}
where $r$ is the running sample correlation coefficient for the SNS up to but not including that batch. The test statistic is compared with the 0.0027 upper quantile of the $\chi^2$ distribution with 2 degrees of freedom, 11.8, to mimic the 3-sigma limits of a Shewhart chart. It is seen in Figure 6.4 that a clear signal is received on batch 11, similar to the results using the two methods illustrated by \citet{qiu_2014}. Upon getting the signal at batch 11, the reference set is frozen at batch 10 for further comparisons using SNS. The result is shown in Figure \ref{fig:c6s1f4}, a graph of $T^2$ for sequential normal scores. 

\begin{figure}[t!]
    \centering
    \includegraphics[width=0.75\textwidth,angle=-90]{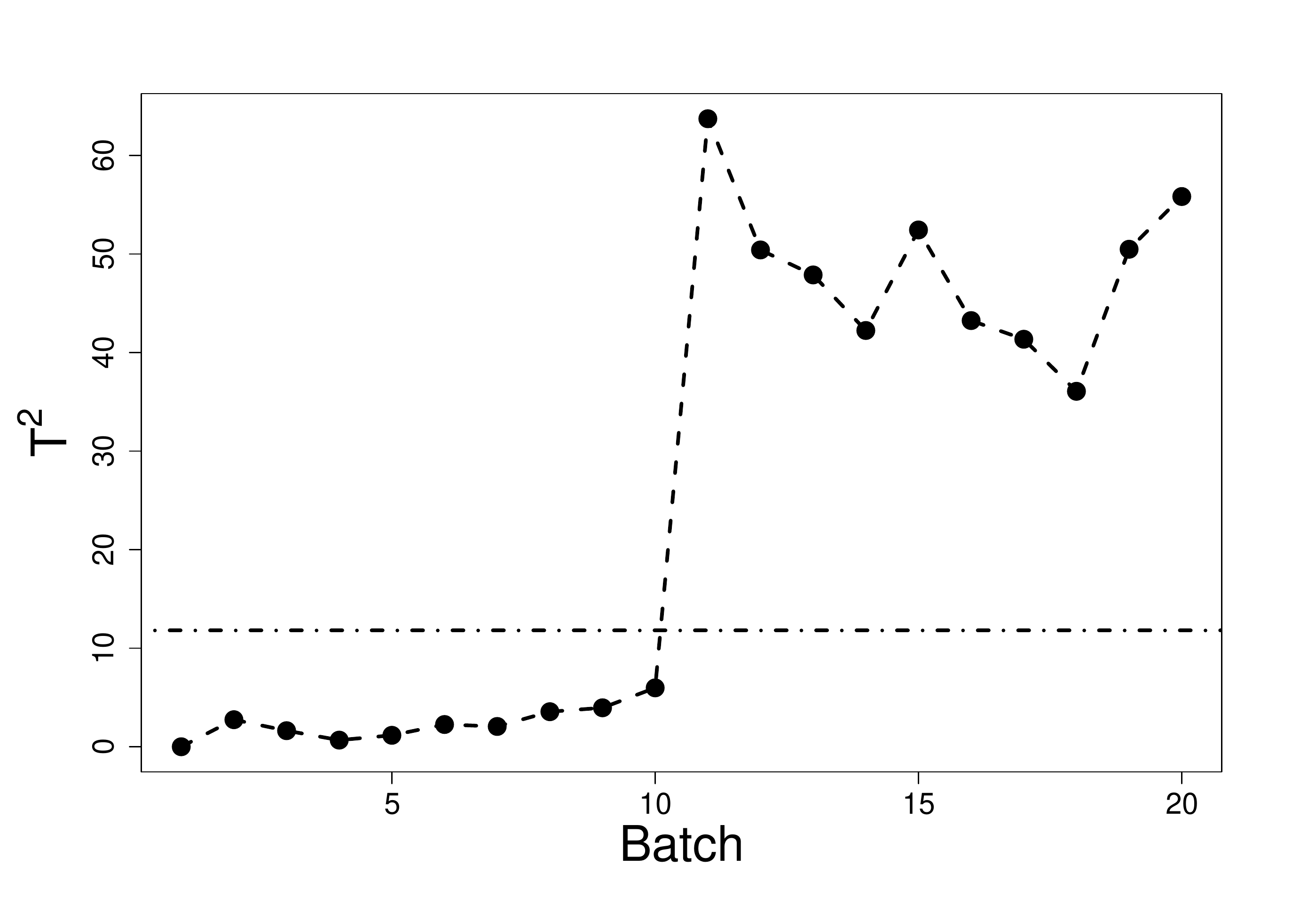}
    \caption{The Hotelling $T^2$ statistic computed on the SNS for ($X,Y$) in Example 6.1.1, showing a signal at batch 11. Subsequent values of $T^2$ use only batches 1-10 as a reference set.}
    \label{fig:c6s1f4}
\end{figure}

Although we compared $T^2$ with the $\chi^2$ distribution with 2 degrees of freedom in Figure 6.4, we are not aware of any theoretical justification for using that distribution. It is well known that normally distributed marginal distributions do not imply a joint multivariate normal distribution. So we compared the empirical distribution function of the $T^2$ values for batches 1 through 10, the in-control batches, with the $\chi^2$ distribution with 2 degrees of freedom (see Figure \ref{fig:c6s1f5}). We used the Kolmogorov goodness-of-fit test (see \citet{conover_1999}, for details) and accepted the null hypothesis of a $\chi^2$ distribution, $p>0.20$.

\begin{figure}[t!]
    \centering
    \includegraphics[width=0.75\textwidth,angle=-90]{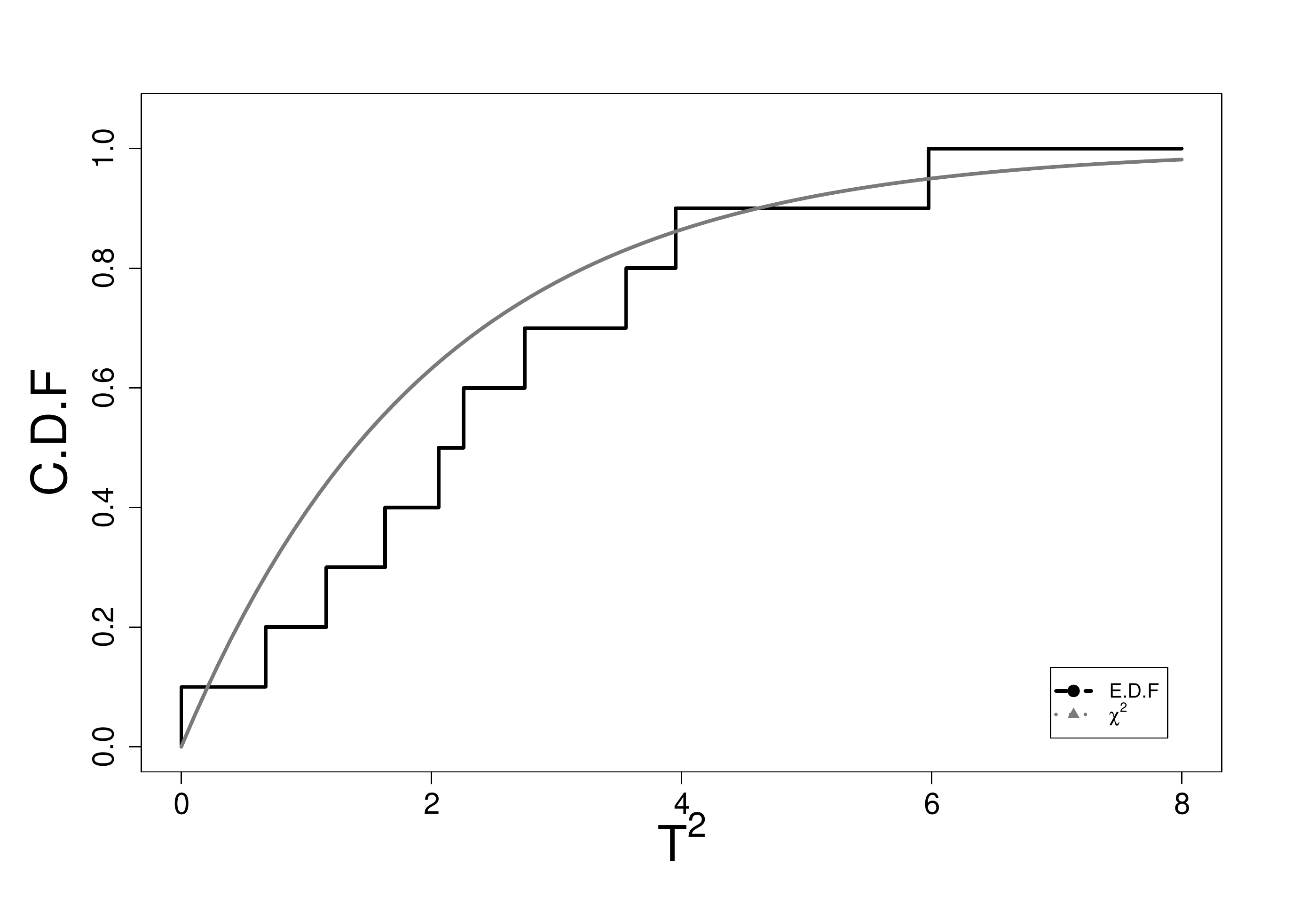}
    \caption{A comparison of the empirical distribution function of the first 10 values of $T^2$ in Example 6.1.1, while the process is in control, with the theoretical $\chi^2$ distribution with 2 degrees of freedom.}
    \label{fig:c6s1f5}
\end{figure}
To be safe, however, we converted the sequence of $T^2$ values to SNS, and calculated the EWMA ($\lambda=0.1$) on these values. The SNS on the values of T$^2$ also gives a signal with its value 1.69 at batch 11, significant at the 0.05 level, and the EWMA ($\lambda=0.1$) gives a strong signal (0.633) at batch 11 ($p<0.0025$).

The two methods illustrated by \citet{qiu_2014} were proposed by \citet{boone_etal_2012}. One method assumes the median is known and uses a sign test within each batch for each of the two variables. The results are used in the above $T^2$ statistic, using the appropriate estimates of the mean and covariance matrix. A large batch size is obviously needed for that test. The graph of the $T^2$ statistic is given in Figure 9.3(c) on page 369 of \citet{qiu_2014}. The statistic exceeds the upper bound 11.8 on batch 11.

The other method assumes not only that the median is known, but that the distribution is symmetric, so the Wilcoxon signed ranks test can be used on each batch. Again the means and covariance matrix are estimated in each batch, so a large batch size is needed. The graph of the $T^2$ statistic is given in Figure 9.3(d) on page 369 of \citet{qiu_2014}. As before, the statistic exceeds the upper bound 11.8 on batch 11. Qui notes, ``The signal by the (signed ranks) chart seems more convincing because its charting statistic takes larger values (than the sign test) when (the batch number is greater than or equal to) 11.'' By that measure, the SNS test statistic is the most convincing of the three tests, as its charting statistic is approximately twice the size of either one.

Our SNS method does not require knowledge of the median, nor an assumption of symmetry, and should work well with even with small batches. This method is easily adapted to use information about a known quantile such as the median; only the SNS are slightly different, as described in earlier sections. Some study of the properties of this method is clearly needed.

\end{myexample}
\section{Multivariate Sequential Normal Scores with a Smaller Batch Size}\label{sec:c6s2}

\begin{myexample}
To see how this SNS multivariate method behaves with smaller batches, a second example is presented. This is the same data set presented in \citet{qiu_2014} on page 375 in Figure 9.5 (a) and (b) to illustrate a nonparametric procedure suggested by \citet{liu_1995}. For this example 100 observations of bivariate data are used in phase I as a reference set, followed by 20 batches with 10 observation in each batch. Liu’s method finds a signal on the 11$^\text{th}$ batch, as illustrated in Figure 9.5(c) on page 375 of \citet{qiu_2014}. We were not able to obtain the 100 observations used as a reference sample, other than to observe them in \citet{qiu_2014}’s Figure 9.5 (a) and (b). Therefore we treat the 20 batches, of 10 observations each, in a self-starting sense. The data are given in Table \ref{tab:c6s2t1} for one variable called $X$, and in Table \ref{tab:c6s2t2} for the other variable, called $Y$.

\begin{table}[t!]
\small
    \caption{The data for $X$ in the bivariate random variable ($X,Y$) used in Example 9.3 in \citet{qiu_2014} and in our Example 6.2.1. The data consist of 20 batches of 10 observations each. The column on the right is the batch average.}
    \centering
    \begin{tabular}{crrrrrrrrrrr}
    Batch & \multicolumn{10}{c}{observations on $X$}                                             & \multicolumn{1}{c}{average} \\
    1     & -1.197 & -0.536 & 1.302 & -0.449 & -2.569 & -0.805 & -0.584 & -1.635 & 1.679 & -0.122 & -0.492 \\
    2     & -0.245 & -0.993 & 0.563 & -1.075 & -0.332 & -0.102 & 1.335 & 4.296 & -1.602 & 1.911 & 0.376 \\
    3     & 0.142 & 0.584 & 2.490 & 0.271 & -0.698 & -2.598 & -1.691 & -1.434 & 1.220 & -0.541 & -0.226 \\
    4     & -0.439 & 0.025 & -0.298 & 1.199 & 2.499 & -3.138 & -1.722 & -0.432 & 0.855 & -1.324 & -0.278 \\
    5     & 0.160 & 1.623 & -1.149 & 1.005 & 0.719 & -0.068 & -0.307 & 0.473 & 1.366 & 1.077 & 0.490 \\
    6     & 0.200 & 0.647 & -0.151 & 0.440 & 0.049 & 0.372 & 0.781 & -2.452 & -1.384 & -0.293 & -0.179 \\
    7     & 0.854 & -3.344 & 0.791 & 0.788 & 0.039 & -1.141 & 0.629 & 1.998 & -0.171 & -1.349 & -0.091 \\
    8     & 0.824 & -3.508 & -3.760 & -0.355 & -0.206 & -0.985 & -0.065 & -0.147 & 1.537 & -0.903 & -0.757 \\
    9     & -0.998 & 1.682 & 1.087 & 0.996 & -0.076 & -1.327 & 10.691 & -0.636 & 2.101 & 0.150 & 1.367 \\
    10    & -0.221 & -1.812 & 1.337 & -0.186 & -1.557 & 2.200 & 0.077 & 0.473 & -1.687 & 1.744 & 0.037 \\
    11    & 2.600 & 1.657 & 2.422 & 1.329 & 2.642 & 1.189 & 3.472 & 1.350 & 2.188 & 2.495 & 2.134 \\
    12    & 2.321 & 2.462 & 1.348 & 2.317 & 1.905 & 1.086 & 0.801 & 2.804 & -2.053 & 2.094 & 1.509 \\
    13    & 3.128 & -0.321 & 2.520 & 2.146 & 1.437 & 2.889 & 1.088 & 1.600 & 1.278 & 3.038 & 1.880 \\
    14    & 3.654 & 2.420 & 0.466 & 2.798 & 2.621 & 1.928 & 3.313 & 3.688 & 1.519 & 1.355 & 2.376 \\
    15    & 1.704 & 2.822 & 2.904 & 2.070 & 0.615 & 3.003 & 1.953 & 2.203 & 1.219 & 2.426 & 2.092 \\
    16    & 2.339 & 1.201 & 2.122 & 4.052 & 0.990 & 3.241 & 0.788 & 3.421 & 2.565 & 1.128 & 2.185 \\
    17    & 2.667 & 1.066 & 2.796 & 2.048 & 1.269 & 2.316 & 2.906 & 0.167 & 2.861 & 0.468 & 1.856 \\
    18    & 2.910 & 1.240 & 2.244 & 2.371 & 0.835 & 3.678 & 1.656 & 1.447 & 3.220 & 1.281 & 2.088 \\
    19    & 2.065 & -0.472 & 3.705 & 2.721 & 2.597 & 2.494 & 2.863 & 2.645 & 2.001 & 1.931 & 2.255 \\
    20    & 2.561 & 3.454 & 1.901 & 2.462 & 1.596 & 2.166 & 1.632 & 1.396 & 1.416 & 1.824 & 2.041 \\
    \end{tabular}
    \label{tab:c6s2t1}
\end{table}

A clearer picture of the variable $X$ in Table 6.1 emerges with Figure \ref{fig:c6s2f1} that graphs the batch averages for $X$. Now it is evident that somewhere around batch 9 or 10 the location parameter increases by about 2 units.

\begin{figure}[t!]
    \centering
    \includegraphics[width=0.75\textwidth,angle=-90]{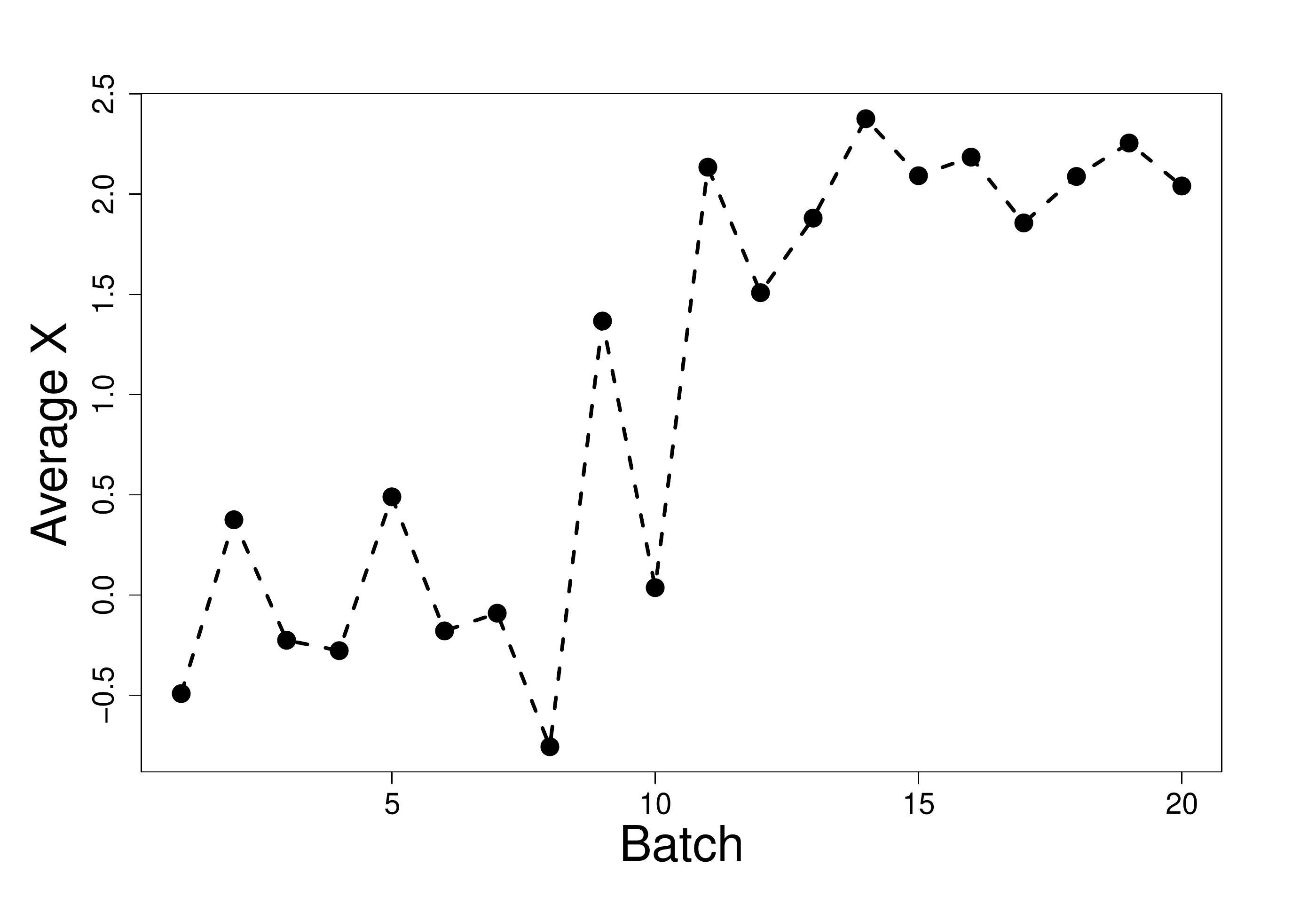}
    \caption{The batch averages for 20 batches of size 10, for the first variable in the bivariate ($X,Y$) as given in Table 6.1.}
    \label{fig:c6s2f1}
\end{figure}

\begin{table}[t!]
\small
    \caption{The data for $Y$ in the bivariate random variable ($X,Y$) used in Example 9.3 in \citet{qiu_2014} and in our Example 6.2.1. The data consist of 20 batches of 10 observations each. The column on the right is the batch average.}
    \centering
    \begin{tabular}{crrrrrrrrrrr}
    Batch & \multicolumn{10}{c}{observations on $Y$}                                             & \multicolumn{1}{l}{average} \\
    1     & -1.060 & -0.447 & 1.826 & 0.933 & -2.584 & -0.594 & -0.069 & -2.966 & -1.278 & 0.274 & -0.597 \\
    2     & 0.468 & -0.876 & 1.128 & -0.237 & -1.384 & -0.513 & 1.463 & 0.683 & -2.032 & -1.010 & -0.231 \\
    3     & 0.783 & -0.212 & 0.847 & -0.038 & 0.791 & 2.023 & 0.750 & -0.342 & 1.469 & 1.623 & 0.769 \\
    4     & -1.242 & 0.156 & 0.975 & 0.296 & 1.582 & 0.379 & -0.084 & 0.457 & -0.422 & 0.209 & 0.231 \\
    5     & -1.625 & 0.286 & -1.193 & 0.946 & -0.181 & 0.636 & 1.793 & -1.669 & -1.113 & 1.009 & -0.111 \\
    6     & 1.130 & 0.134 & -0.625 & -1.227 & 1.407 & 1.689 & 0.715 & -3.279 & -1.670 & -1.033 & -0.276 \\
    7     & -0.447 & -3.784 & -0.295 & 1.584 & 0.402 & -1.339 & 0.843 & -3.063 & -0.807 & -1.343 & -0.825 \\
    8     & 0.000 & -2.270 & -3.867 & 0.464 & 0.231 & -0.382 & 0.569 & -0.344 & 1.641 & -2.084 & -0.604 \\
    9     & -1.232 & 1.149 & 4.433 & 0.792 & -0.215 & -0.202 & 7.456 & -0.474 & 1.178 & 0.666 & 1.355 \\
    10    & 0.600 & 1.159 & 1.450 & 0.129 & -1.590 & 0.485 & 0.976 & -0.561 & -0.951 & 2.074 & 0.377 \\
    11    & -1.807 & 0.912 & 0.374 & 0.074 & 1.445 & -0.492 & 0.332 & -0.867 & 1.700 & 0.658 & 0.233 \\
    12    & 0.898 & -1.068 & -1.524 & -0.044 & 0.326 & -0.874 & 1.388 & 0.164 & -2.152 & -0.360 & -0.325 \\
    13    & 2.001 & -0.032 & 0.760 & 0.683 & 0.698 & 0.838 & -0.378 & -0.460 & 0.024 & 0.446 & 0.458 \\
    14    & 1.436 & 0.235 & -1.346 & -0.247 & 0.429 & -0.212 & 1.783 & 2.150 & -0.294 & -0.408 & 0.353 \\
    15    & -1.101 & 0.673 & 0.679 & 1.178 & -1.933 & 2.246 & -0.121 & 1.463 & 2.892 & 1.165 & 0.714 \\
    16    & 0.563 & 1.062 & 0.394 & 2.492 & 0.248 & 0.101 & -0.017 & 1.267 & 1.593 & 0.840 & 0.854 \\
    17    & -0.823 & -0.649 & 1.404 & -0.381 & -0.224 & 1.113 & 0.551 & 0.452 & -0.172 & 1.557 & 0.283 \\
    18    & 0.514 & 0.592 & 1.294 & -0.081 & 0.125 & 1.669 & 1.055 & 0.013 & -0.188 & -3.352 & 0.164 \\
    19    & -0.548 & -0.573 & 0.023 & 2.480 & -0.790 & 0.793 & -1.893 & -0.004 & 0.017 & 0.249 & -0.025 \\
    20    & 0.618 & 1.572 & 1.161 & -0.866 & -0.014 & -0.852 & -0.638 & -2.119 & -0.727 & -0.353 & -0.222 \\
    \end{tabular}
    \label{tab:c6s2t2}
\end{table}

The batch averages for $Y$ in Table 6.2 appear in Figure \ref{fig:c6s2f2}. Any change in the distribution of $Y$ is not noticeable in Figure \ref{fig:c6s2f2}.

\begin{figure}[t!]
    \centering
    \includegraphics[width=0.75\textwidth,angle=-90]{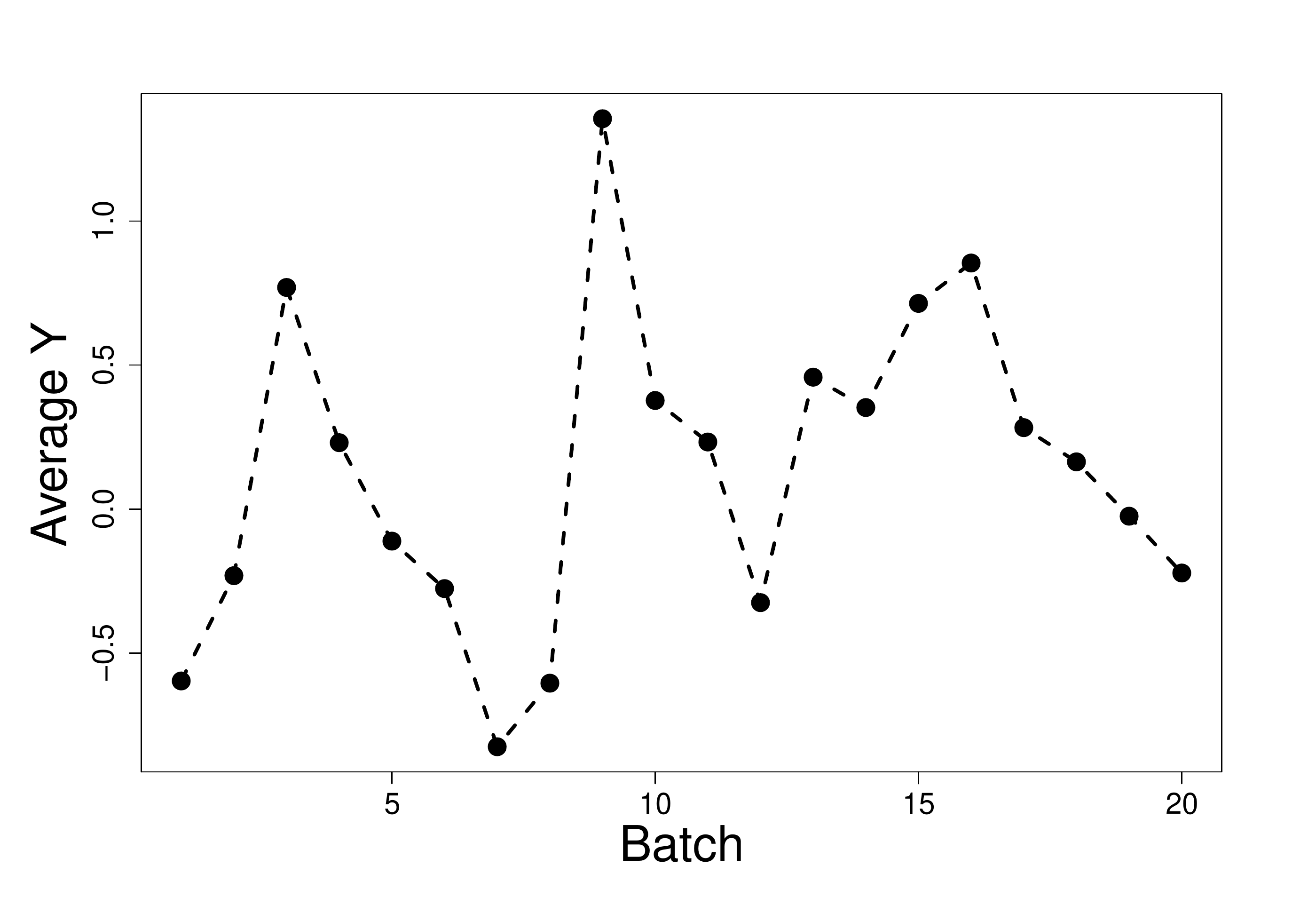}
    \caption{The batch averages for 20 batches of size 10, for the second variable in the bivariate ($X,Y$) as given in Table 6.2.}
    \label{fig:c6s2f2}
\end{figure}

The data are converted to sequential normal scores in the usual way, one variable at a time. Then the correlation coefficient between SNS($X$) and SNS($Y$) is calculated on the observations in a batch and all previous batches, to keep a {\em running} correlation coefficient as new batches are obtained, for use in the correlation matrix. The statistic Hotelling’s $T^2$ appears in Figure 6.8. Details of the computation are the same as in the previous section, so they are not repeated here.

\begin{figure}[t!]
    \centering
    \includegraphics[width=0.75\textwidth,angle=-90]{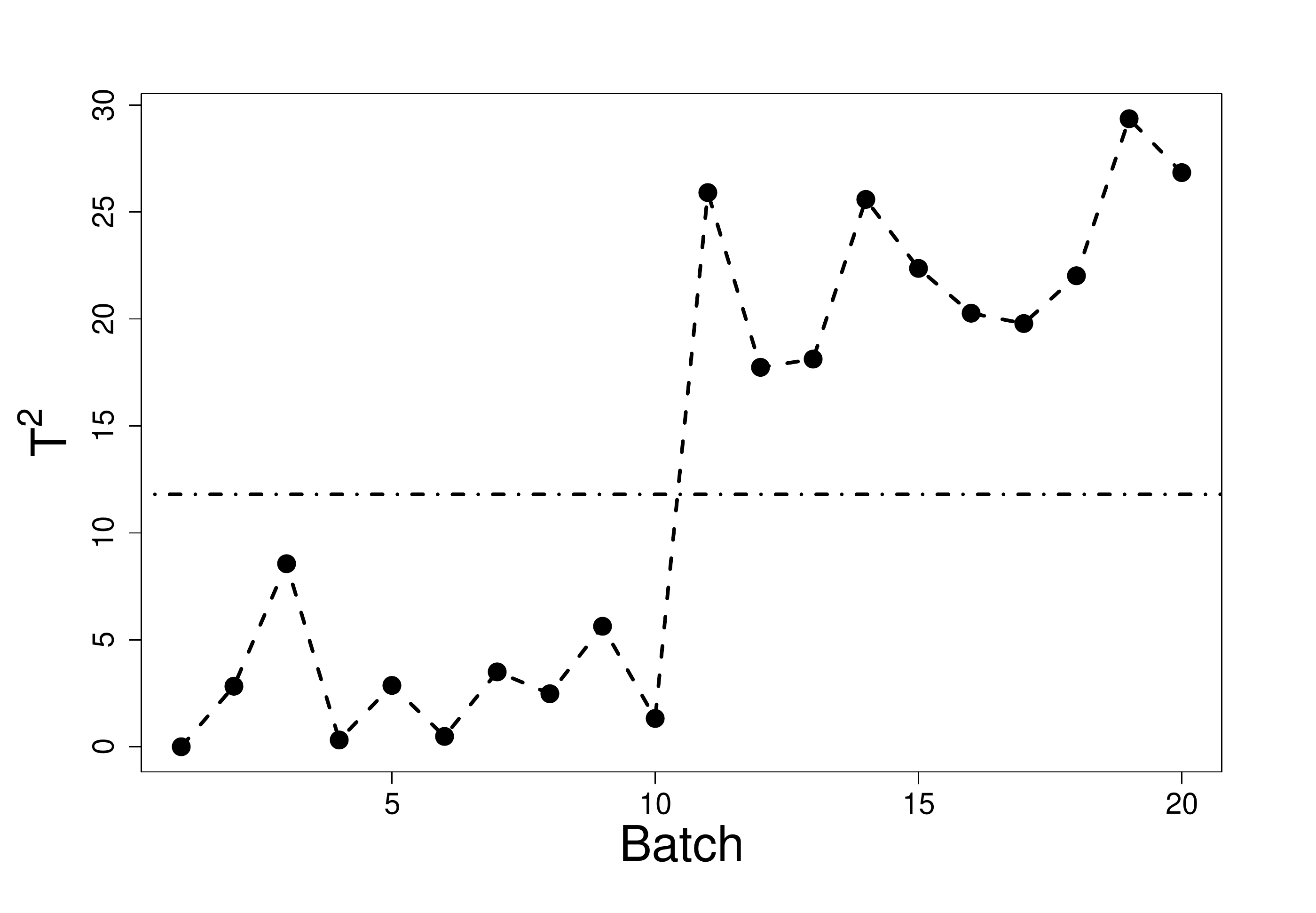}
    \caption{The Hotelling $T^2$ statistic computed on the SNS in our Example 6.2.1 for ($X,Y$) in Tables 6.1 and 6.2, showing a signal at batch 11. Subsequent values of $T^2$ use only batches 1-10 as a reference set.}
    \label{fig:c6s2f3}
\end{figure}

Note that a strong signal appears with sample 11, just as in the example presented by \citet{qiu_2014}. At this point we freeze the previous observations as our reference set and proceed to calculate the test statistic for the remaining batches, relative to the first 10 batches. Figure \ref{fig:c6s2f3} illustrates the results, which are compared with the $\chi^2$ distribution with 2 degrees of freedom, which has an upper 0.0027 quantile of 11.8. It is apparent that this SNS method works well in this example also, even with the smaller batch size. 

Again, as with Example 6.1.1, we are not aware of any theoretical justification for using the $\chi^2$ (2 d.f.) as an approximation for the distribution of $T^2$, so we used the Kolmogorov goodness-of-fit test (see \citet{conover_1999}) to compare the empirical distribution function of $T^2$ with the $\chi^2$ distribution, two degrees of freedom, as shown in Figure 6.9. The result was a failure to reject the null hypothesis ($p>0.2$) so the results stand as valid.

\begin{figure}[t!]
    \centering
    \includegraphics[width=0.75\textwidth,angle=-90]{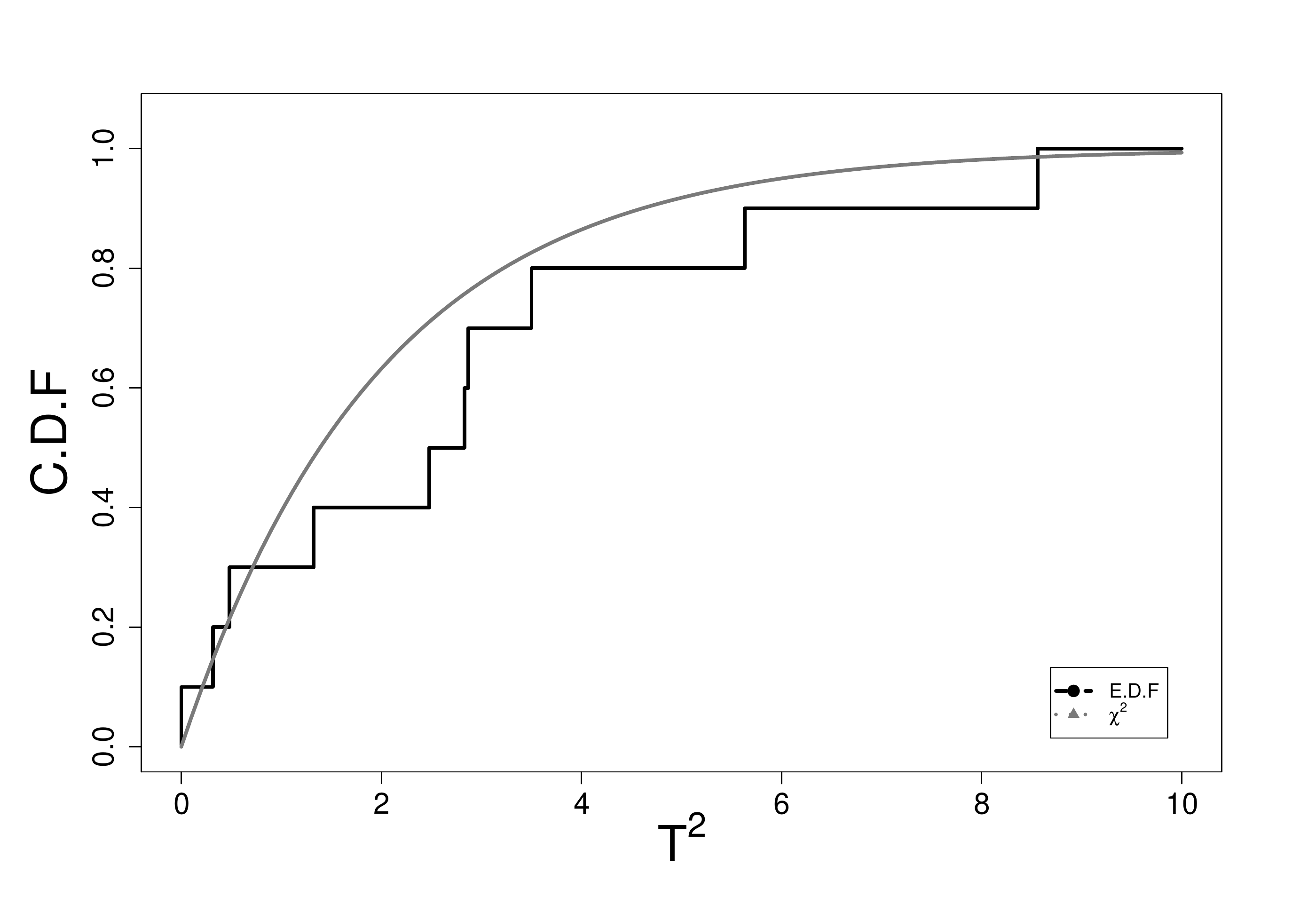}
    \caption{A comparison of the empirical distribution function of $T^2$ for the first 10 observations in Example 6.2.1 vs. the $\chi^2$ distribution with 2 degrees of freedom.}
    \label{fig:c6s2f4}
\end{figure}
However, to be safe, we converted the $T^2$ values to sequential normal scores, and got a signal ($p<0.05$) at batch 11, to match the result \citet{qiu_2014} obtained using the method proposed by \citet{liu_1995}. The SNS approach gives the same result as the nonparametric procedure suggested by \citet{liu_1995}, even though the 100 reference observations used by \citet{qiu_2014} to illustrate Liu’s method were not used.
\end{myexample}

\section{Multivariate Sequential Normal Scores to Detect a Change in Location, Individual Observations}\label{sec:c6s3}

\begin{myexample}
For the third example of this chapter consider the data set in Example 7.1 on page 267 of \citet{qiu_2014}, which is given in Table 7.1 on page 267 of \citet{qiu_2014} and in our Table \ref{tab:c6s3t1}. The data represent 30 observations from a tri-variate random variable ($X_1, X_2, X_3$) in a production process.

\begin{table}[t!]
    \caption{The data used in Example 7.1 of \citet{qiu_2014} and in our Example 6.3.1. The data consist of 30 observations on ($X_1, X_2, X_3$).}
    \centering
    \begin{tabular}{crrrcrrr}
    sample & \multicolumn{1}{c}{$X_1$} & \multicolumn{1}{c}{$X_2$} & \multicolumn{1}{c}{$X_3$} & sample & \multicolumn{1}{c}{$X_1$} & \multicolumn{1}{c}{$X_2$} & \multicolumn{1}{c}{$X_3$} \\
    1     & -0.224 & -0.464 & -0.662 & 16    & 0.217 & -0.260 & -0.005 \\
    2     & -0.082 & -0.203 & 0.682 & 17    & -0.611 & -0.048 & -0.428 \\
    3     & -0.436 & 0.342 & -0.174 & 18    & 0.473 & 0.066 & 0.890 \\
    4     & 0.455 & 0.715 & 1.229 & 19    & -1.130 & -1.214 & -0.064 \\
    5     & 0.437 & 0.237 & -0.377 & 20    & 1.379 & 2.443 & 2.350 \\
    6     & 0.669 & -0.002 & 0.230 & 21    & 0.671 & 0.534 & 1.120 \\
    7     & -0.186 & -0.480 & -0.907 & 22    & 1.467 & 0.945 & 0.402 \\
    8     & 0.672 & 0.384 & 0.903 & 23    & 1.077 & 0.949 & -0.085 \\
    9     & -0.121 & -0.812 & -1.279 & 24    & 2.126 & 3.964 & 3.775 \\
    10    & -0.683 & 0.093 & -0.436 & 25    & 2.237 & 0.959 & 0.175 \\
    11    & 0.063 & 0.193 & -0.028 & 26    & 0.362 & 0.344 & 0.375 \\
    12    & -1.029 & 0.503 & 0.728 & 27    & 2.119 & 1.569 & 1.031 \\
    13    & -0.080 & -0.238 & -0.218 & 28    & 1.147 & 1.003 & 1.417 \\
    14    & 0.943 & 0.272 & 0.831 & 29    & 0.667 & 0.525 & 1.676 \\
    15    & 0.147 & 0.655 & -0.543 & 30    & 1.944 & 0.935 & -0.130 \\
    \end{tabular}%
    \label{tab:c6s3t1}
\end{table}

In Example 7.1 of Qiu a parametric analysis of the same data assumes the first 20 observations are {\em in control} and determines {\em out-of-control} behavior at observation 24, using an upper bound with probability of exceedance of 0.005, corresponding to an ARL of 200. For purposes of comparing our nonparametric SNS method with the parametric method we will also assume the first 20 observations are the in-control observations, and treat the last 10 observations as Phase II, as they did. For the reference set we rank each of the first 20 variables relative to the corresponding variables in the first 20 observations, for each component $X_1$, $X_2$ and $X_3$ separately, and convert it to a sequential normal score as described in Section \ref{sec:c1s1}. The SNS correlation matrix for the reference set of 20 observations has $r_{1,2} = 0.536$, $r_{1,3} = 0.561$, and $r_{2,3} = 0.634$. The determinant of the correlation matrix is 0.377. This correlation matrix $S^2$ is used in the statistic

\begin{align}
    T^2 = (\text{SNS}(X_1), \text{SNS}(X_2), \text{SNS}(X_3))(S^2)^{-1}(\text{SNS}(X_1), \text{SNS}(X_2), \text{SNS}(X_3))'
\end{align}
for all 30 observations in the usual way as for the parametric test, but with the sequential normal scores taking the place of the original data.

The values of $T^2$ are graphed in Figure \ref{fig:c6s3f1}. They do not send a signal at the 0.005 level ($\text{ARL}=200$), which is 12.8 for a chi-squared random variable with 3 degrees of freedom. The empirical distribution function of the first 20 values of $T^2$, in the reference set, is compared with the $\chi^2$ distribution with 3 degrees of freedom in Figure \ref{fig:c6s3f2}. According to the Kolmogorov goodness-of-fit test (\citet{conover_1999}) the variation is within sampling variability ($p>0.2$). However, to be on the safe side, the 30 values of $T^2$ are converted to sequential normal scores in the usual manner. These SNS values also fail to send a signal at the 0.005 level. However, the EWMA ($\lambda=0.1$) show a signal at observation 24 and all observations following observation 24, as can be seen from Figure \ref{fig:c6s3f3}. This is in agreement with the signal \citet{qiu_2014} obtained at observation 24 using a parametric method.

\begin{figure}[t!]
    \centering
    \includegraphics[width=0.75\textwidth,angle=-90]{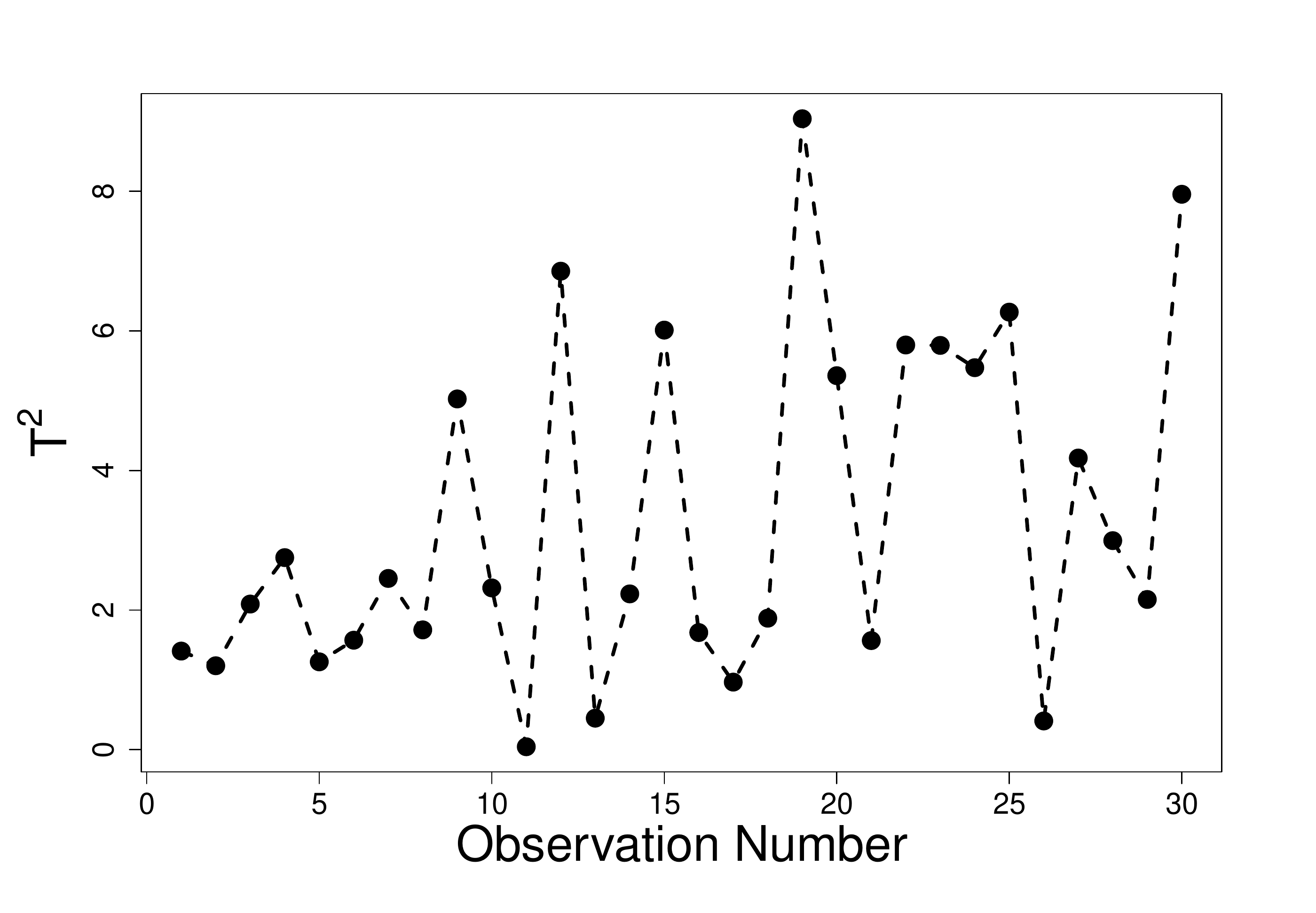}
    \caption{The values of $T^2$ for the 30 observations on ($X_1, X_2, X_3$) in Example 6.3.1, shown in Table 6.3, using observations 1--20 as a reference set, and observations 21--30 as Phase II observations. None of the values reach the upper 0.005 limit 12.8 for a $\chi^2$ distribution with 3 degrees of freedom.}
    \label{fig:c6s3f1}
\end{figure}

\begin{figure}[t!]
    \centering
    \includegraphics[width=0.75\textwidth,angle=-90]{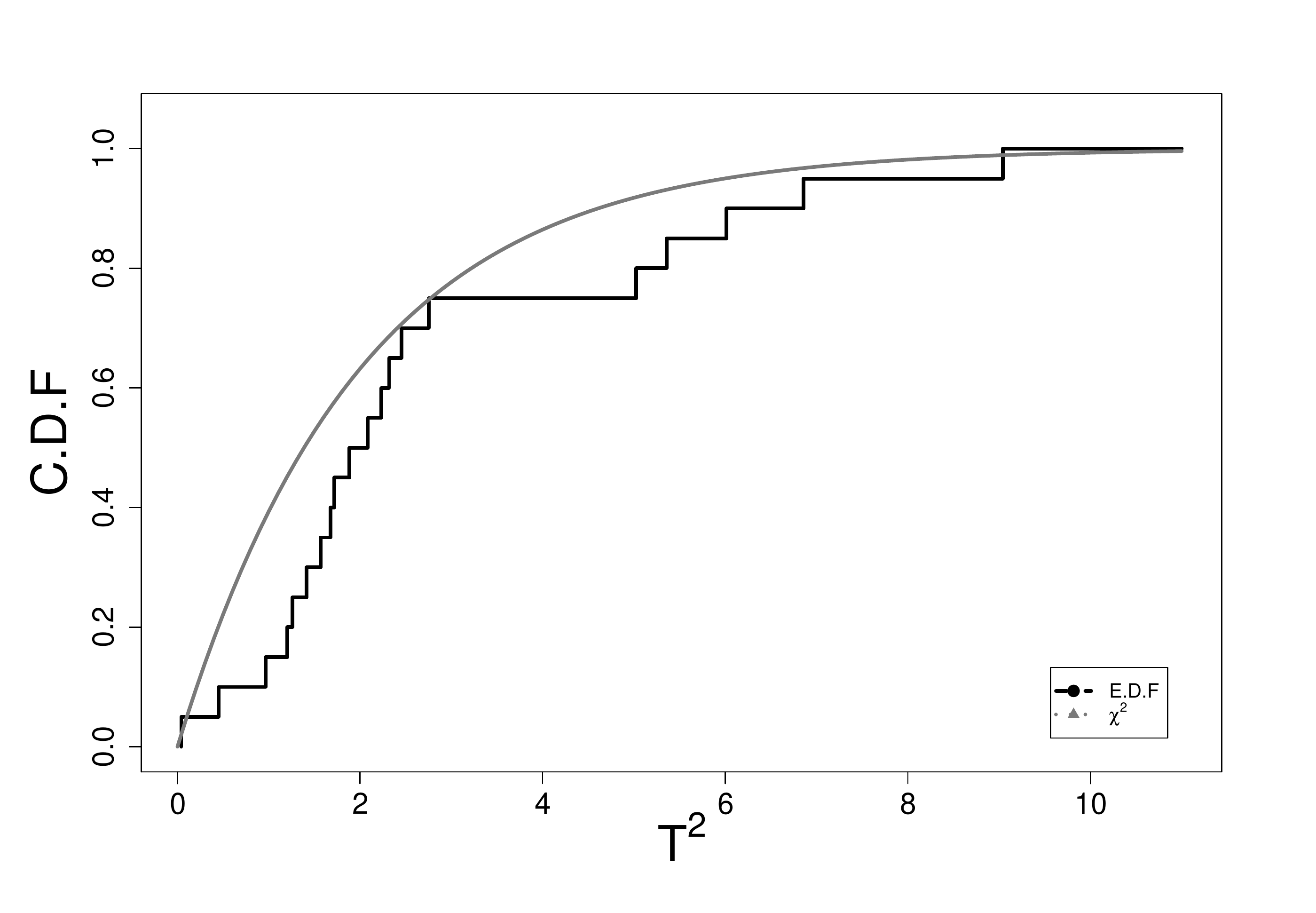}
    \caption{A comparison of the empirical distribution function of the 20 in-control values of $T^2$ in Example 6.3.1 compared to the $\chi^2$ distribution with 3 degrees of freedom.}
    \label{fig:c6s3f2}
\end{figure}

\begin{figure}[t!]
    \centering
    \includegraphics[width=0.75\textwidth,angle=-90]{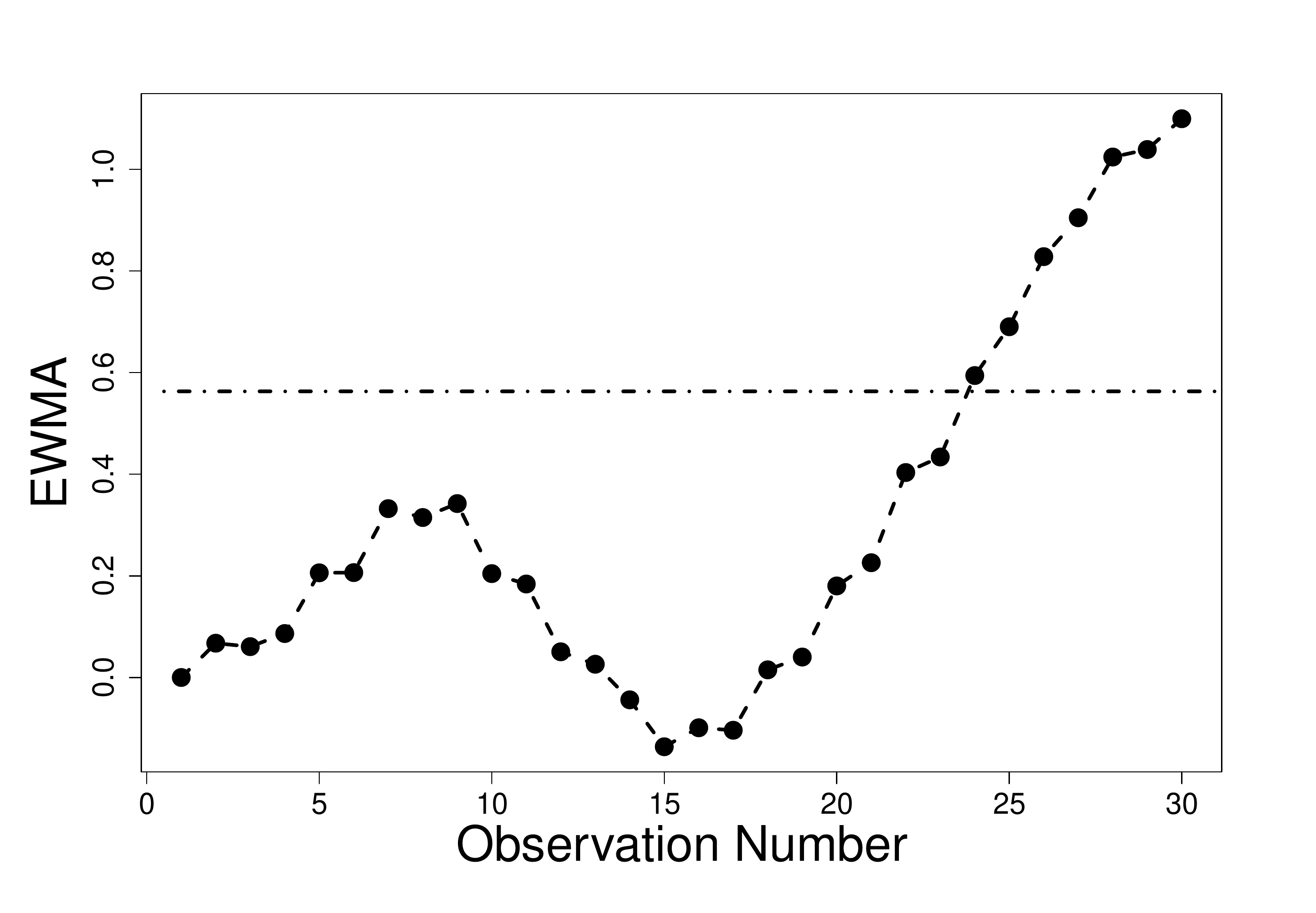}
    \caption{The EWMA ($\lambda=0.1$) of the 30 SNS values from the $T^2$ values in Example 6.3.1 and graphed in Figure \ref{fig:c6s3f1}. This graph sends a signal by crossing the upper limit 0.563 for $\text{ARL}=200$ on observation 24, in agreement with the parametric analysis.}
    \label{fig:c6s3f3}
\end{figure}

\end{myexample}

\section{MSNS to Detect Changes in Location or Scale or Both}\label{sec:c6s4}

Our emphasis thus far in this section has been in detecting a change in location. As discussed earlier,
often a change in scale is of interest, or a change in both location and scale. In Sections \ref{sec:c4s2} and \ref{sec:c4s3} two univariate methods for detecting a change in location or a change in scale were proposed. The method
in Section \ref{sec:c4s2} proposed using SNS$^2$ as a statistic, which appears to be effective. However, when that method is extended to the multivariate case, a problem arises because we obtain a multivariate procedure where the variables are approximately $\chi^2$ with 1 degree of freedom. We are not aware of parametric methods that assume the variables are $\chi^2$ in distribution, so we cannot try to adapt a parametric method to the joint distribution of SNS$^2$ variables.

Therefore we will concentrate on the method in Section \ref{sec:c4s3}, which appears to be equally effective as the method in Section \ref{sec:c4s2}, for detecting changes in location or increases in scale. That method uses SNS which is approximately standard normal instead of SNS$^2$ which is approximately $\chi^2$ (1 d.f.). Later we will present an analysis using the method in Section 4.2.

Each variable in a $p$-variate random variable is converted to sequential normal scores in the manner described in Section \ref{sec:c4s3}. That is, a reasonable estimate of a location parameter is subtracted from each variable, and the result is squared. This squared value is converted to a sequential normal score. Let $X_i$ be a $p$-variate random variable, for $i=1, 2, \ldots,$ and let $\theta$ be the vector of location parameters to be subtracted from $X_i$. Then let $Y_i = (X_i - \theta)^2$ for $i=1, 2, 3, \ldots$. The components of $Y_i$ are converted to sequential normal scores, so we have a $p$-variate vector of sequential normal scores from observation 2 onward. The components of this vector are approximately standard normal in distribution, as in Section \ref{sec:c4s3} for the univariate case.

\begin{myexample}
To illustrate the method, consider the data used in Example 7.4 on page 282 of \citet{qiu_2014}, to illustrate three parametric approaches to this problem. Three tri-variate data sets were generated. All three data sets used the same 10 observations as a reference set. These 10 Phase I observations were generated as multivariate normal observations with mean vector $\mu_0 = (0,0,0)$ and covariance matrix $\Sigma_0 = (\sigma_{i,j})$ where the variances equal 1 and the covariances are $\sigma_{1,2} = 0.8 = \sigma_{2,3}$ and $\sigma_{1,3} = 0.5$. Observations 11 to 30 were generated from different multivariate normal distributions, described as follows. For data set (a) the mean vector was changed to ($1,0,0$) and the covariance matrix was left unchanged, that is, there was only a change in location. For data set (b) the mean vector was unchanged but the covariance matrix was multiplied by 1.1, a change in scale only, and for data set (c) the mean vector was changed to (1,0,0) and the covariance matrix was multiplied by 1.1, both a change in location and a change in scale.

The first 10 observations are given in Table \ref{tab:c6s4t1}. The vector $X_i$ is converted to the vector $Y_i$ as described above. From the first 10 observations on $X$ the vector $\theta$ = (0, 0, 0)' seems to be a reasonable vector to use for location. See the first 10 observations in Figures 6.14 to 616. The 10 observations are assumed to be the reference set, so the variables $Y$ are ranked relative to each other and converted to sequential normal scores, as batch 1 for Phase 1 observations. The correlation matrix is computed on the SNS of the first 10 observations on the three variables, and is used as the correlation matrix in the statistic
\begin{align}
    T_i^2 = (\text{SNS}_i)'\Sigma_{10}^{-1}(\text{SNS}_i).
\end{align}
The correlations in the correlation matrix  $\Sigma$, estimated from the SNS on Y in the reference set, are $r_{1,2} = 0.500$, $r_{1,3} = 0.648$, and $r_{2,3} = 0.699$. The determinant of the correlation matrix is 0.295. These first 10 observations will be the reference set in all of our computations. That means the correlation matrix $\Sigma_{10}$ for observation 10 is frozen for use with the later observations 11--30 in all three cases, as was done in Example 7.4 of \cite{qiu_2014}

The three parametric methods examined in Example 7.4 of \citet{qiu_2014} are as follows. The first is the ordinary parametric Hotelling $T^2$. The second is a multivariate CUSUM method proposed by \citet{healy_1987}, and the third is a COT (CUSUM on $T^2$) method proposed by \citet{crosier_1988}.

\begin{table}[t!]
    \caption{The first 10 3-variable observations serving as the reference data set in the three data sets in Example 7.4 in \citet{qiu_2014} and in the examples in Section 6.4. They appear in columns $X_1$, $X_2$, and $X_3$. The location parameters $\theta$ are assumed to be zero for each variable. The next three columns represent the transformation $Y_i=(X_i-\theta)^2$. These are converted to sequential normal scores in the next three columns. The final column is the Hotelling $T^2$ statistic computed on the SNS vector and the correlation matrix estimated from only the SNS columns of these 10 observations.}
    \centering
    \begin{tabular}{ccccccccccc}
    sample & $X_1$    & $X_2$    & $X_3$    & $Y_1$    & $Y_2$    & $Y_3$    & $SNS_1$ & $SNS_2$ & $SNS_3$ & $T^2$\\
    1     & -0.533 & -0.385 & -0.443 & 0.284 & 0.148 & 0.196 & 1.036 & -0.385 & -0.126 & 2.464 \\
    2     & 0.167 & -0.052 & 0.263 & 0.028 & 0.003 & 0.069 & -0.674 & -1.645 & -0.385 & 4.005 \\
    3     & 0.068 & -0.138 & -0.133 & 0.005 & 0.019 & 0.018 & -1.645 & -1.036 & -1.036 & 2.814 \\
    4     & 0.492 & 0.680 & 1.232 & 0.242 & 0.462 & 1.518 & 0.674 & 1.645 & 1.036 & 2.742 \\
    5     & -0.071 & 0.306 & 0.053 & 0.005 & 0.094 & 0.003 & -1.036 & -0.674 & -1.645 & 3.149 \\
    6     & 0.229 & 0.413 & 0.199 & 0.052 & 0.171 & 0.040 & -0.385 & 0.385 & -0.674 & 1.893 \\
    7     & -0.400 & -0.396 & -0.789 & 0.160 & 0.157 & 0.623 & 0.126 & -0.126 & 0.126 & 0.112 \\
    8     & 0.387 & 0.640 & 0.898 & 0.150 & 0.410 & 0.806 & -0.126 & 1.036 & 0.674 & 1.733 \\
    9     & -0.404 & -0.529 & -1.318 & 0.163 & 0.280 & 1.737 & 0.385 & 0.674 & 1.645 & 3.850 \\
    10    & -1.430 & -0.408 & 0.880 & 2.045 & 0.166 & 0.774 & 1.645 & 0.126 & 0.385 & 3.631 \\
    \end{tabular}
    \label{tab:c6s4t1}
\end{table}
The 10 values of $T^2$   obtained from the sample correlation matrix of the SNS values for the reference set are compared with the theoretical chi-squared distribution with 3 degrees of freedom in Figure \ref{fig:c6s4f1}. A Kolmogorov goodness-of-fit test accepts the null hypothesis of a possible sample from that distribution, $p>0.2$ (see \citet{conover_1999}).

\begin{figure}[t!]
    \centering
    \includegraphics[width=0.75\textwidth,angle=-90]{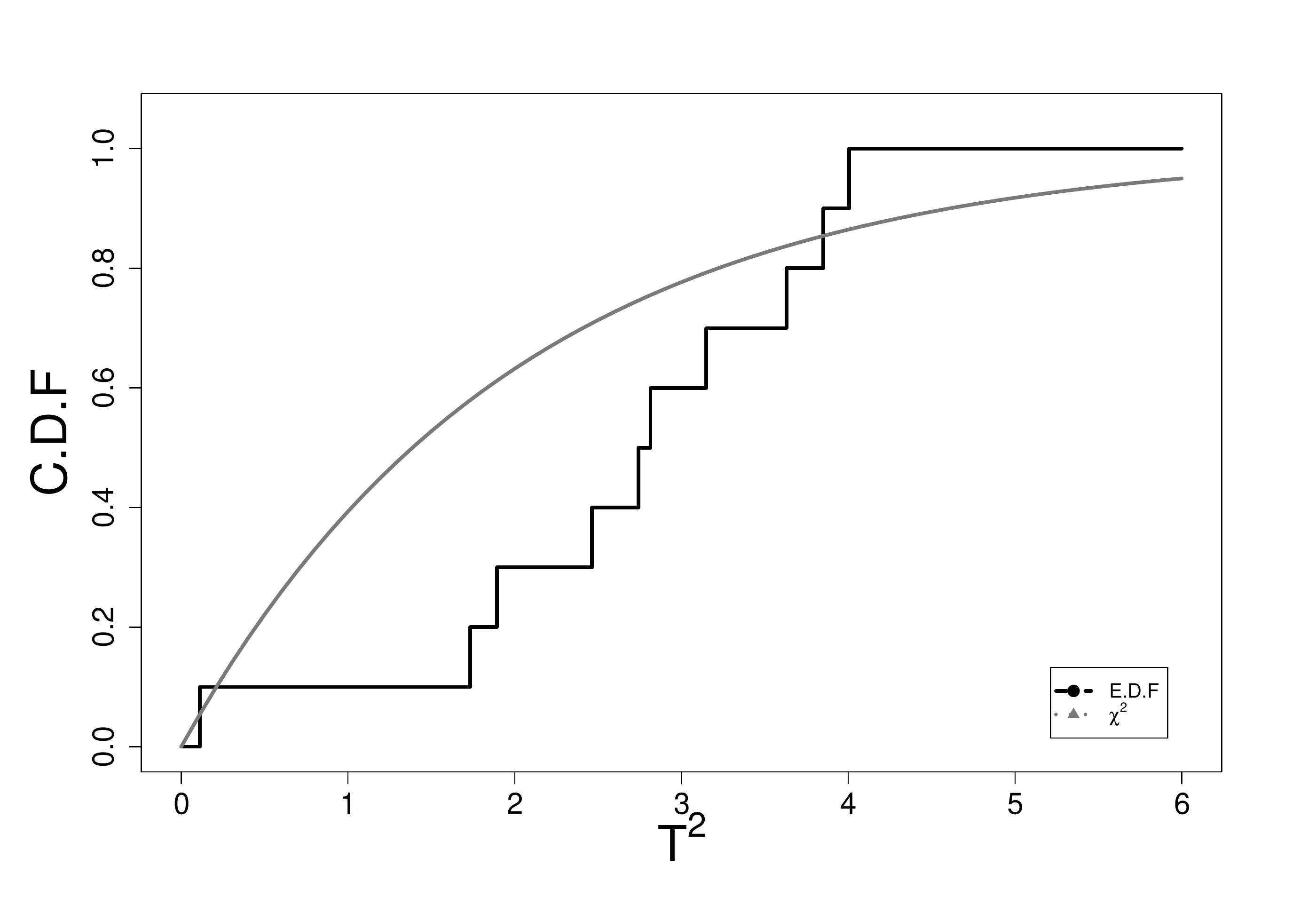}
    \caption{The empirical distribution function of the first 10 values of $T^2$ in Example 6.4.1, compared with the chi-squared distribution function with 3 degrees of freedom.}
    \label{fig:c6s4f1}
\end{figure}

For data set (a), observations 11-30 were generated from $N(\mu_1,\Sigma_0)$ where $\mu_1 = (1, 0, 0)'$  as described above. They appear in Table \ref{tab:c6s4t2}. There is a signal of change using the nonparametric SNS method, at observation 20. Graphs of the 30 observations in data set (a) are given in Figures \ref{fig:c6s4f2} to \ref{fig:c6s4f4}. The shift in location parameter for $X_1$ is obvious, but there is no apparent change in distribution for $X_2$ and $X_3$.

For comparison, it is interesting to note how the three parametric methods fared with this data set. As reported by \citet{qiu_2014} the parametric $T^2$ did not signal at all, the parametric CUSUM signaled at observation 14, and the parametric COT signaled at observation 11, exactly where the change occurred. This can be observed in Figure 7.7, parts (a), (b) and (c), on page 283 of \citet{qiu_2014}.

\begin{table}[t!]
    \caption{Observations 11--30 in data set (a), used in Example 6.4.1, appear in columns 2--4, the transformed values $Y$ are in columns 5--7, and the SNS computed on the components of $Y$ are in columns 8--10. The far right column gives the value of $T^2$ computed on the vector of SNS values, using $\Sigma_{10}$ from Table \ref{tab:c6s4t1} as the covariance (correlation in this case) matrix. The upper 0.005 limit for $T^2$ is 12.8.}
    \centering
    \begin{tabular}{crrrrrrrrrr}
    \textbf{sample} & \multicolumn{1}{c}{\textbf{$X_1$}} & \multicolumn{1}{c}{\textbf{$X_2$}} & \multicolumn{1}{c}{\textbf{$X_3$}} & \multicolumn{1}{c}{\textbf{$Y_1$}} & \multicolumn{1}{c}{\textbf{$Y_2$}} & \multicolumn{1}{c}{\textbf{$Y_3$}} & \multicolumn{1}{c}{\textbf{$SNS_1$}} & \multicolumn{1}{c}{\textbf{$SNS_2$}} & \multicolumn{1}{c}{\textbf{$SNS_3$}} & \multicolumn{1}{c}{\textbf{$T^2$}} \\
    11    & 1.206 & -0.194 & -0.241 & 1.454 & 0.038 & 0.058 & 1.097 & -0.748 & -0.473 & 4.167 \\
    12    & 3.231 & 1.959 & 0.492 & 10.439 & 3.838 & 0.242 & 1.691 & 1.691 & 0.000 & 9.684 \\
    13    & 0.637 & 0.140 & -0.184 & 0.406 & 0.020 & 0.034 & 1.097 & -0.748 & -0.748 & 5.109 \\
    14    & 0.177 & -0.497 & 1.074 & 0.031 & 0.247 & 1.153 & -0.473 & 0.473 & 0.748 & 2.138 \\
    15    & 1.270 & -0.185 & -2.002 & 1.613 & 0.034 & 4.008 & 1.097 & -0.748 & 1.691 & 10.202 \\
    16    & 0.322 & -0.219 & 0.302 & 0.104 & 0.048 & 0.091 & -0.230 & -0.748 & -0.230 & 0.729 \\
    17    & 1.710 & 0.434 & -0.688 & 2.924 & 0.188 & 0.473 & 1.691 & 0.473 & 0.000 & 5.145 \\
    18    & 0.481 & 0.342 & 1.306 & 0.231 & 0.117 & 1.706 & 0.473 & -0.473 & 1.097 & 4.237 \\
    19    & 1.356 & 0.798 & 1.737 & 1.839 & 0.637 & 3.017 & 1.097 & 1.691 & 1.691 & 3.368 \\
    20    & 1.972 & 1.432 & 0.133 & 3.889 & 2.051 & 0.018 & 1.691 & 1.691 & -1.097 & \bf{21.232} \\
    21    & 1.164 & -0.310 & -0.098 & 1.355 & 0.096 & 0.010 & 1.097 & -0.473 & -1.097 & 6.869 \\
    22    & 2.228 & 1.375 & 1.160 & 4.964 & 1.891 & 1.346 & 1.691 & 1.691 & 0.748 & 5.324 \\
    23    & -0.511 & -1.410 & -1.890 & 0.261 & 1.988 & 3.572 & 0.748 & 1.691 & 1.691 & 3.633 \\
    24    & -0.016 & 1.060 & 1.565 & 0.000 & 1.124 & 2.449 & -1.691 & 1.691 & 1.691 & 17.277 \\
    25    & 0.583 & -0.675 & -0.287 & 0.340 & 0.456 & 0.082 & 1.097 & 1.097 & -0.230 & 5.364 \\
    26    & 2.467 & 1.322 & 0.754 & 6.086 & 1.748 & 0.569 & 1.691 & 1.691 & 0.000 & 9.684 \\
    27    & 0.399 & -0.404 & 0.037 & 0.159 & 0.163 & 0.001 & 0.000 & 0.000 & -1.691 & 7.281 \\
    28    & 2.137 & 1.089 & 1.339 & 4.567 & 1.186 & 1.793 & 1.691 & 1.691 & 1.691 & 3.888 \\
    29    & 1.304 & -0.364 & 0.234 & 1.700 & 0.132 & 0.055 & 1.097 & -0.473 & -0.473 & 3.741 \\
    30    & -0.575 & -1.801 & -1.668 & 0.331 & 3.244 & 2.782 & 1.097 & 1.691 & 1.691 & 3.368 \\
    \end{tabular}

    \label{tab:c6s4t2}
\end{table}

\begin{figure}[t!]
    \centering
    \includegraphics[width=0.75\textwidth,angle=-90]{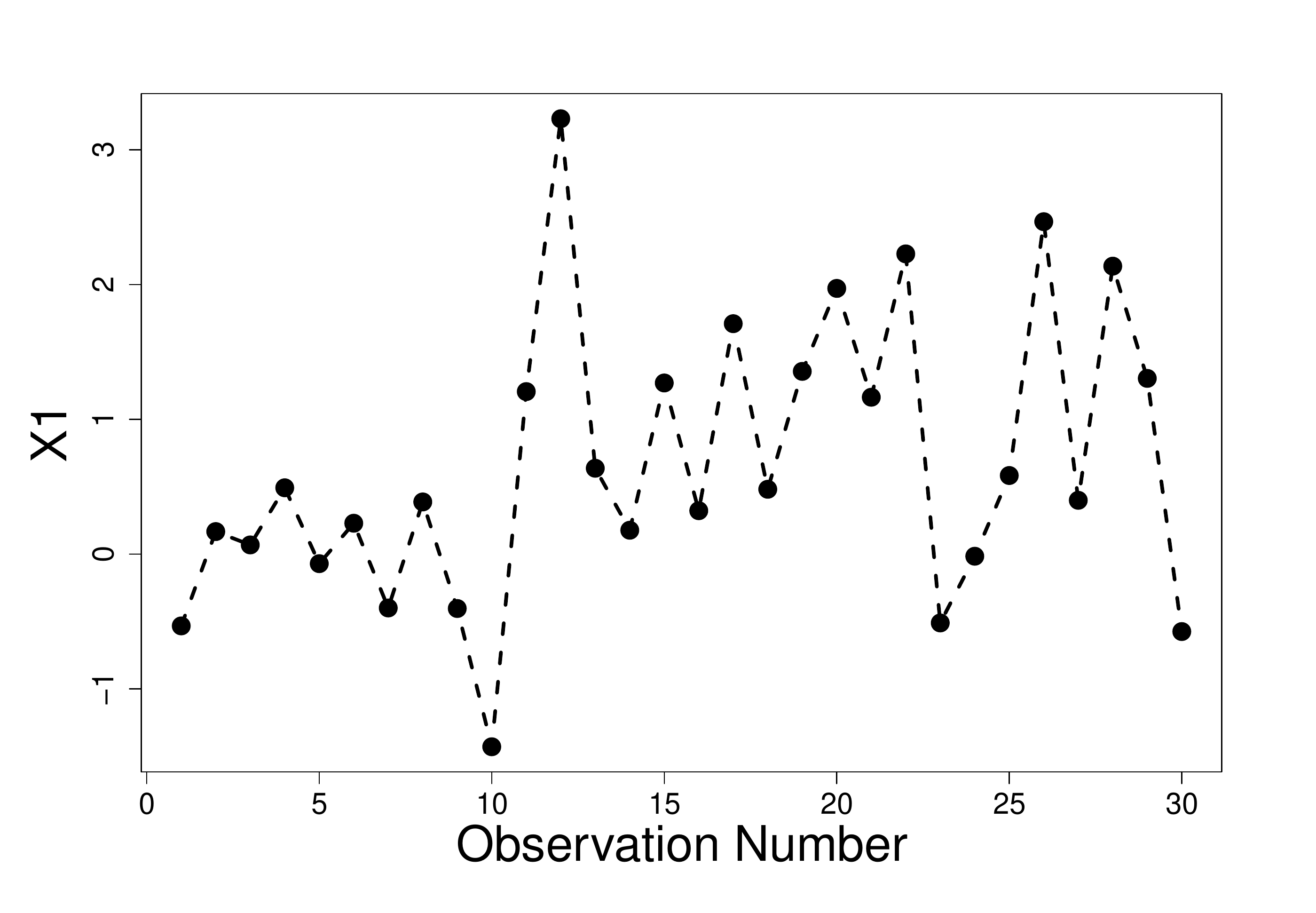}
    \caption{The variable $X_1$ in a sequence of 30 observations on the multivariate ($X_1, X_2, X_3$) in data set (a), given in Tables 6.4 and 6.5.}
    \label{fig:c6s4f2}
\end{figure}

\begin{figure}[t!]
    \centering
    \includegraphics[width=0.75\textwidth,angle=-90]{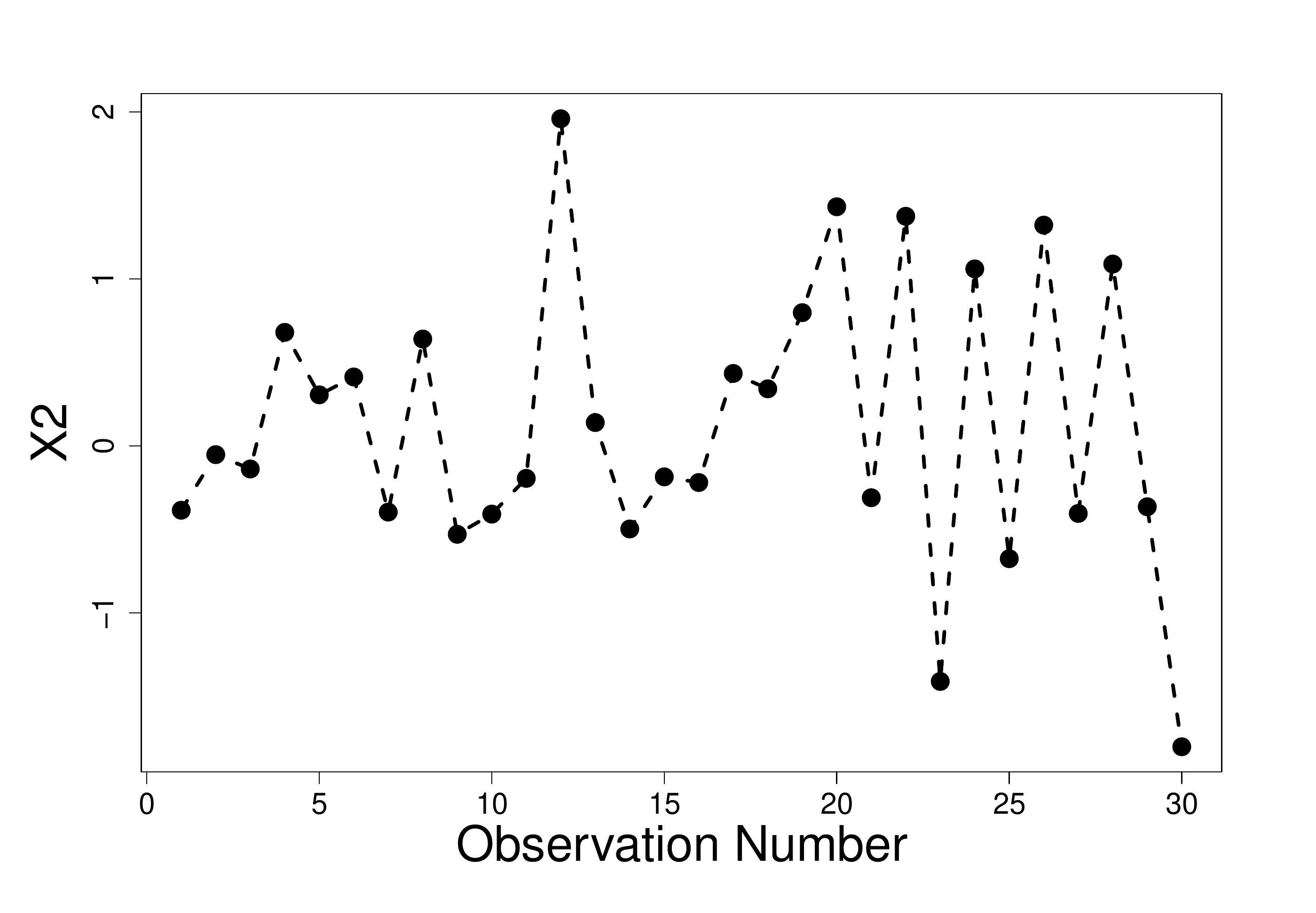}
    \caption{The variable $X_2$ in a sequence of 30 observations on the multivariate ($X_1, X_2, X_3$) in data set (a), given in Tables 6.4 and 6.5.}
    \label{fig:c6s4f3}
\end{figure}

\begin{figure}[t!]
    \centering
    \includegraphics[width=0.75\textwidth,angle=-90]{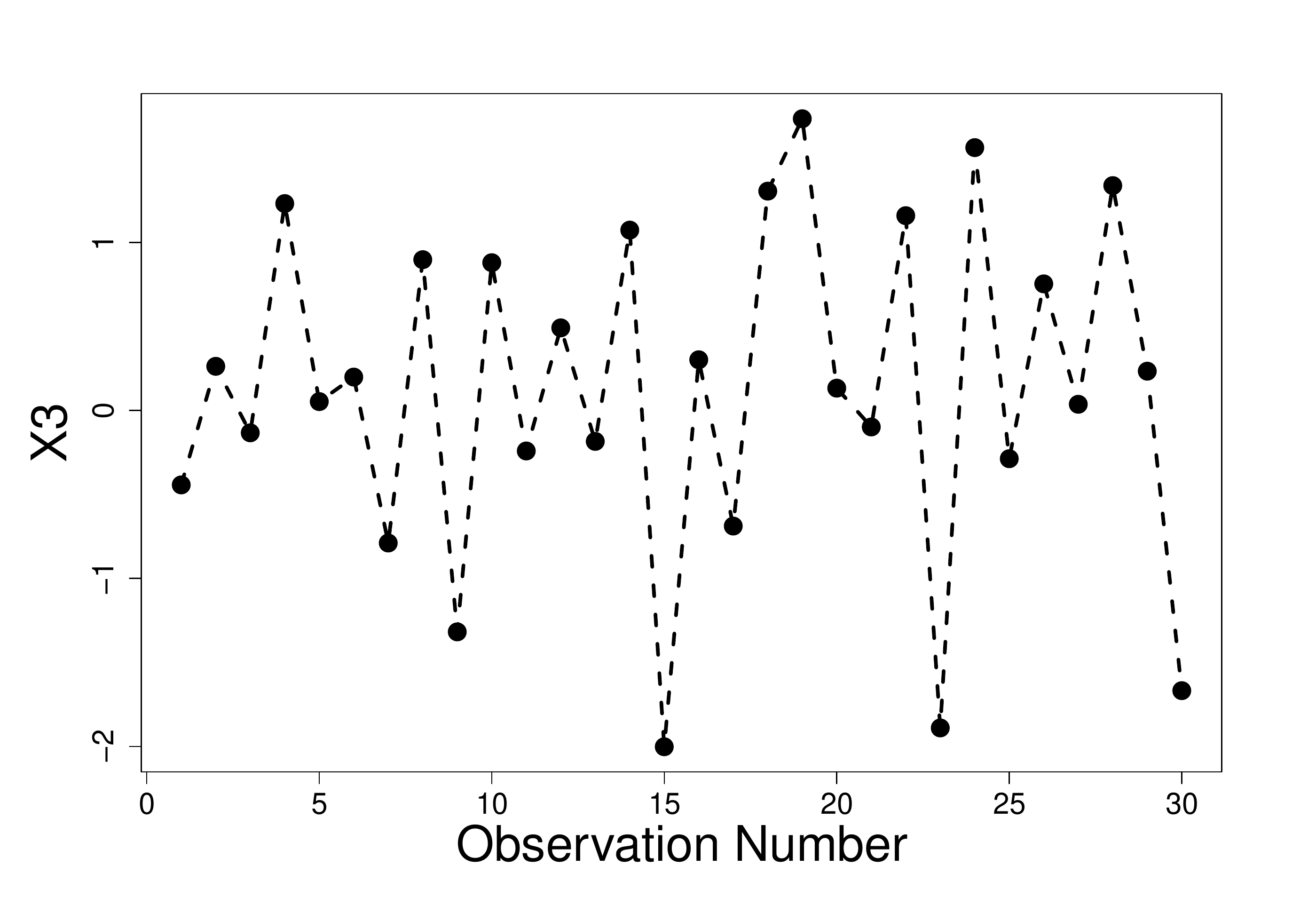}
    \caption{The variable $X_3$ in a sequence of 30 observations on the multivariate ($X_1, X_2, X_3$) in data set (a), given in Tables 6.4 and 6.5.}
    \label{fig:c6s4f4}
\end{figure}

For the data set (b), where the location parameter did not change but the covariance matrix was changed to $\Sigma_1 = 1.1\Sigma_0$ , the observations are given in Table \ref{tab:c6s4t3}. A signal is given at p=0.005 at observation 20. For comparison, \citet{qiu_2014} reports that the parametric $T^2$ did not signal at all, the parametric CUSUM gave only one signal, at observation 24, and the parametric COT signaled at observation 12. This can be observed in Figure 7.7, parts (d), (e), and (f), on page 283 of \citet{qiu_2014}.

Figures \ref{fig:c6s4f5} to \ref{fig:c6s4f7} for the data in this case show no apparent change in location, but a possible increase in spread

\begin{table}[t!]
    \caption{Observations 11--30 in data set (b), used in Example 6.4.1, appear in columns 2--4, the transformed values $Y$ are in columns 5--7, and the SNS computed on the components of $Y$ are in columns 8--10. The far right column gives the value of $T^2$ computed on the vector of SNS values, using $\Sigma_{10}$ from Table \ref{tab:c6s4t1} as the covariance (correlation in this case) matrix. The upper 0.005 limit for $T^2$ is 12.8.}
    \centering
    \begin{tabular}{crrrrrrrrrr}
    \textbf{sample} & \multicolumn{1}{c}{\textbf{$X_1$}} & \multicolumn{1}{c}{\textbf{$X_2$}} & \multicolumn{1}{c}{\textbf{$X_3$}} & \multicolumn{1}{c}{\textbf{$Y_1$}} & \multicolumn{1}{c}{\textbf{$Y_2$}} & \multicolumn{1}{c}{\textbf{$Y_3$}} & \multicolumn{1}{c}{\textbf{$SNS_1$}} & \multicolumn{1}{c}{\textbf{$SNS_2$}} & \multicolumn{1}{c}{\textbf{$SNS_3$}} & \multicolumn{1}{c}{\textbf{$T^2$}} \\
    11    & 0.216 & -0.204 & -0.253 & 0.047 & 0.042 & 0.064 & -0.473 & -0.748 & -0.473 & 0.593 \\
    12    & 2.339 & 2.054 & 0.516 & 5.471 & 4.219 & 0.266 & 1.691 & 1.691 & 0.000 & 9.684 \\
    13    & -0.381 & 0.147 & -0.193 & 0.145 & 0.022 & 0.037 & -0.230 & -0.748 & -0.748 & 0.789 \\
    14    & -0.863 & -0.521 & 1.127 & 0.745 & 0.271 & 1.270 & 1.097 & 0.473 & 0.748 & 1.225 \\
    15    & 0.283 & -0.194 & -2.100 & 0.080 & 0.038 & 4.410 & -0.230 & -0.748 & 1.691 & 12.432 \\
    16    & -0.711 & -0.229 & 0.317 & 0.506 & 0.052 & 0.100 & 1.097 & -0.748 & -0.230 & 3.657 \\
    17    & 0.745 & 0.455 & -0.722 & 0.555 & 0.207 & 0.521 & 1.097 & 0.473 & 0.000 & 2.363 \\
    18    & -0.545 & 0.359 & 1.369 & 0.297 & 0.129 & 1.874 & 1.097 & -0.473 & 1.691 & 8.258 \\
    19    & 0.373 & 0.837 & 1.822 & 0.139 & 0.701 & 3.320 & -0.230 & 1.691 & 1.691 & 6.626 \\
    20    & 1.019 & 1.502 & 0.140 & 1.038 & 2.256 & 0.020 & 1.097 & 1.691 & -0.748 & \bf{13.442} \\
    21    & 0.172 & -0.325 & -0.103 & 0.030 & 0.106 & 0.011 & -0.473 & -0.473 & -1.097 & 1.449 \\
    22    & 1.288 & 1.443 & 1.216 & 1.659 & 2.082 & 1.479 & 1.097 & 1.691 & 0.748 & 3.671 \\
    23    & -1.585 & -1.479 & -1.982 & 2.512 & 2.187 & 3.928 & 1.691 & 1.691 & 1.691 & 3.888 \\
    24    & -1.066 & 1.112 & 1.642 & 1.136 & 1.237 & 2.696 & 1.097 & 1.691 & 1.691 & 3.368 \\
    25    & -0.437 & -0.708 & -0.301 & 0.191 & 0.501 & 0.091 & 0.473 & 1.691 & -0.230 & 7.114 \\
    26    & 1.538 & 1.387 & 0.790 & 2.365 & 1.924 & 0.624 & 1.691 & 1.691 & 0.230 & 8.041 \\
    27    & -0.630 & -0.424 & 0.039 & 0.397 & 0.180 & 0.002 & 1.097 & 0.473 & -1.691 & 15.436 \\
    28    & 1.193 & 1.142 & 1.404 & 1.423 & 1.304 & 1.971 & 1.097 & 1.691 & 1.691 & 3.368 \\
    29    & 0.318 & -0.382 & 0.245 & 0.101 & 0.146 & 0.060 & -0.230 & -0.473 & -0.473 & 0.277 \\
    30    & -1.652 & -1.889 & -1.749 & 2.729 & 3.568 & 3.059 & 1.691 & 1.691 & 1.691 & 3.888 \\
    \end{tabular}

    \label{tab:c6s4t3}
\end{table}

\begin{figure}[t!]
    \centering
    \includegraphics[width=0.75\textwidth,angle=-90]{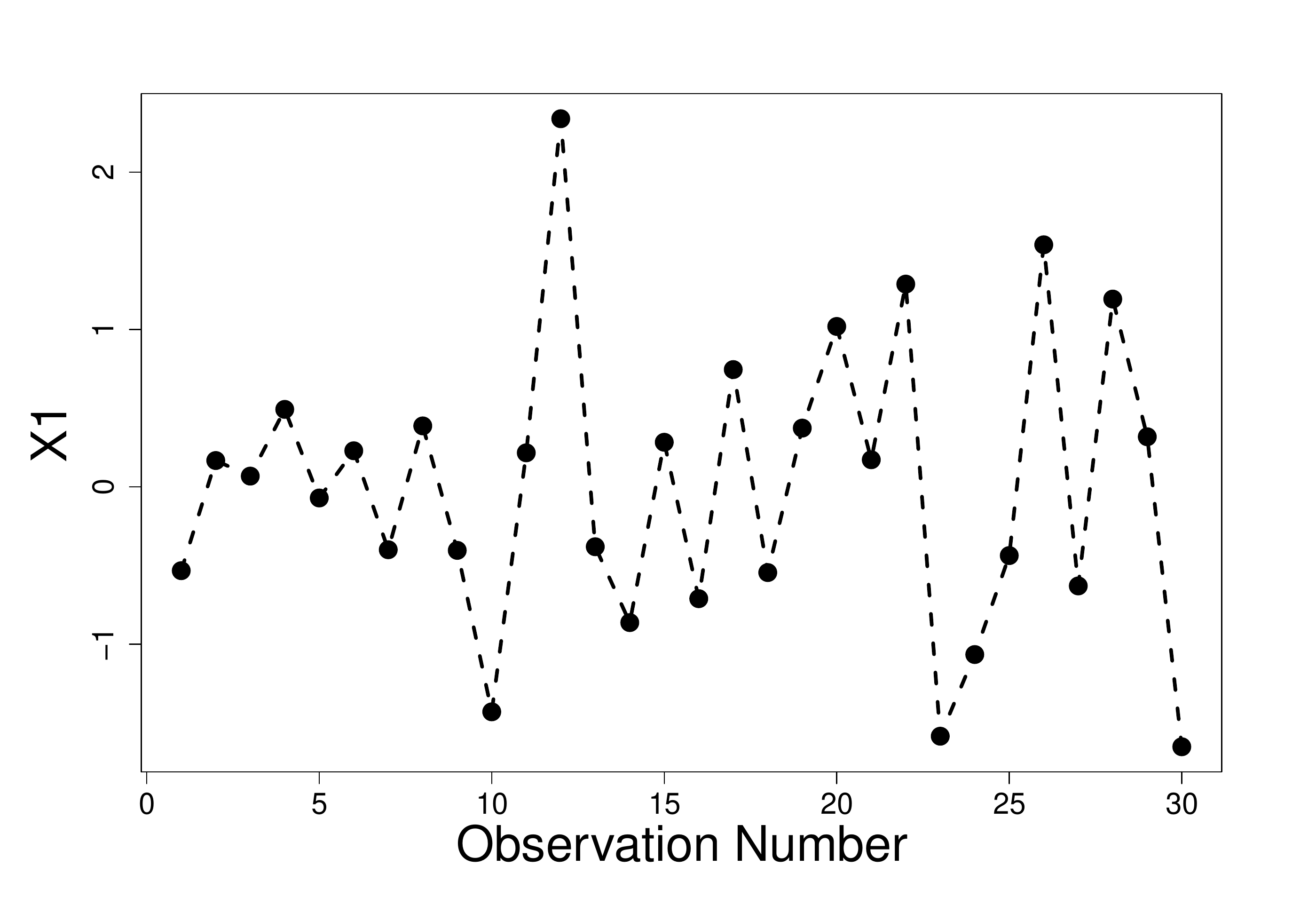}
    \caption{The variable $X_1$ in a sequence of 30 observations on the multivariate ($X_1, X_2, X_3$) in data set (b), given in Tables 6.4 and 6.6.}
    \label{fig:c6s4f5}
\end{figure}

\begin{figure}[t!]
    \centering
    \includegraphics[width=0.75\textwidth,angle=-90]{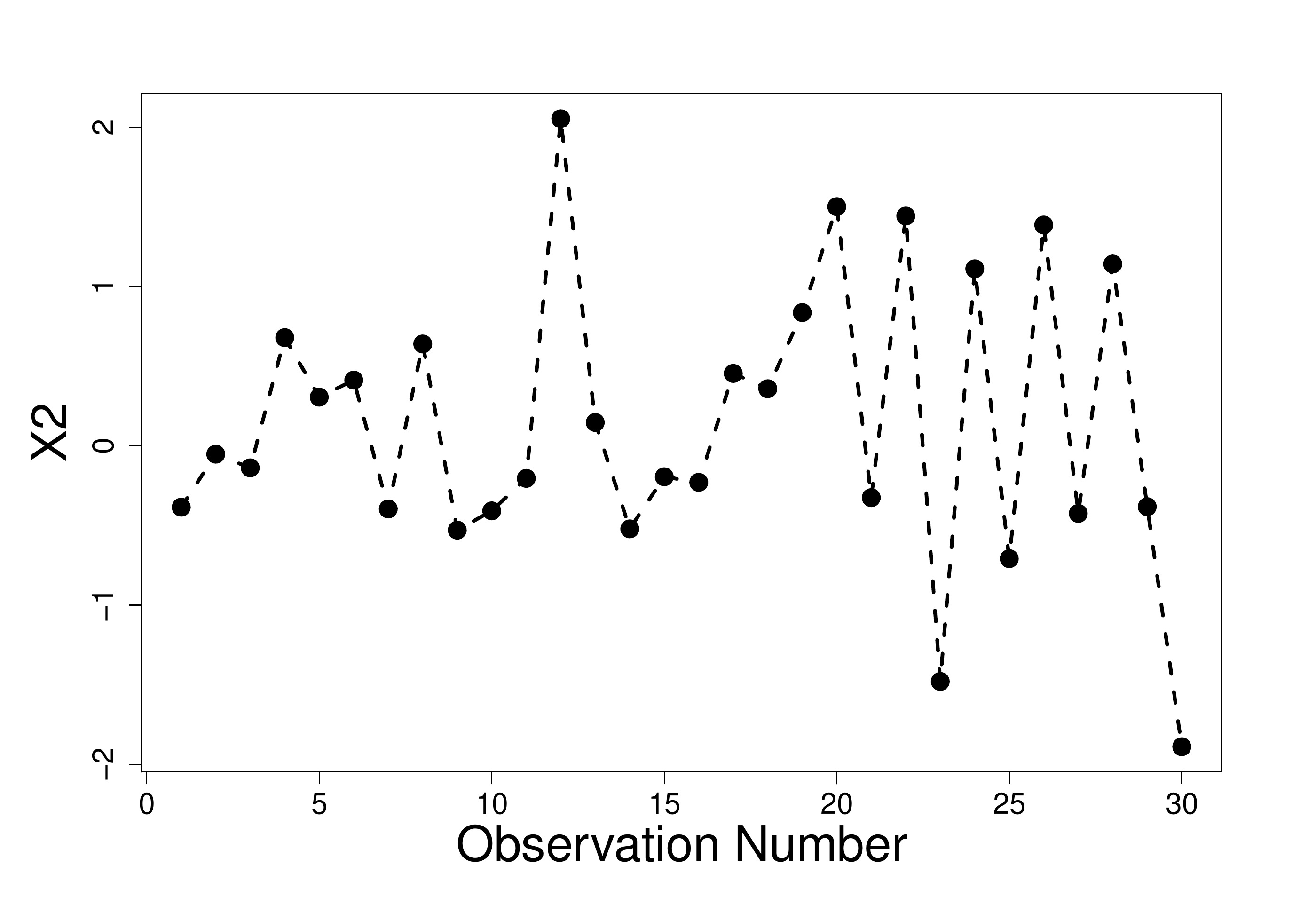}
    \caption{The variable $X_2$ in a sequence of 30 observations on the multivariate ($X_1, X_2, X_3$) in data set (b), given in Tables 6.4 and 6.6.}
    \label{fig:c6s4f6}
\end{figure}

\begin{figure}[t!]
    \centering
    \includegraphics[width=0.75\textwidth,angle=-90]{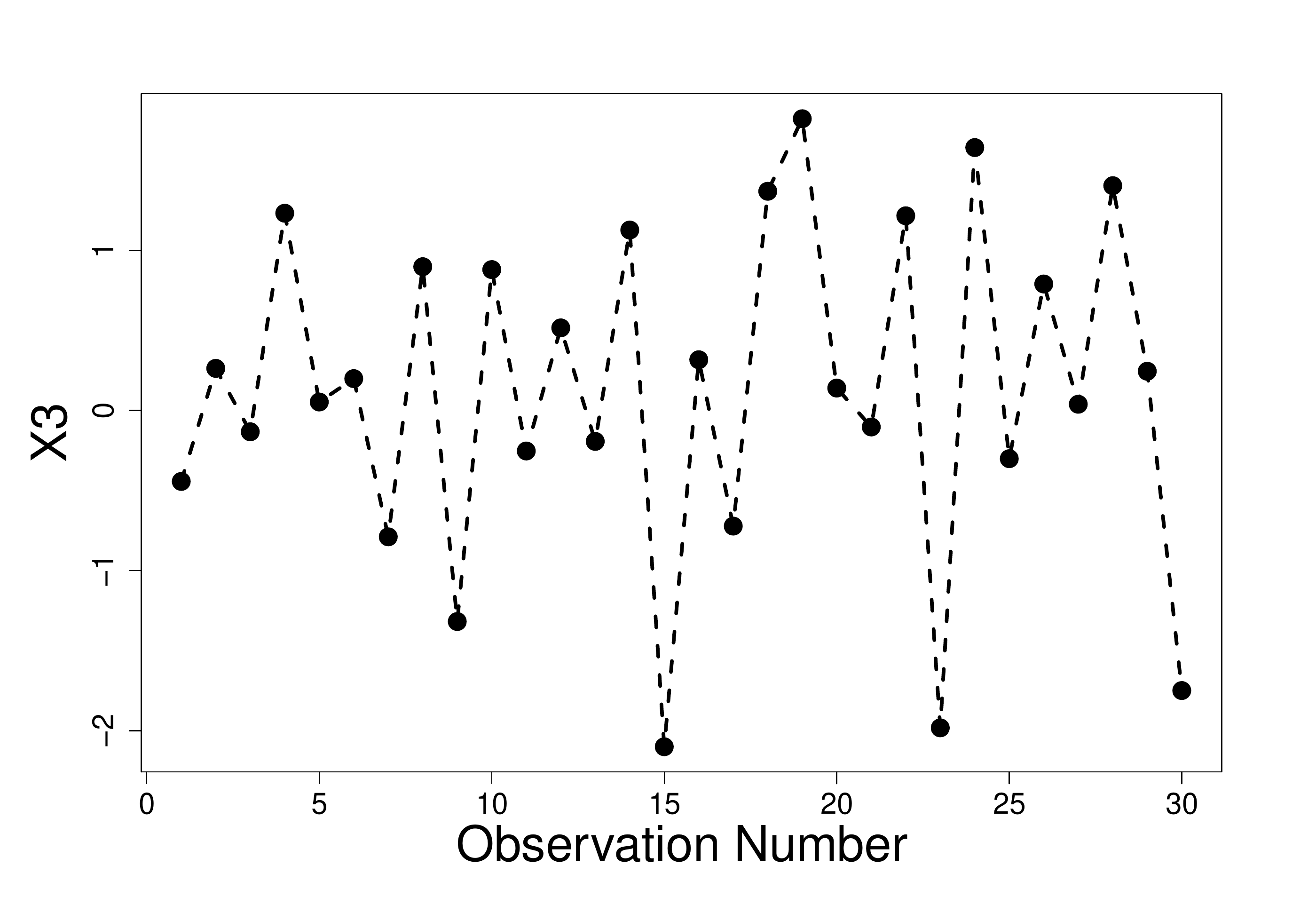}
    \caption{The variable $X_3$ in a sequence of 30 observations on the multivariate ($X_1, X_2, X_3$) in data set (b), given in Tables 6.4 and 6.6.}
    \label{fig:c6s4f7}
\end{figure}

The third data set used both the location change in data set (a) and the covariance matrix change in data set (b) for observations 11--30. The results are given in Table \ref{tab:c6s4t4}. The nonparametric SNS method shows its first signal at the 0.005 level with observation 20.
\begin{table}[t!]
    \caption{Observations 11--30 in data set (c), used in Example 6.4.1, appear in columns 2--4, the transformed values $Y$ are in columns 5--7, and the SNS computed on the components of $Y$ are in columns 8--10. The far right column gives the value of $T^2$ computed on the vector of SNS values, using $\Sigma_{10}$ from Table \ref{tab:c6s4t1} as the covariance (correlation in this case) matrix. The upper 0.005 limit for $T^2$ is 12.8.}
    \centering
    \begin{tabular}{crrrrrrcccr}
    \textbf{sample} & \multicolumn{1}{c}{\textbf{$X_1$}} & \multicolumn{1}{c}{\textbf{$X_2$}} & \multicolumn{1}{c}{\textbf{$X_3$}} & \multicolumn{1}{c}{\textbf{$Y_1$}} & \multicolumn{1}{c}{\textbf{$Y_2$}} & \multicolumn{1}{c}{\textbf{$Y_3$}} & \multicolumn{1}{c}{\textbf{$SNS_1$}} & \multicolumn{1}{c}{\textbf{$SNS_2$}} & \multicolumn{1}{c}{\textbf{$SNS_3$}} & \multicolumn{1}{c}{\textbf{$T^2$}} \\
    11    & 1.216 & -0.204 & -0.253 & 1.479 & 0.042 & 0.064 & 1.097 & -0.748 & -0.473 & 4.167 \\
    12    & 3.339 & 2.054 & 0.516 & 11.149 & 4.219 & 0.266 & 1.691 & 1.691 & 0.000 & 9.684 \\
    13    & 0.619 & 0.147 & -0.193 & 0.383 & 0.022 & 0.037 & 1.097 & -0.748 & -0.748 & 5.109 \\
    14    & 0.137 & -0.521 & 1.127 & 0.019 & 0.271 & 1.270 & -0.748 & 0.473 & 0.748 & 3.178 \\
    15    & 1.283 & -0.194 & -2.100 & 1.646 & 0.038 & 4.410 & 1.097 & -0.748 & 1.691 & 10.202 \\
    16    & 0.289 & -0.229 & 0.317 & 0.084 & 0.052 & 0.100 & -0.230 & -0.748 & -0.230 & 0.729 \\
    17    & 1.745 & 0.455 & -0.722 & 3.045 & 0.207 & 0.521 & 1.691 & 0.473 & 0.000 & 5.145 \\
    18    & 0.455 & 0.359 & 1.369 & 0.207 & 0.129 & 1.874 & 0.473 & -0.473 & 1.691 & 8.601 \\
    19    & 1.373 & 0.837 & 1.822 & 1.885 & 0.701 & 3.320 & 1.097 & 1.691 & 1.691 & 3.368 \\
    20    & 2.019 & 1.502 & 0.140 & 4.076 & 2.256 & 0.020 & 1.691 & 1.691 & -0.748 & \bf{16.894} \\
    21    & 1.172 & -0.325 & -0.103 & 1.374 & 0.106 & 0.011 & 1.097 & -0.473 & -1.097 & 6.869 \\
    22    & 2.288 & 1.443 & 1.216 & 5.235 & 2.082 & 1.479 & 1.691 & 1.691 & 0.748 & 5.324 \\
    23    & -0.585 & -1.479 & -1.982 & 0.342 & 2.187 & 3.928 & 1.097 & 1.691 & 1.691 & 3.368 \\
    24    & -0.066 & 1.112 & 1.642 & 0.004 & 1.237 & 2.696 & -1.691 & 1.691 & 1.691 & 17.277 \\
    25    & 0.563 & -0.708 & -0.301 & 0.317 & 0.501 & 0.091 & 1.097 & 1.691 & -0.230 & 8.768 \\
    26    & 2.538 & 1.387 & 0.790 & 6.441 & 1.924 & 0.624 & 1.691 & 1.691 & 0.230 & 8.041 \\
    27    & 0.370 & -0.424 & 0.039 & 0.137 & 0.180 & 0.002 & -0.230 & 0.473 & -1.691 & 9.100 \\
    28    & 2.193 & 1.142 & 1.404 & 4.809 & 1.304 & 1.971 & 1.691 & 1.691 & 1.691 & 3.888 \\
    29    & 1.318 & -0.382 & 0.245 & 1.737 & 0.146 & 0.060 & 1.097 & -0.473 & -0.473 & 3.741 \\
    30    & -0.652 & -1.889 & -1.749 & 0.425 & 3.568 & 3.059 & 1.097 & 1.691 & 1.691 & 3.368 \\
    \end{tabular}
    \label{tab:c6s4t4}
\end{table}

Figures \ref{fig:c6s4f8} to \ref{fig:c6s4f10} of the raw data clearly show the shift in location of $X_1$, and indicate a possible increase in spread for all three variables. For comparison, the parametric $T^2$ barely signaled at observation 12 and nowhere else, the parametric CUSUM signaled at observation 14, and the parametric COT signaled at observation 11, where the actual change took place. This can be observed in Figure 7.7, parts (g), (h) and (i), on page 283 of \citet{qiu_2014}.

\begin{figure}[t!]
    \centering
    \includegraphics[width=0.75\textwidth,angle=-90]{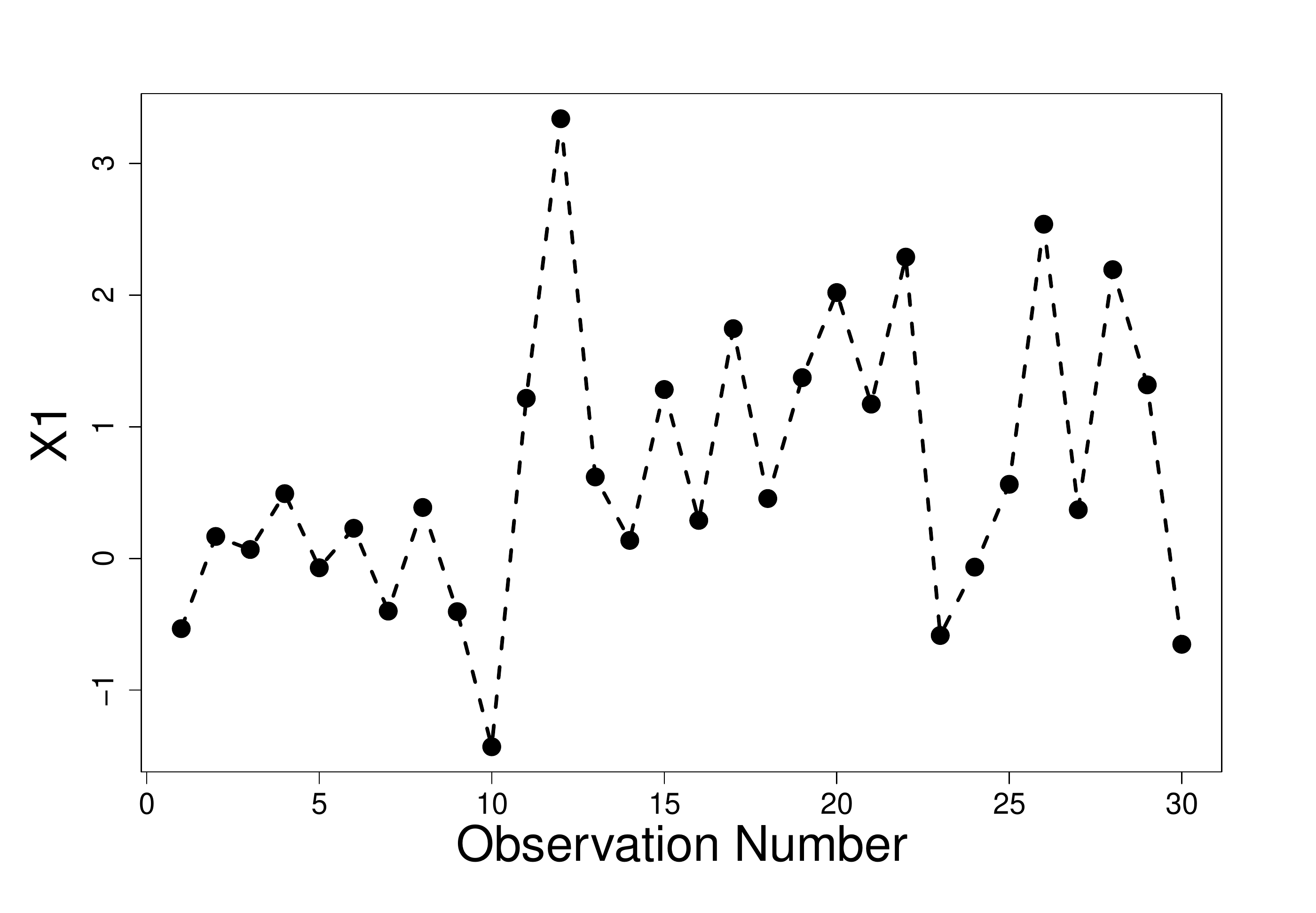}
    \caption{The variable $X_1$ in a sequence of 30 observations on the multivariate ($X_1, X_2, X_3$) in data set (c), given in Tables 6.4 and 6.7.}
    \label{fig:c6s4f8}
\end{figure}

\begin{figure}[t!]
    \centering
    \includegraphics[width=0.75\textwidth,angle=-90]{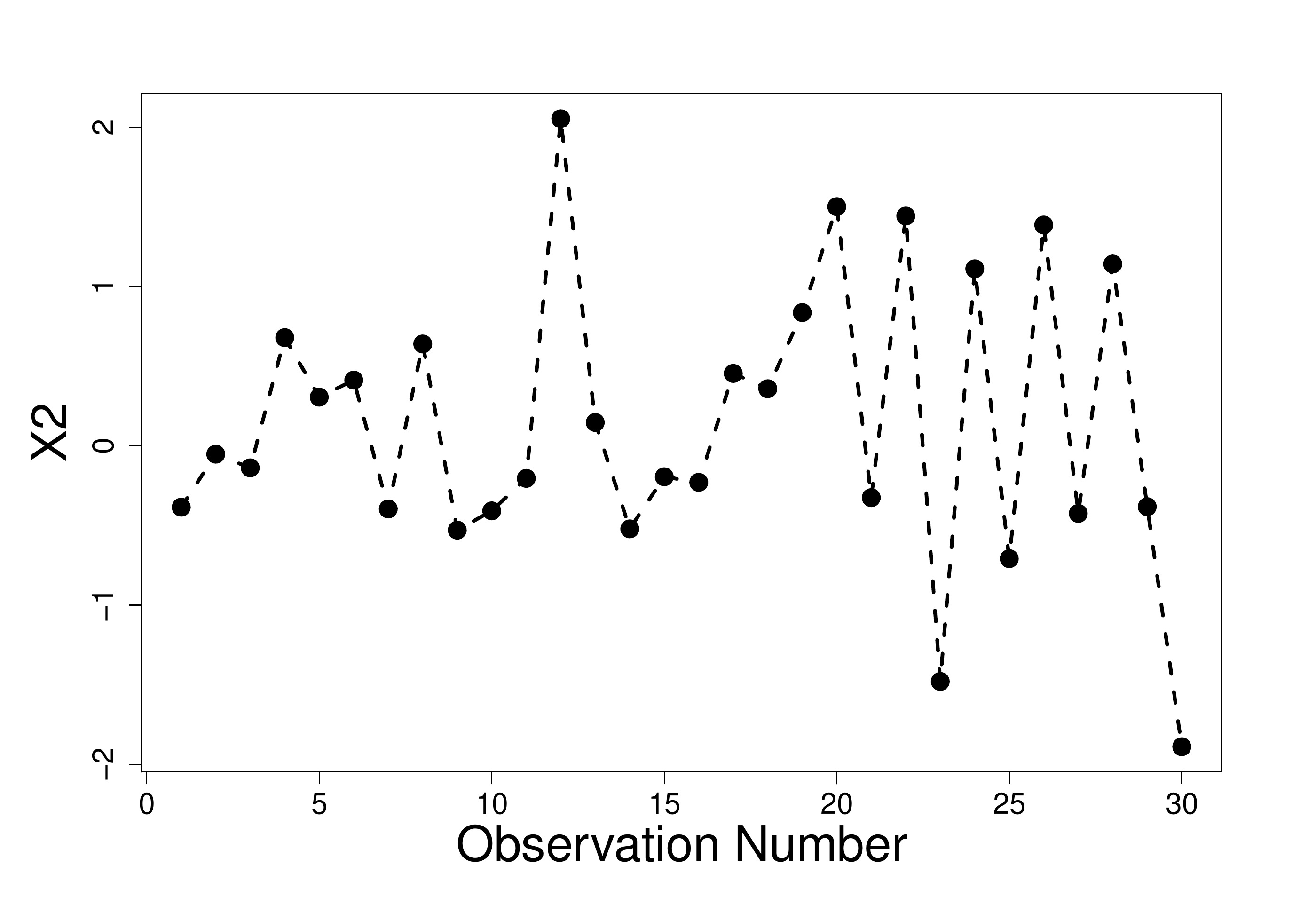}
    \caption{The variable $X_2$ in a sequence of 30 observations on the multivariate ($X_1, X_2, X_3$) in data set (c), given in Tables 6.4 and 6.7.}
    \label{fig:c6s4f9}
\end{figure}

\begin{figure}[t!]
    \centering
    \includegraphics[width=0.75\textwidth,angle=-90]{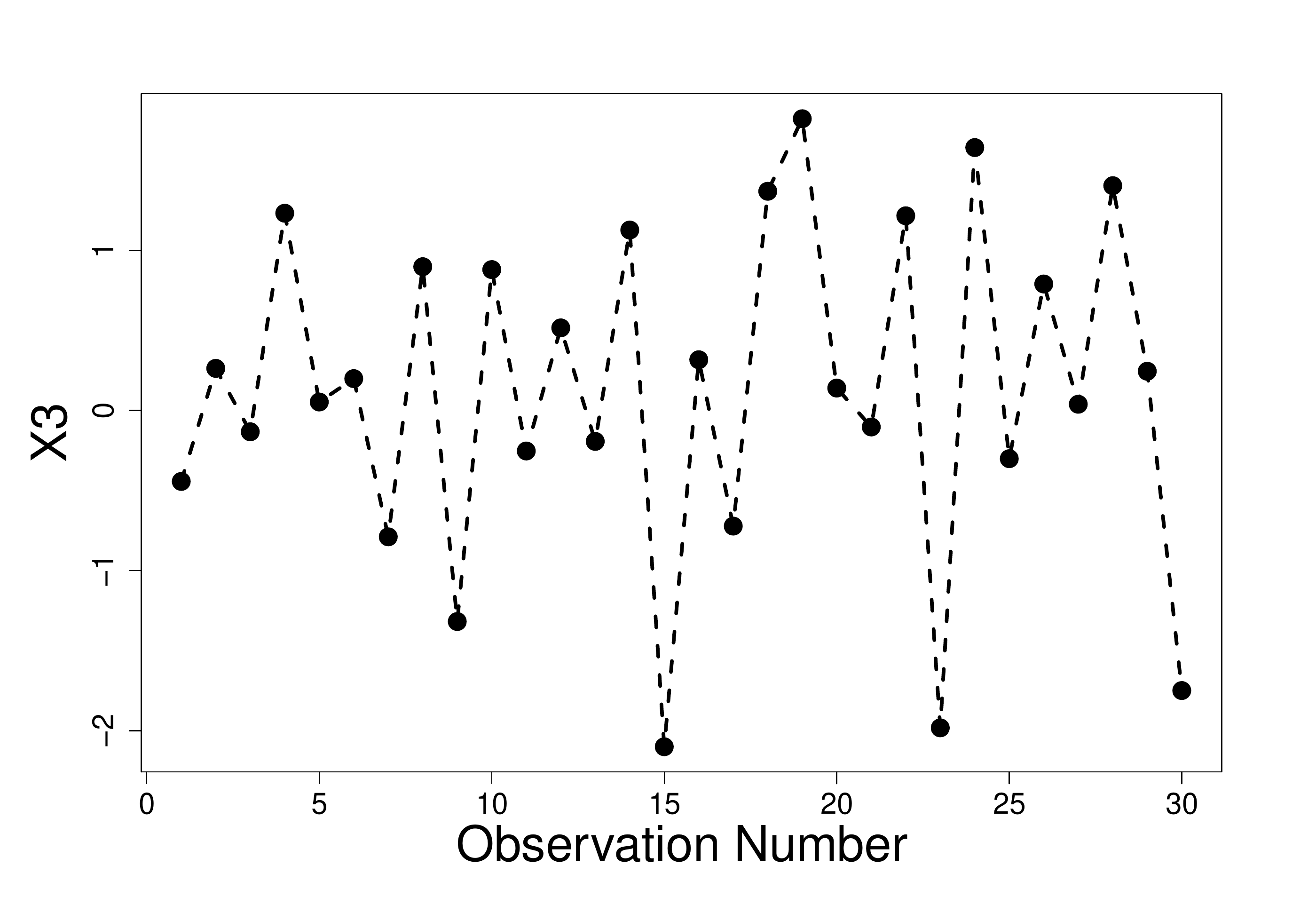}
    \caption{The variable $X_3$ in a sequence of 30 observations on the multivariate ($X_1, X_2, X_3$) in data set (c), given in Tables 6.4 and 6.7.}
    \label{fig:c6s4f10}
\end{figure}

In summary the nonparametric SNS method was effective in signaling a change in location, spread, or both location and spread for this data set, competing well with three parametric methods designed to be effective for the known changes in location and spread that produced the data. Also, it appears that a nonparametric modification of the parametric COT method may be an effective add-on to the nonparametric SNS method for detecting changes that may be too subtle to be detected using only $T^2$.

A comparison of the three sets of $T^2$ values for data sets (a), (b), and (c) in Table \ref{tab:c6s4t5} shows strong similarities as expected. They are exactly the same for observations 1--10 as expected, and show only slight variations for observations 11--30, because the data for observations 11--30 are based on one set of generated data, modified in three different ways for the three different scenarios. They all send the first signal at observation 20. However, we compared these $T^2$ values with the $\chi^2$ distribution with 3 degrees of freedom, not knowing if that is the correct comparison. So we need to revert to a nonparametric approach for identifying the first signal and the change point estimate.

\begin{sidewaystable}[t!]
    \caption{The $T^2$ statistics from Tables 6.5-6.7 for the multivariate SNS method applied to three data sets used in Example 7.4 of \citet{qiu_2014} and in our Example 6.4.1.}
    \centering
    \begin{tabular}{c|rrrr|rrrr|rrrr|}
    \multicolumn{1}{c}{Sample} & \multicolumn{1}{c}{$T^2(a)$} & \multicolumn{1}{c}{SNS} & \multicolumn{1}{c}{EWMA} & \multicolumn{1}{c}{$T(a)$} & \multicolumn{1}{c}{$T^2(b)$} & \multicolumn{1}{c}{SNS} & \multicolumn{1}{c}{EWMA} & \multicolumn{1}{c}{$T(b)$} & \multicolumn{1}{c}{$T^2(c)$} & \multicolumn{1}{c}{SNS} & \multicolumn{1}{c}{EWMA} & \multicolumn{1}{c}{$T(c)$} \\
\cline{2-13}    1     & 2.464 & 0.000 &       &       & 2.464 & 0.000 &       &       & 2.464 & 0.000 &       &  \\
    2     & 4.005 & 0.674 &       &       & 4.005 & 0.674 &       &       & 4.005 & 0.674 &       &  \\
    3     & 2.814 & 0.000 &       &       & 2.814 & 0.000 &       &       & 2.814 & 0.000 &       &  \\
    4     & 2.742 & -0.319 &       &       & 2.742 & -0.319 &       &       & 2.742 & -0.319 &       &  \\
    5     & 3.149 & 0.524 &       &       & 3.149 & 0.524 &       &       & 3.149 & 0.524 &       &  \\
    6     & 1.893 & -1.383 &       &       & 1.893 & -1.383 &       &       & 1.893 & -1.383 &       &  \\
    7     & 0.112 & -1.465 &       &       & 0.112 & -1.465 &       &       & 0.112 & -1.465 &       &  \\
    8     & 1.733 & -0.887 &       &       & 1.733 & -0.887 &       &       & 1.733 & -0.887 &       &  \\
    9     & 3.850 & 0.967 &       &       & 3.850 & 0.967 &       &       & 3.850 & 0.967 &       &  \\
    10    & 3.631 & 0.674 & 0.000 &       & 3.631 & 0.674 & 0.000 &       & 3.631 & 0.674 & 0.000 &  \\
    11    & 4.167 & 1.691 & 0.169 & \textbf{2.272} & 0.593 & -1.097 & -0.110 & 1.683 & 4.167 & 1.691 & 0.169 & \textbf{2.383} \\
    12    & 9.684 & 1.732 & 0.325 & 1.731 & 9.684 & 1.732 & 0.074 & 2.201 & 9.684 & 1.732 & 0.325 & 1.838 \\
    13    & 5.109 & 1.198 & 0.413 & 1.192 & 0.789 & -0.869 & -0.020 & 1.623 & 5.109 & 1.198 & 0.413 & 1.290 \\
    14    & 2.138 & -0.674 & 0.304 & 0.881 & 1.225 & -0.674 & -0.085 & 2.064 & 3.178 & 0.090 & 0.380 & 0.976 \\
    15    & 10.202 & 1.834 & 0.457 & 1.346 & 12.432 & 1.834 & 0.107 & \textbf{2.440} & 10.202 & 1.834 & 0.526 & 1.150 \\
    16    & 0.729 & -1.318 & 0.279 & 0.778 & 3.657 & 0.579 & 0.154 & 1.860 & 0.729 & -1.318 & 0.341 & 0.548 \\
    17    & 5.145 & 1.049 & 0.356 & 1.555 & 2.363 & -0.299 & 0.108 & 1.772 & 5.145 & 1.049 & 0.412 & 1.412 \\
    18    & 4.237 & 0.674 & 0.388 & 1.333 & 8.258 & 1.085 & 0.206 & 2.042 & 8.601 & 1.085 & 0.479 & 1.188 \\
    19    & 3.368 & 0.000 & 0.349 & 1.283 & 6.626 & 0.899 & 0.275 & 1.772 & 3.368 & 0.000 & 0.431 & 0.932 \\
    20    & \textbf{21.232} & 1.960 & 0.510 & 1.583 & \textbf{13.442} & 1.960 & 0.444 & 1.577 & \text{16.894} & 1.960 & \textbf{0.584} & 1.317 \\
    21    & 6.869 & 0.967 & 0.556 & 0.886 & 1.449 & -0.792 & 0.320 & 0.914 & 6.869 & 0.792 & \textbf{0.605} & 0.442 \\
    22    & 5.324 & 0.825 & \textbf{0.583} & 0.665 & 3.671 & 0.410 & 0.329 & 1.470 & 5.324 & 0.674 & \textbf{0.612} & 0.245 \\
    23    & 3.633 & 0.000 & 0.525 & 0.502 & 3.888 & 0.575 & 0.354 & 1.519 & 3.368 & -0.164 & 0.534 &  \\
    24    & \textbf{17.277} & 1.534 & \textbf{0.626} & 1.130 & 3.368 & 0.052 & 0.324 & 1.512 & \textbf{17.277} & 2.037 & \textbf{0.685} &  \\
    25    & 5.364 & 0.772 & \textbf{0.640} &       & 7.114 & 0.915 & 0.383 & 1.862 & 8.768 & 0.915 & \textbf{0.708} &  \\
    26    & 9.684 & 1.020 & \textbf{0.678} &       & 8.041 & 0.942 & 0.439 & 1.782 & 8.041 & 0.674 & \textbf{0.704} &  \\
    27    & 7.281 & 0.828 & \textbf{0.693} &       & \textbf{15.436} & 2.085 & \textbf{0.603} & 1.818 & 9.100 & 0.967 & \textbf{0.731} &  \\
    28    & 3.888 & -0.045 & \textbf{0.620} &       & 3.368 & -0.090 & 0.534 &       & 3.888 & -0.045 & \textbf{0.653} &  \\
    29    & 3.741 & -0.174 & 0.540 &       & 0.277 & -1.628 & 0.318 &       & 3.741 & -0.174 & 0.570 &  \\
    30    & 3.368 & -0.431 & 0.443 &       & 3.888 & 0.431 & 0.329 &       & 3.368 & -0.385 & 0.475 &  \\
\cline{2-13}
    \end{tabular}

    \label{tab:c6s4t5}
\end{sidewaystable}

We convert each $T^2$ to a sequential normal score, shown in Table 6.8. None of the SNS values exceed the 0.005 limit 2.58 because of the small number of observations. We then apply the EWMA approach, beginning with the first Phase II observation. This gives a signal ($> 0.563$ for $\text{ARL}=200$) in every case. For data set (a) the first signal is at observation 22, and the change point estimate is at observation 11, the actual change point. For data set (b) the first EWMA signal is at observation 27, and the change point estimate is at observation 15. For data set (c) the first EWMA signal is at observation 20 with the change point estimate again at observation 11, the actual change point.
\end{myexample}
Thus the extension of the SNS method, introduced in Section \ref{sec:c4s3}, to the multivariate case is an effective nonparametric competitor to the parametric methods demonstrated by Example 7.4 in \cite{qiu_2014}. Now we will demonstrate an extension of the $SNS^2$ method, introduced in Section 4.2, to the multivariate case, and show that it also is an effective nonparametric competitor. It differs from the previous nonparametric method in that no $a~priori$ estimate of location is needed to use this method.

\begin{myexample}
For this example the SNS are found on the $X_i$ rather than on the $Y_i$ as they were in the previous example. Then they are squared to get $SNS^2$ as described in Section \ref{sec:c4s2}. These $SNS^2$ values tend to be large if the location parameter changes in either direction or if the spread increases, as discussed in Section \ref{sec:c4s2}. However, their use in a multivariate analysis based on $T^2$ is limited because the parametric methods based on $T^2$ assume multivariate normality, not a multivariate distribution based on chi-squared random variables. Therefore an additional step, transforming the approximate chi-squared random variables to approximate standard normal random variables, is required. That is, the $SNS^2$ values for each component $X_i$ are ranked, converted to rankits, and converted to a new set of $SNS$ values, which now are approximately standard normal. The correlation matrix is estimated based on the first 10 observations, and used to compute $T^2$ on the remaining observations, as in the previous method. The results are given in Table \ref{tab:c6s4t9}.

\begin{table}[t]
\caption{ Data set to illustrate our second nonparametric method. Columns 2-4 are the original observations from Example 7.4 in \cite{qiu_2014}. Columns 5-7 are the squared values of the SNS computed on columns 2-4 respectively. Columns 8-10 are new $SNS$ computed on columns 5-7.}
\label{tab:c6s4t9}
\begin{tabular}{llllllllll}
Obs. & $X_1$ & $X_2$ & $X_3$ & $SNS_{X_1}^2$ & $SNS_{X_2}^2$ & $SNS_{X_1}^2$ & $SNS_1$ & $SNS_2$ & $SNS_3$ \\
1    & -0.53 & -0.39 & -0.44 & 1.07          & 0.15          & 0.45          & 0.35    & -0.60   & -0.11   \\
2    & 0.17  & -0.05 & 0.26  & 0.15          & 0.02          & 0.15          & -0.60   & -1.10   & -0.60   \\
3    & 0.07  & -0.14 & -0.13 & 0.02          & 0.02          & 0.15          & -1.10   & -1.69   & -0.60   \\
4    & 0.49  & 0.68  & 1.23  & 2.71          & 2.71          & 2.71          & 0.75    & 0.75    & 0.75    \\
5    & -0.07 & 0.31  & 0.05  & 0.02          & 0.15          & 0.02          & -1.69   & -0.60   & -1.69   \\
6    & 0.23  & 0.41  & 0.20  & 0.45          & 0.45          & 0.02          & -0.11   & -0.11   & -1.10   \\
7    & -0.40 & -0.40 & -0.79 & 0.15          & 0.45          & 1.07          & -0.60   & -0.11   & 0.35    \\
8    & 0.39  & 0.64  & 0.90  & 1.07          & 1.07          & 1.07          & 0.35    & 0.35    & 0.35    \\
9    & -0.40 & -0.53 & -1.32 & 0.45          & 2.71          & 2.71          & -0.11   & 1.10    & 1.10    \\
10   & -1.43 & -0.41 & 0.88  & 2.71          & 1.07          & 0.45          & 1.10    & 0.35    & -0.11   \\
11   & 1.21  & -0.19 & -0.24 & 2.86          & 0.05          & 0.22          & 1.69    & -0.75   & -0.23   \\
12   & 3.23  & 1.96  & 0.49  & 2.86          & 2.86          & 0.22          & 1.69    & 1.69    & -0.23   \\
13   & 0.64  & 0.14  & -0.18 & 2.86          & 0.05          & 0.22          & 1.69    & -0.75   & -0.23   \\
14   & 0.18  & -0.50 & 1.07  & 0.22          & 1.20          & 1.20          & -0.23   & 0.75    & 0.75    \\
15   & 1.27  & -0.19 & -2.00 & 2.86          & 0.05          & 2.86          & 1.69    & -0.75   & 1.69    \\
16   & 0.32  & -0.22 & 0.30  & 0.56          & 0.05          & 0.22          & 0.23    & -0.75   & -0.23   \\
17   & 1.71  & 0.43  & -0.69 & 2.86          & 0.56          & 0.56          & 1.69    & 0.23    & 0.23    \\
18   & 0.48  & 0.34  & 1.31  & 1.20          & 0.22          & 2.86          & 0.75    & -0.23   & 1.69    \\
19   & 1.36  & 0.80  & 1.74  & 2.86          & 2.86          & 2.86          & 1.69    & 1.69    & 1.69    \\
20   & 1.97  & 1.43  & 0.13  & 2.86          & 2.86          & 0.00          & 1.69    & 1.69    & -1.69   \\
21   & 1.16  & -0.31 & -0.10 & 2.86          & 0.05          & 0.05          & 1.69    & -0.75   & -0.75   \\
22   & 2.23  & 1.38  & 1.16  & 2.86          & 2.86          & 1.20          & 1.69    & 1.69    & 0.75    \\
23   & -0.51 & -1.41 & -1.89 & 0.56          & 2.86          & 2.86          & 0.23    & 1.69    & 1.69    \\
24   & -0.02 & 1.06  & 1.57  & 0.00          & 2.86          & 2.86          & -1.69   & 1.69    & 1.69    \\
25   & 0.58  & -0.68 & -0.29 & 2.86          & 2.86          & 0.22          & 1.69    & 1.69    & -0.23   \\
26   & 2.47  & 1.32  & 0.75  & 2.86          & 2.86          & 0.22          & 1.69    & 1.69    & -0.23   \\
27   & 0.40  & -0.40 & 0.04  & 1.20          & 0.56          & 0.05          & 0.75    & 0.23    & -0.75   \\
28   & 2.14  & 1.09  & 1.34  & 2.86          & 2.86          & 2.86          & 1.69    & 1.69    & 1.69    \\
29   & 1.30  & -0.36 & 0.23  & 2.86          & 0.05          & 0.05          & 1.69    & -0.75   & -0.75   \\
30   & -0.58 & -1.80 & -1.67 & 1.20          & 2.86          & 2.86          & 0.75    & 1.69    & 1.69   
\end{tabular}
\end{table}

As in Example 6.4.1 the three data sets from Example 7.4 on page 282 of \cite{qiu_2014} are used to illustrate this second method. As before, the first 10 observations are considered the reference set, and all computations on $T^2$ use the correlation matrix based on the $SNS$ values from these first 10 observations (columns 8-10 in Table \ref{tab:c6s4t9}). The correlations in the matrix are $r_{1,2}$ = 0.622,  $r_{1,3}$ = 0.598  and $r_{2,3} = 0.676$, and the determinant is 0.302. The resulting $T^2$ is given in Table \ref{tab:c6s4t10} for data set (a), showing significance at $p=0.005$ (the value used in Qiu’s example) at observation 15, with an estimated change point at the actual change point, observation 11, based on SNS values computed on $T^2$, also shown in Table 6.10.

Similar computations were made for data sets (b) and (c), not shown here, with the results shown in Table \ref{tab:c6s4t10}. Data sets (a) and (c), both involving a shift in location, show similar results. Data set (b), with only a change in scale, does not show significance ($>12.8$ for a chi-squared random variable, 3 d.f., at $p = 0.005$) until observation 20, with an estimated change point at observation 15. Note that the two new nonparametric methods illustrated in the two examples in this section have the same estimates for change points, even though the second method shows greater sensitivity in $T^2$ values. Also note that at the 5\% level both methods show significant $T^2$ values (greater than 7.8) at observations 11 or 12 for all three data sets, close to the actual change point which was observation 11.

\begin{table}[h]
\caption{The $T^2$ values from the three data sets used in Example 7.4 of Qiu (2014) and in our Example 6.4.2, illustrated as the second nonparametric multivariate SNS test. Also given are the SNS values of each $T^2$ and the estimated change point, based on significance of $T^2$ at the 0.005 level, shown in bold face.}
\label{tab:c6s4t10}
\begin{tabular}{@{}cccccccccc@{}}
\toprule
Obs. & $T^2(a)$ & SNS   & $T(a)$  & $T^2(b)$ & SNS   & $T(b)$ & $T^2(c)$ & SNS   & $T(c)$ \\ \midrule
1    & 1.24      & 0.00  &              & 1.24      & 0.00  &              & 1.24      & 0.00  &              \\
2    & 1.24      & 0.67  & 0.46         & 1.24      & 0.67  & 0.36         & 1.24      & 0.67  & 0.51         \\
3    & 3.48      & 0.97  & 0.16         & 3.48      & 0.97  & 0.02         & 3.48      & 0.97  & 0.23         \\
4    & 0.74      & -1.15 & -0.20        & 0.74      & -1.15 & -0.37        & 0.74      & -1.15 & -0.11        \\
5    & 4.84      & 1.28  & 0.75         & 4.84      & 1.28  & 0.51         & 4.84      & 1.28  & 0.86         \\
6    & 2.08      & 0.21  & 0.24         & 2.08      & 0.21  & -0.01        & 2.08      & 0.21  & 0.37         \\
7    & 1.16      & -0.79 & 0.35         & 1.16      & -0.79 & 0.06         & 1.16      & -0.79 & 0.51         \\
8    & 0.16      & -1.53 & 0.99         & 0.16      & -1.53 & 0.60         & 0.16      & -1.53 & 1.16         \\
9    & 2.93      & 0.59  & 2.01         & 2.93      & 0.59  & 1.44         & 2.93      & 0.59  & 2.21         \\
10   & 2.13      & 0.39  & 1.97         & 2.13      & 0.39  & 1.31         & 2.13      & 0.39  & 2.20         \\
11   & 8.38      & 1.69  & \bf{2.08}         & 0.74      & -1.10 & 1.29         & 8.38      & 1.69  & \bf{2.35}         \\
12   & 8.46      & 1.73  & 1.49         & 8.46      & 1.73  & 1.95         & 8.46      & 1.73  & 1.80         \\
13   & 8.38      & 1.02  & 0.81         & 0.95      & -0.62 & 1.35         & 8.38      & 1.02  & 1.19         \\
14   & 1.74      & -0.27 & 0.51         & 0.74      & -0.79 & 1.84         & 3.67      & 0.46  & 1.00         \\
15   &\bf{13.17}     & 1.83  & 1.44         & 9.45      & 1.83  & \bf{2.47} & \bf{13.17}     & 1.83  & 1.39         \\
16   & 1.36      & -0.40 &              & 2.97      & 0.58  & 1.85         & 1.36      & -0.40 &              \\
17   & 4.24      & 0.46  &              & 4.24      & 0.82  & 1.88         & 4.24      & 0.46  &              \\
18   & 6.48      & 0.67  &              & 6.48      & 1.09  & 1.81         & 6.48      & 0.67  &              \\
19   & 3.79      & 0.27  &              & 5.21      & 0.90  & 1.61         & 3.79      & 0.27  &              \\
20   & 22.06     & 1.96  &              & \bf{22.06}     & 1.96  & 1.65         & 22.06     & 1.96  &              \\
21   & 9.65      & 1.18  &              & 0.84      & -0.79 &              & 9.65      & 1.18  &              \\
22   & 4.21      & 0.17  &              & 4.21      & 0.41  &              & 4.21      & 0.17  &              \\
23   & 5.21      & 0.45  &              & 3.79      & 0.33  &              & 3.83      & 0.11  &              \\
24   & 18.76     & 1.53  &              & 3.83      & 0.37  &              & 18.76     & 1.53  &              \\
25   & 8.46      & 0.84  &              & 6.58      & 1.08  &              & 8.46      & 0.84  &              \\
26   & 8.46      & 0.80  &              & 8.46      & 1.20  &              & 8.46      & 0.80  &              \\
27   & 3.01      & -0.28 &              & 4.32      & 0.48  &              & 3.51      & -0.28 &              \\
28   & 3.79      & -0.09 &              & 3.79      & 0.09  &              & 3.79      & -0.09 &              \\
29   & 9.65      & 1.09  &              & 1.74      & -0.45 &              & 9.65      & 1.09  &              \\
30   & 3.83      & -0.04 &              & 3.79      & 0.13  &              & 3.83      &       &              \\ \bottomrule
\end{tabular}
\end{table}

\end{myexample}

\section{Sequential Normal Scores in Profile Monitoring}\label{sec:c6s5}

Closely related to multivariate SPC is a relatively new area called profile monitoring. Consider a new brand of cake mix, and an experiment is designed to determine how long it should bake in the oven to give the highest quality end product. A batch of cake batter is divided into 10 smaller pans and placed in a pre-heated oven. At 15 minutes one pan is removed from the oven and evaluated on the basis of taste, appearance and quality. We will consider only taste $Y_1$ but a similar analysis can be conducted for appearance ($Y_2$) and for quality ($Y_3$). Every three minutes another pan is removed and evaluated, until all 10 pans have been removed and evaluated. The graph of time in the oven $t$ versus $Y_1$ appears unimodal, with low scores at early times and later times, and a maximum around 27 minutes. Then another batch of cake batter is put into the oven, and the experiment is repeated, for a total of 20 replications. Although there is an effort to keep the evaluation times the same from replication to replication, there is always some variation there, but the exact time of removal from the oven is the time recorded.

We could consider $Y_1$ at each time point as input variables for all 10 time points at each replicate in a multivariate SPC, but we are more interested in the overall profile of Y values, so we fit a parabola $y=a+bt+ct^2$ to the data at each replicate, where $t=time$. The three variables we are interested in are the estimates of the three coefficients in the parabola, $a$, $b$, and $c$. Fitting the data to a parabola is not a statistical problem, it is simply mathematics, and a method such as least squares can be used, or any other good method.

For each of the 20 replicates the values for $a$, $b$, and $c$ are estimated. Each replicate may have slightly different ingredients in the cake batter, and the process is monitored to see if there is a change in the baking profile, as measured by the estimates of $a$, $b$, and $c$. A parametric method involves several assumptions which may be difficult to verify, so we are considering a nonparametric method based on sequential normal scores, as illustrated in Example 6.5.1.

\begin{myexample}
In Example 10.1 on page 409 of \cite{qiu_2014} a linear profile $y = a + bx$ is fit to five values of x for 20 replicates, and estimates of the two coefficients $a$, $b$, and the variance of the residuals are obtained using the least-squares method. The same five values of $x = 0.2, 0.4, 0.6, 0.8$, and $1.0$ are used in each replicate. The resulting five $y$ values for each replicate, and the estimates for $a$, $b$, and $\sigma^2$ are calculated for each replicate. These appear in Table \ref{tab:c6s5t1}.

\begin{table}[t!]
    \caption{The data from Table 10.1 on page 410 in Qiu (2014) and used in his Example 10.1 as well as our Example 6.5.1. The last three columns are the estimates of the intercept, slope, and residual variance from a fit of each replicate to a straight line, and appear in his Figures 10.1(a), (b), and (c).}
    \centering
    \begin{tabular}{crrrrrrrr}
    \textbf{X values} & \multicolumn{1}{c}{\textbf{0.2}} & \multicolumn{1}{c}{\textbf{0.4}} & \multicolumn{1}{c}{\textbf{0.6}} & \multicolumn{1}{c}{\textbf{0.8}} & \multicolumn{1}{c}{\textbf{1}} &       &       &  \\
    \textbf{Replicate} & \multicolumn{5}{c}{\textbf{Y values}} & \multicolumn{1}{c}{\textbf{$\hat{a}$}} & \multicolumn{1}{c}{\textbf{$\hat{b}$}} & \multicolumn{1}{c}{\textbf{$\hat{\sigma}^2$}} \\
    1     & 2.209 & 2.308 & 1.914 & 2.500 & 3.147 & 1.795 & 1.034 & 0.140 \\
    2     & 2.395 & 1.796 & 2.418 & 1.987 & 2.872 & 1.950 & 0.573 & 0.190 \\
    3     & 2.751 & 2.778 & 2.481 & 3.294 & 3.371 & 2.408 & 0.878 & 0.092 \\
    4     & 2.245 & 1.923 & 2.502 & 3.263 & 3.241 & 1.635 & 1.666 & 0.109 \\
    5     & 1.902 & 1.307 & 2.263 & 1.740 & 2.367 & 1.507 & 0.681 & 0.180 \\
    6     & 2.013 & 2.056 & 2.164 & 2.749 & 2.873 & 1.647 & 1.207 & 0.028 \\
    7     & 1.273 & 2.361 & 3.084 & 2.892 & 2.310 & 1.603 & 1.303 & 0.437 \\
    8     & 1.482 & 2.581 & 1.720 & 2.638 & 2.674 & 1.487 & 1.221 & 0.237 \\
    9     & 2.743 & 2.019 & 2.186 & 3.217 & 2.516 & 2.313 & 0.372 & 0.280 \\
    10    & 2.186 & 2.516 & 2.449 & 2.461 & 3.328 & 1.919 & 1.115 & 0.084 \\
    11    & 2.900 & 3.033 & 3.984 & 4.469 & 3.753 & 2.685 & 1.571 & 0.249 \\
    12    & 3.493 & 2.749 & 3.566 & 3.077 & 3.645 & 3.116 & 0.316 & 0.180 \\
    13    & 2.481 & 2.972 & 2.885 & 3.570 & 4.033 & 2.078 & 1.851 & 0.043 \\
    14    & 3.708 & 3.568 & 3.059 & 3.681 & 2.877 & 3.843 & -0.774 & 0.117 \\
    15    & 3.290 & 2.485 & 2.776 & 3.291 & 2.755 & 2.999 & -0.132 & 0.168 \\
    16    & 3.686 & 2.460 & 3.085 & 2.874 & 4.261 & 2.804 & 0.782 & 0.586 \\
    17    & 3.396 & 3.089 & 3.656 & 3.758 & 3.720 & 3.129 & 0.658 & 0.048 \\
    18    & 3.179 & 3.530 & 4.410 & 2.808 & 3.463 & 3.524 & -0.077 & 0.469 \\
    19    & 2.892 & 3.104 & 3.335 & 3.978 & 3.797 & 2.616 & 1.342 & 0.040 \\
    20    & 2.390 & 2.397 & 3.746 & 3.474 & 4.114 & 1.867 & 2.263 & 0.153 \\
    \end{tabular}

    \label{tab:c6s5t1}
\end{table}

A parametric multivariate SPC method is applied in Figure 10.1(d) of Qiu (2014) and sends a strong signal at batch 11 and several subsequent batches. Qiu (2014) reports that the parametric method results in a fit $y = 2 + x$ from in-control observations that are not available in our example. The residual variance is assumed to be 0.25. The only data available to us is the Phase II observations in Table \ref{tab:c6s5t1}, so we will use the self-starting mode of MSNS and look for changes in location and increases in scale. The calculations are the same as in Example 6.4.1. That is, the estimates for $a$, $b$, and $\sigma^2$ are converted to $Y$ values by subtracting reference values and squaring the difference. The reference values for $(a, b, \sigma^2$ ) will be assumed to be (2, 1, 0.25) in agreement with the information in Example 10.1, and in agreement with the initial observations in Figure 10.1 in Qiu (2014) and our Table \ref{tab:c6s5t1}. Those values for Y are transformed to sequential normal scores, and the multivariate $T^2$ is computed as described in Section \ref{sec:c6s4}. The results appear in Table \ref{tab:c6s5t2}.

\begin{sidewaystable}[t!]
    \caption{The transformed variables $Y=(X-\theta)^2$ in Example 6.5.1, their sequential normal scores, the running correlations, and the determinant of the correlation matrix. $T^2$ is computed using the sample correlation matrix for the previous observations, and then converted to sequential normal scores.}
    \centering
    \begin{tabular}{crrrcccccccccc}\textbf{Rep.}
    & \multicolumn{1}{c}{\textbf{$Y(a)$}} & \multicolumn{1}{c}{\textbf{$Y(b)$}}
    & \multicolumn{1}{c}{\textbf{$Y(\sigma^2)$}} & \multicolumn{1}{c}{\textbf{$\text{SNS}_a$}} & \multicolumn{1}{c}{\textbf{$\text{SNS}_b$}} & \multicolumn{1}{c}{\textbf{$text{SNS}_{\sigma^2}$}} & \multicolumn{1}{c}{\textbf{$r_{1,2}$}} & \multicolumn{1}{c}{\textbf{$r_{1,3}$}} & \multicolumn{1}{c}{\textbf{$r_{2,3}$}} &\multicolumn{1}{c}{\textbf{Det.}} & \multicolumn{1}{c}{\textbf{$T^2$}} & \multicolumn{1}{c}{\textbf{SNS}}   & \multicolumn{1}{c}{\textbf{EWMA}} \\
    1     & 0.042 & 0.001 & 0.012 & 0.000 & 0.000 & 0.000 &       &       &       &       &       &       &  \\
    2     & 0.002 & 0.183 & 0.004 & -0.674 & 0.674 & -0.674 & -1.000 & 1.000 & -1.000 &       &       &       &  \\
    3     & 0.167 & 0.015 & 0.025 & 0.967 & 0.000 & 0.967 & -0.810 & 1.000 & -0.810 &       &       &       &  \\
    4     & 0.133 & 0.444 & 0.020 & 0.319 & 1.150 & 0.319 & -0.319 & 1.000 & -0.319 &       &       &       &  \\
    5     & 0.243 & 0.101 & 0.005 & 1.282 & 0.000 & -0.524 & -0.475 & 0.381 & -0.086 & 0.653 &       &       &  \\
    6     & 0.125 & 0.043 & 0.049 & -0.210 & -0.210 & 1.383 & -0.256 & 0.039 & -0.361 & 0.810 & 2.842 & 0.000 & 0.000 \\
    7     & 0.158 & 0.092 & 0.035 & 0.366 & 0.000 & 0.792 & -0.260 & 0.050 & -0.396 & 0.783 & 0.905 & -0.674 & -0.067 \\
    8     & 0.263 & 0.049 & 0.000 & 1.534 & -0.157 & -1.534 & -0.370 & -0.359 & -0.084 & 0.705 & 5.137 & 0.967 & 0.036 \\
    9     & 0.098 & 0.394 & 0.001 & -0.589 & 0.967 & -0.967 & -0.508 & -0.143 & -0.251 & 0.621 & 2.201 & -0.319 & 0.001 \\
    10    & 0.007 & 0.013 & 0.027 & -1.036 & -1.036 & 0.674 & -0.011 & -0.238 & -0.338 & 0.827 & 4.399 & 0.524 & 0.053 \\
    11    & 0.469 & 0.326 & 0.000 & 1.691 & 0.748 & -1.691 & 0.131 & -0.423 & -0.426 & 0.670 & 4.720 & 0.674 & 0.115 \\
    12    & 1.246 & 0.468 & 0.005 & 1.732 & 1.732 & -0.319 & 0.341 & -0.409 & -0.377 & 0.679 & 6.715 & 1.465 & 0.250 \\
    13    & 0.006 & 0.724 & 0.043 & -1.198 & 1.769 & 1.198 & 0.053 & -0.502 & -0.130 & 0.735 & 8.979 & 1.534 & 0.379 \\
    14    & 3.398 & 3.149 & 0.018 & 1.803 & 1.803 & 0.090 & 0.200 & -0.454 & -0.104 & 0.762 & 8.029 & 0.967 & \textbf{0.437} \\
    15    & 0.997 & 1.281 & 0.007 & 0.967 & 1.282 & -0.168 & 0.223 & -0.455 & -0.111 & 0.753 & 2.274 & -0.674 & 0.326 \\
    16    & 0.646 & 0.048 & 0.113 & 0.776 & -0.579 & 1.863 & 0.184 & -0.367 & -0.245 & 0.804 & 7.355 & 0.748 & 0.368 \\
    17    & 1.274 & 0.117 & 0.041 & 1.352 & 0.000 & 0.821 & 0.148 & -0.317 & -0.263 & 0.833 & 3.834 & -0.319 & 0.300 \\
    18    & 2.323 & 1.160 & 0.048 & 1.383 & 1.085 & 1.085 & 0.177 & -0.257 & -0.215 & 0.876 & 6.171 & 0.396 & 0.309 \\
    19    & 0.379 & 0.117 & 0.044 & 0.267 & 0.000 & 0.899 & 0.185 & -0.266 & -0.234 & 0.864 & 1.093 & -1.242 & 0.154 \\
    20    & 0.018 & 1.594 & 0.009 & -0.935 & 1.440 & -0.319 & 0.084 & -0.207 & -0.256 & 0.894 & 3.642 & -0.341 & 0.105 \\
    \end{tabular}
    \label{tab:c6s5t2}
\end{sidewaystable}

For these few observations, the nonparametric SNS method fails to signal a change using $T^2$ or using its sequential normal scores. However the EWMA ($\lambda=0.1$), not shown here, on those SNS shows a weak signal (control limit for in-control ARL of 50 was exceeded) at observation 14. A change point estimate, also not shown here, based on this weak signal estimated a change point at observation 11, in agreement with the parametric method. Follow-up EWMA analyses on the SNS of the Y vector shows a strong (control limit for in-control ARL of 200 was exceeded) signal at observation 17 for $a$ and at observation 14 for $b$. There was no signal of a change for the residual variance. In practice the number of observations will be much larger than we are using here to illustrate the method.

\end{myexample}
\chapter{Summary and Conclusions}\label{ch:c7}

Sequential normal scores provide a useful transformation of data from an unknown distribution to a standard normal distribution. If the original observations are independent and identically distributed, then the sequential normal scores are independent, and approximately standard normal in distribution.

This provides an opportunity for parametric methods, developed to have good theoretical properties, to be adapted to the sequential normal scores for a nonparametric method with good properties. Several examples are provided in this paper of how sequential normal scores can be applied to a wide variety of situations involving sequences of independent random variables, to test for changes in location or scale, or both, in univariate and multivariate data. These methods seem to have good theoretical properties, but that can be the subject of much future research.
\bibliographystyle{plainnat}
\bibliography{mybiblio}

\end{document}